\documentclass[aps,prx,citeautoscript,reprint,showpacs,superscriptaddress,twocolumn,longbibliography]{revtex4-2}

\usepackage{3D_preamble}
\usepackage{outlines}

\usetikzlibrary{external}
\begin{document}

\date{\today}

\title{Entanglement Dynamics in Monitored Kitaev Circuits: \texorpdfstring{\\}{} Loop Models, Symmetry Classification, and Quantum Lifshitz Scaling}

\author{Kai Klocke}
\thanks{These two authors contributed equally.}
\affiliation{Department of Physics, University of California, Berkeley, California 94720, USA}
\author{Daniel Simm}
\thanks{These two authors contributed equally.}
\affiliation{Institut f\"ur Theoretische Physik, Universit\"at zu K\"oln, D-50937 Cologne, Germany}
\author{Guo-Yi Zhu}
\affiliation{Institut f\"ur Theoretische Physik, Universit\"at zu K\"oln, D-50937 Cologne, Germany}
\affiliation{Advanced Materials Thrust, The Hong-Kong University of Science and Technology (Guangzhou), Guangzhou, China}
\author{Simon Trebst}
\author{Michael Buchhold}
\affiliation{Institut f\"ur Theoretische Physik, Universit\"at zu K\"oln, D-50937 Cologne, Germany}

\begin{abstract}
    Quantum circuits offer a versatile platform for simulating digital quantum dynamics and uncovering novel states of non-equilibrium quantum matter. 
    One principal example are measurement-induced phase transitions arising from non-unitary dynamics in monitored circuits, 
    which employ mid-circuit measurements as an essential building block next to standard unitary gates.
    Although a comprehensive understanding of the dynamics in generic circuits is still evolving, we contend that
    monitored quantum circuits give rise to \emph{robust phases of dynamic matter},
    which -- akin to Hamiltonian ground state phases -- can be categorized based on circuit symmetries and spatial dimensionality.
    To illustrate this concept, we focus on measurement-only quantum circuits within symmetry classes BDI and D, 
    which are measurement-only circuit adaptations of the paradigmatic Kitaev and Yao-Kivelson models, 
    embodying particle-hole-symmetric Majorana fermions with or without time-reversal.     
    We establish a general framework -- Majorana loop models -- for both symmetry classes (in arbitrary spatial dimensions) 
    to provide access to the phenomenology of the entanglement dynamics in these circuits, 
    displaying both an area-law phase of localized Majorana loops and a delocalized, highly entangled Majorana liquid phase. 
    The two phases are separated by a continuous transition displaying quantum Lifshitz scaling, 
    albeit with critical exponents of two distinct universality classes.
    The loop model framework provides not only analytical understanding of these universality classes in terms of non-linear sigma models,
    but also allows for highly efficient numerical techniques capable of simulating excessively large circuits with up to $10^8$ qubits. 
    We utilize this framework to accurately determine universal probes that distinguish both the entangled phases and the critical points 
    of the two symmetry classes. Our work thereby further solidifies the concept of emergent circuit phases and their phase transitions.
\end{abstract}

\maketitle

%%%%%%%%%%%%%%%%%%%%%%%%%%%%%%%%%%%%%%%%%%%%%%%%%%%%%%%%%%%%%%%%%%%%%%%%%%%%

%%%%%%%%%%%%%%%%%%%%%%%%%%%%%%%%%%%%%%%%%%%%%%%%%%%%%%%%%%%%%%%%%%%%%%%%%%%%
\section{Introduction}\label{sec:intro}
%%%%%%%%%%%%%%%%%%%%%%%%%%%%%%%%%%%%%%%%%%%%%%%%%%%%%%%%%%%%%%%%%%%%%%%%%%%%

%%Entanglement characterizes phases of matter in and out of equilibrium
Many-body entanglement has emerged as a fundamental concept in understanding and classifying quantum phases of matter \cite{zeng2019quantum}.
In an equilibrium setting, it can serve as a tool to discern ground states of gapped from gapless Hamiltonians, to identify topological order, and to distinguish between generic and non-generic excited states \cite{KitaevPreskill2006,LevinWen2006,Chen2010}. Recently, this notion of many-body entanglement has been extended beyond static Hamiltonian states to characterize dynamically generated quantum states, in particular those arising in monitored quantum circuits. Analogous to Hamiltonian systems, robust \emph{entanglement structures}, such as for instance area-law or volume-law scaling, arise in quantum circuits and can be used to characterize the generated quantum states \cite{Fisher2022reviewMIPT,Potter21review}.

%%Symmetry classification of entanglement phases in monitored circuits
The robustness of these entanglement structures against small variations in microscopic circuit parameters defines \emph{entanglement phases of matter}, akin to phases of matter in thermal equilibrium, which depend only on global properties, such as underlying symmetries, the range of quantum gates and measurements, and the spatial geometry. This perspective of defining entanglement phases in quantum circuits is intricately connected to the concept of universality in statistical mechanics -- stating that macroscopic features of a many-body system are governed by a few global properties, such as symmetries, interaction range, and dimensionality -- and it anticipates a \emph{symmetry classification of entanglement phases} in monitored quantum circuits.

%%Define circuit notion of symmetry and apply to BDI and D, mention topological?
Along these lines it has been recognized that in generic quantum circuits with measurements, the presence or absence of a U$(1)$-symmetry, associated with the conservation of the total particle number or magnetization, dramatically alters the potential phases \cite{Jian2023measurementinduced,Fava2023,Poboiko2023}. However, a comprehensive classification of circuit phases in terms of their symmetries is presently absent. Here we provide a first step in this direction, analyzing two symmetry classes of measurement-only circuits in two spatial dimensions. In addition, we provide a powerful framework for the analytical and numerical analysis of said circuits and symmetry classes in arbitrary spatial dimensions.

\begin{figure}[t] 
   \centering
   \includegraphics{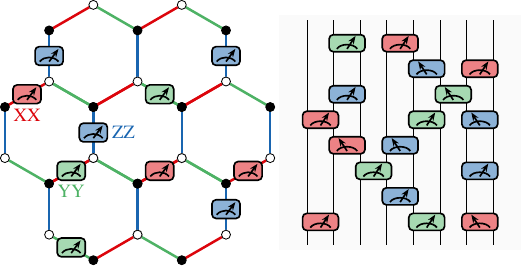}
   \caption{{\bf Monitored Kitaev models.} 
   Left: Spatial geometry of the circuit in which bond-directional $XX$, $YY$, and $ZZ$
   	   parity checks are performed at random with probabilities $K_x$, $K_y$, and 
	   $K_z$, respectively.
   Right: Circuit representation where the qubit parity checks are performed at
   	     discrete time time steps.
	}
   \label{fig:circuit_introduction}
\end{figure}

In a quantum circuit with mid-circuit measurements, \emph{monitored circuit} for short, the global symmetries are inherited from the symmetries of  the combined set of generators of the dynamics, i.e., from all measurement projectors and quantum gates that evolve the wave function.  In this work, we focus on systems characterized by particle-hole symmetry, expressed through operators that are even in Majorana fermions. Specifically, our circuits are  measurement-only circuit analogs \cite{Vijay23Kitaev,Ippoliti23kitaev,Zhu23structuredVolumeLaw}  of the paradigmatic Kitaev model \cite{Kitaev2006} (and variations thereof \cite{YaoKivelson2007,Schmidt10honeycomb}), where frustration arises not from non-commuting Hamiltonian terms but from non-commuting two-qubit parity check measurements, see Fig.~\ref{fig:circuit_introduction} for an illustration. 
In terms of symmetry, these ``Kitaev circuits" adhere to $C^2=1$ particle-hole symmetry and, depending on the underlying spatial geometry of the circuit, fall into two categories with regard to time reversal symmetry. For bipartite spatial geometries (such as the honeycomb geometry) the system exhibits $T^2=1$ time-reversal symmetry and falls within symmetry class BDI of the ten-fold way classification of free-fermion systems \cite{AltlandZirnbauer_1997}.
For non-bipartite spatial geometries (such as the next-nearest neighbor Kitaev~\cite{Kitaev2006} or the decorated honeycomb geometry which Yao and Kivelson have first explored as Hamiltonian model~\cite{YaoKivelson2007}), time-reversal symmetry in the ground state is broken and the system belongs to symmetry class D.
For the Hamiltonian systems, this distinction with regard to time-reversal symmetry has crucial impact on their (spin liquid) ground states. The Yao-Kivelson model, for instance, allows for the formation of a \emph{chiral} spin liquid which is absent in the conventional honeycomb Kitaev model.
Whether a similar distinction arises also for the quantum states stabilized by the corresponding measurement-only circuits in symmetry class BDI and D is a central question that we address in this manuscript.

For both symmetry classes, e.g., the Kitaev honeycomb circuit or its generalizations including next-nearest neighbor couplings or the Yao-Kivelson geometry, we find that when the \emph{measurement-induced frustration} is large, i.e., when non-commuting operators are measured frequently, each circuit creates an entangled Majorana liquid state. It displays a distinctive subsystem entanglement entropy $S(A)\sim L\log(L)$ for a subsystem $A=L\times L$ of linear dimension $L$. In contrast, when the measurement-induced frustration is low, a weakly entangled area-law state is realized. The two phases are separated by a critical line, which displays a characteristic quantum Lifshitz scaling behavior~\cite{Fradkin04RK}, previously associated with quantum dimer models~\cite{Stephan_2012, Stephan_2013}, Dirac fermions and certain $(2+1)$-dimensional conformal field theories (CFTs)~\cite{Inglis_2013,Chen_2015}.
But while the general entanglement phase diagrams appear almost identical for the two systems, their underlying symmetry does manifest itself, primarily at the phase transition between the two principal entanglement phases which fall into distinct universality classes whose critical exponents we determine.

\begin{table}[t]
	\begin{tabular}{c|c|c}
	lattice / circuit 	& bipartite		& non-bipartite \\		
	geometry	& (e.g. honeycomb) & (e.g. Yao-Kivelson) \\
	\hline\hline
	symmetry class 	& BDI 		& D \\
	\hline\hline
        \multicolumn{3}{l}{}\\[-1.7ex]
	\multicolumn{3}{l}{\it loop model} \\[1mm]
	\hline
	loop symmetry					& orientable	& non-orientable \\
	field theory (sigma model)		&	$\mathbb{CP}^{n-1}$		& $\mathbb{RP}^{n-1}$ \\
	\hline\hline
        \multicolumn{3}{l}{}\\[-1.7ex]
	\multicolumn{3}{l}{\it quantum circuit$^\dagger$} : loop model with fugacity $n=1$ \\[1mm]
	\hline
	entanglement scaling& \multicolumn{2}{c}{$\sqrt{\mathcal{D}} \cdot L\log(L)$}	 \\	
	dynamics 	(asymptotic loops)		& $P(\ell)\sim $ const.&  $P(\ell)\sim (\mathcal{L}-\ell)^{-\frac{1}{2}}$\\	
	\hline\hline
        \multicolumn{3}{l}{}\\[-1.7ex]
	\multicolumn{3}{l}{\it quantum Hamiltonian$^*$}: loop model with fugacity $n=\sqrt{2}$ \\[1mm]
	\hline
	entanglement scaling& $L + \log L$ 		& $L-\gamma_{\rm topo}$ \\		
	Majorana spectrum	& gapless Dirac  	& gapped Chern \\
	spin liquid			& Z$_2$		& chiral (Ising TQFT) \\
	\hline
	\end{tabular}
	\label{tab:LoopCircuitDictionary}
	\caption{{\bf Majorana loop models.}
		Summary of a number of key distinctions for Majorana loop modesl on bipartite vs.\ non-bipartite lattice geometries,
		and their realization in quantum circuits or Hamiltonian systems.
		Asterisks: 
		(*) indicates results for {\it isotropically coupled} Kitaev honeycomb and Yao-Kivelson models,
		while ($\dagger$) indicates results for {\it isotropic probabilities}.
		}
\end{table}

To unveil this physics, we introduce \emph{Majorana loop models} in $(d+1)$ dimensions as a joint theoretical framework which covers both the Hamiltonian and measurement cases. These loop models naturally encapsulate the dynamics of Majorana world lines, distinguishing between symmetry classes BDI and D through the {\it orientability} or {\it non-orientability} of world lines~\cite{Klocke_2023}. Moreover, they differentiate between Hamiltonian ground states and monitored systems using an internal parameter -- the loop fugacity (which, in topological terms, is also referred to as $d$-isotopy parameter)~\cite{Fendley_2006}. 
The fugacity $n$ has an intuitive interpretation: it determines whether all world lines appear with equal probability, yielding measurement-only circuits ($n=1$) or whether world line configurations corresponding to high energy states are suppressed, yielding Hamiltonian ground states ($n=\sqrt{2}$)~\cite{Nahum_2011_3D, Nahum_2013_3D, Nahum_2015_DQC}.

%%%%%%%%%%%%%%%%%%%%%%%%%%%%%%%%%%%%%%%%%%%%%%%%%%%%%%%%%%%%%%%%%%%%%%%%%%%%
\begin{figure*}[t]
    \label{fig:Overview}
    \centering
    \includegraphics{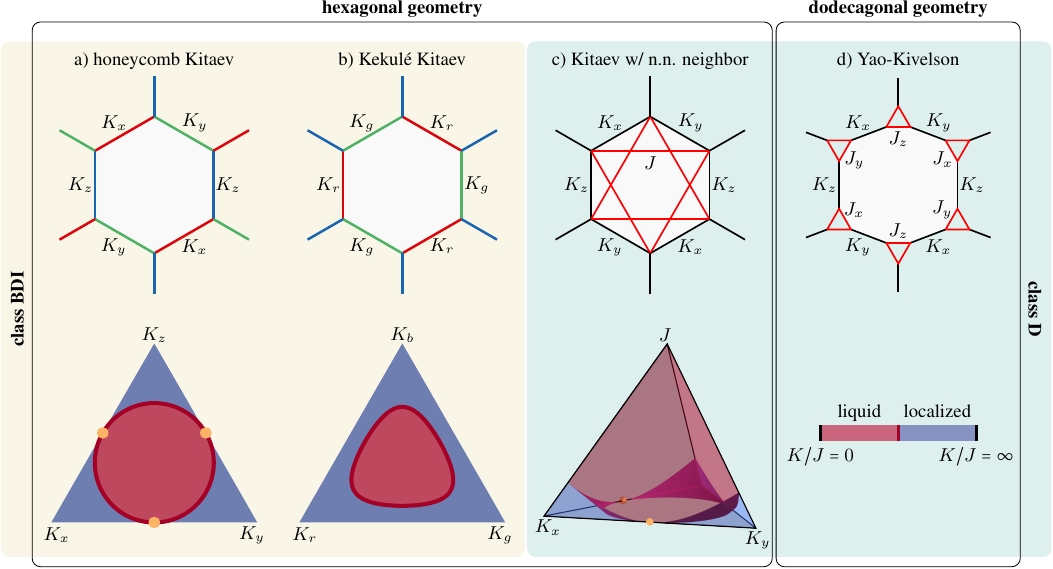}
    \caption{{\bf Kitaev circuits and entanglement phase diagrams.} 
    	The top row shows the family of ``Kitaev circuits" considered in this manuscript. 
	They are all made up of bond-directional two-qubit parity checks $XX$, $YY$, and $ZZ$ sampled
	with probabilities $K_x$, $K_y$, and $K_z$ in varying circuit geometries. 
	From left to right, this is (a) the conventional honeycomb model,
	(b) a Kekul\'e Kitaev model where the parity check assignment of the bonds of the honeycomb lattice forms a Kekul\'e pattern,
	(c) the honeycomb model augmented by next-nearest neighbor parity checks (rendering the lattice geometry non-bipartite), 
	and (d) the Yao-Kivelson circuit on the decorated honeycomb model where every vertex is replaced by a triangle (red) 
	with three different types of bond-directional parity checks.
	The bottom row shows the entanglement phase diagrams as a function of the sampling probabilities of the various $XX$, $YY$, and $ZZ$ parity checks. 
	Localized phases with area-law entanglement are shown in blue, 
	while delocalized phases with liquid-like $L \log(L)$-entanglement scaling are marked in red.
	The dark red contour marks the phase boundary, i.e.\ the quantum Lifshitz-type localization transition between the two principal phases, 
	whose critical behavior we discuss in the manuscript. 
	The orange circles mark special points at which the theory is captured 
	by percolation (in a dimensional reduction).
    }
    \label{fig:overview_circuit_diagram}
\end{figure*}
%%%%%%%%%%%%%%%%%%%%%%%%%%%%%%%%%%%%%%%%%%%%%%%%%%%%%%%%%%%%%%%%%%%%%%%%%%%%

The loop framework gives access to both analytical arguments and efficient numerical simulation techniques for monitored Majorana circuits. This allows for a detailed and thorough analysis of the statistical mechanics of circuits in symmetry classes BDI and D in two spatial dimensions and a detailed exploration of Lifshitz criticality in a setting of free, monitored fermions. It is enabled through new, optimized simulation techniques for loop models, which yield a numerical complexity scaling as $\mathcal{O}(N\log(t))$ for $N\propto L^2$ numbers of qubits and circuit depth $t$. In numerical simulations, system sizes of $10^8$ qubits can be realized, reaching the largest system sizes currently accessible for monitored circuits and raising the standard for the quantitative determination of critical exponents and correlation functions in two spatial dimensions. In addition, we simulate the monitored Kitaev and Yao-Kivelson circuits of interest in this manuscript also in the conventional stabilizer representation. We employ state-of-the-art Clifford circuit simulations techniques \cite{Gidney2021stimfaststabilizer,QuantumClifford} that reach system sizes of $10^4$ qubits and critically analyze what kind of qualitative and quantitative insight they afford.

In summary, we discuss measurement-only Kitaev and Yao-Kivelson models as representatives of symmetry classes BDI and D in monitored circuits. We establish a joint framework for both classes -- Majorana loop models -- and utilize it to accurately determine the long-wavelength properties of a localized area-law and an entangled Majorana liquid phase, and the measurement induced phase transition separating them. This is a rigorous step towards the symmetry classification of  monitored circuits in arbitrary dimensions. In addition, we establish that the measurement-induced phase transition in both symmetry classes is of the quantum Lifshitz-type, uncovering a hitherto unexplored link between loop models in $(2+1)$ dimensions, monitored free fermions, and $(2+1)$-dimensional CFTs.

The paper is organized as follows: Below we provide an overview of the main results for the universal long-distance behavior of both circuits from the loop model and the Clifford simulations. In Sec.~\ref{sec:KitaevModels}, similarities and differences between Hamiltonian and measurement dynamics are discussed, introducing the Kitaev and Yao-Kivelson models both in the Hamiltonian as well as in the measurement-only circuit framework. In Sec.~\ref{sec:loop_models}, the framework of Majorana loop models is introduced for class BDI and class D and for Hamiltonians and circuits in arbitrary dimensions. In addition, we provide a dictionary between the loop framework and circuit observables. The analytical and numerical results for the loop model framework are presented in Sec.~\ref{sec:results}. These are complemented by a selection of observables obtained in high-performance Clifford simulations in Sec.~\ref{sec:clifford_simulations}. A discussion and conclusion is presented in Sec.~\ref{sec:discussion}.

%%%%%%%%%%%%%%%%%%%%%%%%%%%%%%%%%%%%%%%%%%%%%%%%%%%%%%%%%%%%%%%%%%%%%%%%%%%%
\subsection*{Overview of main results}
%%%%%%%%%%%%%%%%%%%%%%%%%%%%%%%%%%%%%%%%%%%%%%%%%%%%%%%%%%%%%%%%%%%%%%%%%%%%
The measurement-induced dynamics of monitored circuits (discussed in this manuscript) share several similarities with the energetics-induced dynamics of their Hamiltonian counterparts.
In both cases, frustration is induced by the presence of non-commuting operators and the statistical mechanics of the stationary state can be mapped to a suitable $(2+1)$-dimensional loop model. While, however, the physics of Kitaev and Yao-Kivelson Hamiltonians is typically understood in terms of their Majorana band structure, their circuit counterparts feature a different link to many-body quantum physics. As a guiding picture, the Majorana loop model representation of both the measurement-only Kitaev and the Yao-Kivelson model in $d$ spatial dimensions anticipate a link between the measurement-only circuits and the physics of diffusive \emph{Majorana metals} -- highly entangled liquid phases of Majorana fermions undergoing Brownian motion on a lattice -- and \emph{disorder-induced localization} in $(d+1)$ dimensions. The transition between the liquid and the localized phase is induced by the degree of measurement-induced frustration, which we will use as an organizing principle in the following.

%%%%%%%%%%%%%%%%%%%%%%%%%%%%%%%%%%%%%%%%%%%%%%%%%%%%%%%%%%%%%%%%%%%%%%%%%%%%
\subsubsection*{\texorpdfstring{Large measurement-induced frustration:\\ entangled Majorana liquid state}{Large measurement-induced frustration}}
%%%%%%%%%%%%%%%%%%%%%%%%%%%%%%%%%%%%%%%%%%%%%%%%%%%%%%%%%%%%%%%%%%%%%%%%%%%%

When all allowed bond operators of nearest neighbor qubits (or free Majorana fermions) in the honeycomb or decorated honeycomb lattice are measured with equal probability, i.e., at the isotropic point in the phase diagram, the measurement-induced frustration is maximal. This pushes the wave function into an entangled state with the subsystem entanglement of an $A=L\times L$ square growing as 
	\[S(A)\sim L\log(L)\,,\] 
characteristic of quantum liquids with a nodal (Fermi) surface in two dimensions \cite{Wolf2006,Klich2006,Li2006,Swingle2010}. The liquid state is robust against small variations in the circuit parameters, i.e., away from the point of maximum frustration, defining an extended highly entangled phase at large measurement frustration, as illustrated in the phase diagrams of Fig.~\ref{fig:overview_circuit_diagram}.

For measurements drawn randomly in space-time, this liquid phase is reminiscent of diffusive metals arising in weakly disordered Fermi systems. This picture is strengthened in the loop model framework: large measurement-frustration causes the endpoints of Majorana loops to perform an isotropic random walk in $(d+1)$-dimensional space-time. This yields an entanglement structure akin to a disordered metal with diffusion constant $\mathcal{D}$, see Sec.~\ref{sss:diffusion_coeff}.

%%%%%%%%%%%%%%%%%%%%%%%%%%%%%%%%%%%%%%%%%%%%%%%%%%%%%%%%%%%%%%%%%%%%%%%%%%%%
\begin{table*}[t]
    \centering
    \begin{tabular}{p{30mm}|p{27mm}|p{27mm}|p{27mm}|p{27mm}|p{30mm}}
        {\bf critical exponent}                       & \multicolumn{1}{c|}{$\nu$}                             &  \multicolumn{1}{c|}{$\eta$ }            &  \multicolumn{1}{c|}{$d_f$} &  \multicolumn{1}{c|}{$\beta$} &  \multicolumn{1}{c}{reference} \\[1mm]
        \hline\hline
        \multicolumn{5}{l}{}                                                                                             \\[-1.7ex]
        \multicolumn{2}{l}{{\it loop model} (this work)}            & \multicolumn{4}{r}{$10^8$ qubits}                                \\[1mm]
        \hline
        BDI / orientable				& $\quad 0.9987 \pm 0.0007$  &  $\quad -0.084 \pm 0.004$ & $\quad 2.5383 \pm 0.0011$& $\quad 0.4590 \pm 0.0020$ & \multicolumn{1}{c}{Figs.~\ref{fig:HC_span_FSS}, \ref{fig:HC_fractal_dim}, \ref{fig:HC_span_length}, \ref{fig:critical_watermelon}} \\
        D / non-orientable 			&  $\quad 0.9403 \pm 0.0006$  &  $\quad -0.066 \pm 0.007$ & $\quad 2.5263 \pm 0.0011$  & $\quad 0.4400 \pm 0.0040$  & \multicolumn{1}{c}{Figs.~\ref{fig:HC_span_FSS}, \ref{fig:HC_fractal_dim}, \ref{fig:HC_span_length}, \ref{fig:critical_watermelon}} \\
        \hline
        \multicolumn{5}{l}{}                                                                                             \\[-1.7ex]
        \multicolumn{2}{l}{{\it loop model} (literature)} & \multicolumn{4}{r}{$10^6$ sites}                                 \\[1mm]
        \hline
        BDI / orientable	                         & $\quad 0.999\pm 0.002$                  &  $\quad -0.068 \pm 0.018^* $     & $\quad 2.534 \pm 0.009$     & $\quad 0.4650 \pm 0.0033^\dagger$ & \multicolumn{1}{c}{Ref.~\cite{Ortuno_2009}} \\
        D / non-orientable			& $\quad 0.918 \pm 0.005$                & $\quad -0.091 \pm 0.009$ & $\quad 2.546\pm 0.005^*$ & $\quad 0.4168 \pm 0.0051^\dagger$ & \multicolumn{1}{c}{Ref.~\cite{Serna_2021}}  \\
        \hline\hline
        \multicolumn{5}{l}{}                                                                                             \\[-1.7ex]
        \multicolumn{2}{l}{{\it Clifford simulations} (this work)}  & \multicolumn{4}{r}{$10^4$ qubits}                                \\[1mm]
        \hline
        BDI / orientable				& $\quad\,\,\,\, 0.96 \pm 0.01$                   & \multicolumn{1}{c|}{---} &  \multicolumn{1}{c|}{---} &  \multicolumn{1}{c|}{---} &  \multicolumn{1}{c}{Fig.~\ref{fig:CliffordExponents}} \\
        D / non-orientable			& $\quad\,\,\,\, 0.91 \pm 0.02$                   &  \multicolumn{1}{c|}{---} &  \multicolumn{1}{c|}{---} &  \multicolumn{1}{c|}{---} & \multicolumn{1}{c}{Fig.~\ref{fig:CliffordExponents}} \\
        \hline\hline
    \end{tabular}
    \caption{{\bf Critical exponents.} 
    Shown are the critical exponents of the localization transition in symmetry class BDI and D, 
    comparing results from loop model simulations (with up to $10^8$ qubits) to previous results in the literature \cite{Ortuno_2009,Serna_2021} 
    (with up to $10^6$ qubits) as well as results from Clifford simulations (with up to $10^4$ qubits).
    The correlation length exponent $\nu$ is extracted from spanning number (see Fig.~\ref{fig:HC_span_FSS}). 
    The anomalous dimension $\eta$ is extracted from the scaling of the length of spanning loops (see Fig.~\ref{fig:HC_span_length})
    and, alternatively, from the scaling of watermelon correlators (see Fig.~\ref{fig:critical_watermelon}). 
    The fractal dimension $d_f$ is extracted from the bulk loop lengths (see Fig.~\ref{fig:HC_fractal_dim}). 
    The critical exponent $\beta$ is calculated from the watermelon correlator in Fig.~\ref{fig:critical_watermelon}. 
    Refs.~\cite{Ortuno_2009,Serna_2021} have determined \emph{either} the anomalous or fractal dimension and calculated the other
    via the scaling relation  $\eta = 5-2d_f$, indicated by the asterisk (*). The critical exponent $\beta$ in both works was obtained via the hyperscaling relation $\beta =  \nu(3-d_f)$, indicated by the asterisk ($\dagger$). We note that the literature values for class D deviate from the ones obtained in this work. We analyze this discrepancy and confirm our results by comparing different scaling ansätze and lattice geometries in Appendix~\ref{app:nonorientable_geometries}.
    The critical exponent $\nu$ from Clifford simulations is obtained from a scaling collapse of the tripartite mutual information $I_3$ in Fig.~\ref{fig:CliffordExponents}. A systematic deviation from the more accurate loop model exponents towards smaller values is observed. 
    }
    \label{tab:crit_exp}
\end{table*}
%%%%%%%%%%%%%%%%%%%%%%%%%%%%%%%%%%%%%%%%%%%%%%%%%%%%%%%%%%%%%%%%%%%%%%%%%%%%
The long-distance properties of the random walkers are governed by a non-linear sigma model \nlsm in class $\mathbb{CP}^{n-1}$ for symmetry class BDI or $\mathbb{RP}^{n-1}$ for symmetry class D, see Eq.~\eqref{eq:NonLinS} below. The replica limit is determined by the fugacity $n$ of the underlying loop model, i.e., $n=1$ for the circuit. 
Both \nlsms are \emph{single-parameter theories} and depend only on the diffusion constant $\mathcal{D}$, which consequently determines the behavior of physical properties at large distances. For instance, the entanglement entropy in the Majorana liquid phase is 
\[
	S(A)\sim\sqrt{\mathcal{D}}\cdot L\log(L)\,.
\] 
The effective diffusion constant $\mathcal{D}$ at large distances emerges from the \nlsm under renormalization group transformations. Deep in the liquid phase, the renormalization of $\mathcal{D}$ is negligible and we find reasonable agreement when approximating $\mathcal{D}=\mathcal{D}_{\text{mic}}$, where $\mathcal{D}_{\text{mic}}$ is the microscopic diffusion constant from the measurement probabilities, see Fig.~\ref{fig:diffusion_vs_c}(a). 

We show that the physical properties in the circuit, such as, e.g., entanglement, mutual information, and correlation functions, can be inferred from the distribution function $P(\ell)$ of Majorana loops of length $\ell$ in $(d+1)$-dimensional space-time. For a two-dimensional system with linear dimensions $L_x, L_y$ in the $x$- and $y$-direction, this distribution separates into two parts: a \emph{Brownian regime} for $\ell<L_xL_y$, and a \emph{Poisson-Dirichlet tail} at space-time volume filling distances $\ell>L_xL_y$, see Fig.~\ref{fig:HC_BulkLoops}. The {Brownian regime} features a loop distribution $P(\ell)\sim \ell^{-\frac{5}{2}}$, which does not distinguish between symmetry class BDI and D. In contrast, the Poisson-Dirichlet tail differentiates between both symmetry classes: it displays an asymptotic behavior $P(\ell) \sim \ell^{-1}(\mathcal{L}-\ell)^{\theta-1}$, where $\mathcal{L}$ is the maximum distance of a single loop in space-time and $\theta=1$ for class BDI and $\theta=\tfrac12$ for class D, respectively. Both symmetry classes thus realize space-time filling loops in the Majorana liquid phase. The tails of the distribution are not detectable at the space time boundary and therefore do not discriminate different wave functions at final times. Instead, they reflect different dynamics in the circuit, which can be detected by out-of-time-ordered correlators (OTOCs), observable via circuit-ancilla measurement schemes, see Fig.~\ref{fig:ancilla_scheme}. Such measurements allow to universally discriminate the dynamics of Majorana liquids in different symmetry classes. 

%%%%%%%%%%%%%%%%%%%%%%%%%%%%%%%%%%%%%%%%%%%%%%%%%%%%%%%%%%%%%%%%%%%%%%%%%%%%
\subsubsection*{\texorpdfstring{Small measurement-induced frustration:\\ localized  Majorana world lines}{Small measurement-induced frustration}}
%%%%%%%%%%%%%%%%%%%%%%%%%%%%%%%%%%%%%%%%%%%%%%%%%%%%%%%%%%%%%%%%%%%%%%%%%%%%
For small measurement-induced frustration, i.e., when a large fraction of measurements can be satisfied simultaneously, the Majorana world lines rarely undergo a measurement-induced motion. Instead, they remain localized in space with an average localization length $\xi$, akin to disorder-induced localization of a Fermi liquid. The loop distribution $P(\ell)$ thus acquires an exponential cutoff, $P(\ell)\sim \ell^{-\frac{5}{2}}\exp(-\ell/\xi)$, suppressing the buildup of correlations and entanglement over distances larger than $\xi$. Consequently, the subsystem entanglement of a contractible region $L\times L$ obeys an area law \[S(A)\sim L\log\xi - \ln 2 + \cdots\] on distances $L>\xi$. 
The negative subleading term is a universal topological correction contributed by the gauge flux, as the complementary part of the fractionalized degrees of freedom from the qubits. Such a topological state is in the same universality class as the toric code wavefunction~\cite{Kitaev2003}, and the circuit can potentially serve as a (dynamical) quantum error correction code -- when the frustration is tuned towards zero, the state resembles an instantaneous state of the Hastings-Haah Floquet code~\cite{Haah21honeycomb,Hastings22honeycomb, Nat23CSSfloquetcode, Zhu23qubit}.

%%%%%%%%%%%%%%%%%%%%%%%%%%%%%%%%%%%%%%%%%%%%%%%%%%%%%%%%%%%%%%%%%%%%%%%%%%%%
\subsubsection*{\texorpdfstring{Critical measurement-induced frustration:\\ localization-delocalization phase transition}{Critical measurement-induced frustration}}
%%%%%%%%%%%%%%%%%%%%%%%%%%%%%%%%%%%%%%%%%%%%%%%%%%%%%%%%%%%%%%%%%%%%%%%%%%%%
The Majorana liquid and the area-law phase are separated by a phase transition at a line of critical measurement-frustration. In the Majorana loop framework, this is a localization-delocalization transition, which takes place when the diffusion constant in the non-linear sigma model renormalizes to zero. Precisely at the transition, the Majorana fermion entanglement obeys an area law but with a peculiar subsystem-dependence that reveals quantum Lifshitz criticality. Despite obeying an area law, correlation functions and the loop distribution function are not bound by a finite correlation length but display algebraic, scale invariant decay, which is faster than the conventional $\sim 1/(\text{distance})^2$. 

The fact that the \nlsm is a single-parameter theory, with the only microscopic parameter entering being the microscopic diffusion constant $\mathcal{D}_{\text{mic}}$, allows us to identify the location of the critical line of the phase transition. It is located at $\mathcal{D}_{\text{mic}}=\mathcal{D}_{\text{mic},c}$ for some critical value $\mathcal{D}_{\text{mic},c}$.
In symmetry class BDI, we can extract the value of $\mathcal{D}_{\text{mic},c}=3/16$ exactly from the high symmetry points of the phase diagram (the yellow circles in Fig.~\ref{fig:overview_circuit_diagram}(a)). This enables us to analytically determine, \emph{without any free parameter}, the phase boundary in the Kitaev honeycomb geometry, see Fig.~\ref{fig:diffusion_vs_c}(b), and, tracing the same value of $\mathcal{D}_{\text{mic},c}$, also allows us to determine the phase boundary of the  Kekulé Kitaev model, see Fig.~\ref{fig:diffusion_vs_c}(c).
This connection to the universal value of $\mathcal{D}_{\text{mic},c}$ in these two BDI models not only illustrates the universality of the loop framework, but it also explains the previously obtained 
circular structure of the phase diagram for the honeycomb geometry \cite{Vijay23Kitaev,Ippoliti23kitaev,Zhu23structuredVolumeLaw}.

The universal critical behavior at long wavelengths is determined by the $\mathbb{CP}^{n-1}$ ($\mathbb{RP}^{n-1}$) non-linear sigma model for symmetry class BDI (D) with the replica limit $n\to1$ in $d+1$ dimensions. The corresponding universality classes have been discussed in the context of dense (fully-packed) polymers~\cite{Jacobsen_1998, Jacobsen_1999, Candu_2009}, localization in gapless spin-singlet superconductors with disorder~\cite{Ortuno_2009}, quantum magnets and spin-ice systems~\cite{Jaubert_2011, Khemani_2012, Chern_2013, Albuquerque_2012, Chern_2013_colloid}.  We numerically extract four different universal exponents using both the loop model framework with simulations of up to $10^8$ qubits and, using the stabilizer framework, for Clifford circuits with up to $10^4$ qubits. These results are summarized in Tab.~\ref{tab:crit_exp}, where we also compare our estimates with previous loop model simulations of both symmetry classes (which reached system sizes equivalent to $10^6$ qubits~\cite{Serna_2021}).

The critical exponents and their interpretation in the loop model and in the quantum circuit can be summarized as:
\begin{itemize}
    \item The correlation length exponent $\nu$ determines the average loop length $\langle \ell \rangle$ 
    and the spanning number between the final and the initial state in the loop framework. 
    In the circuit, the former sets the average entanglement entropy in the localized phase 
    and the latter controls the late-time residual entropy during dynamical purification, see Fig.~\ref{fig:HC_span_FSS}.
    \item The anomalous dimension $\eta$ describes the power-law scaling of loop correlations in the \emph{bulk} of space-time. 
    	In particular, for the two-point watermelon correlator we have
	\begin{equation} 
              G_2(r) \sim\left\{ \begin{array}{ll}\exp(-r/\xi) & \text{localized phase} \\ r^{-(1+\eta)}&\text{critical point} \\
             \text{const.}    & \text{liquid phase}\end{array}\right. .
          \end{equation}
          In the circuit this corresponds to the scaling of a two-point OTOC, see Fig.~\ref{fig:critical_watermelon}. 
          Alternatively, the anomalous dimension can be extracted through the hyperscaling relation $\tau = (11-\eta)/(5-\eta)$ 
          through the Fisher exponent $\tau$. 
          The latter determines the loop length in $(d+1)$-dimensional space time  $P(\ell) \sim \ell^{-\tau}$, see Figs.~\ref{fig:HC_fractal_dim} and 
          \ref{fig:HC_span_length}.
    \item The fractal dimension $d_f$, which is related to $\eta,\tau$ by hyperscaling relations $d_f = (5-\eta)/2=3/(\tau-1)$, 
    describes the structure of the loops in space-time. As with $\eta$, these correlations are reflected in circuit OTOCs in the bulk of space-time. 
    See Fig.~\ref{fig:HC_fractal_dim} for the scaling at the transition and Figs.~\ref{fig:HC_BulkLoops}, \ref{fig:open_loops_FBC} 
    for scaling in the critical phase.
    \item The order parameter exponent $\beta$ determines the probability that a given space-time point falls on a macroscopic loop. 
    Analogous to quantum magnets, extensive loops correspond to macroscopic ordering of the spins, e.g., in a ferromagnet. 
    This is determined by the two-leg watermelon correlator in the limit of large separations, $G_2(r \gg 1, \Delta) \propto \Delta^{2\beta}$, 
    where $\Delta$ is the distance to the critical point in the liquid phase, see Fig.~\ref{fig:critical_watermelon}. 
    This determines circuit OTOCs at long space-time distances.
\end{itemize}

Together with the universal OTOC ratios in the Majorana liquid phases, this provides a complete characterization of the two symmetry classes of measurement-only circuits.

%%%%%%%%%%%%%%%%%%%%%%%%%%%%%%%%%%%%%%%%%%%%%%%%%%%%%%%%%%%%%%%%%%%%%%%%%%%%
\section{Kitaev models, circuits, and symmetries}
\label{sec:KitaevModels}
%%%%%%%%%%%%%%%%%%%%%%%%%%%%%%%%%%%%%%%%%%%%%%%%%%%%%%%%%%%%%%%%%%%%%%%%%%%%

To set the stage we briefly review the fundamental traits of Kitaev physics -- the fractionalization of quantum mechanical degrees of freedom and the subsequent formation of long-range entanglement~\cite{Kitaev2006,Hermanns2018}. We do this first in the context of the well-known honeycomb Kitaev model, i.e.\ in the Hamiltonian model in the parlance of this manuscript, where the ground states are quantum spin liquids with a characteristic and defining entanglement structure. 
In a second step we turn to the quantum circuit variant of interest in this manuscript. We then turn to an elementary symmetry classification of these systems and the conclusions one can draw from this with regard to the formation of long-range entanglement.

\begin{figure}[t] 
   \centering
   \includegraphics{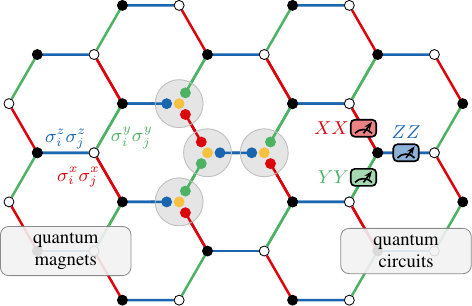}
   \caption{{\bf From quantum magnets to quantum circuits.} 
   Left: For quantum magnets, the Kitaev Hamiltonian with its characteristic bond-directional Ising-like exchange terms 
   is a paradigmatic spin model that is exactly solvable and exhibits gapless and gapped quantum spin liquid ground states.
   The analytical solution (for any trivalent lattice geometry) employs a Majorana fermionization that faithfully captures 
   the fractionalization of elementary spin-1/2 (or qubit) degrees of freedom into itinerant Majorana fermions (yellow) and
   static \bbZ gauge fields (represented here as Majorana bilinears on the bonds).
   Right: In its measurement-only quantum circuit adaptation, the Kitaev model consists of bond-directional 
   two-qubit parity checks whose non-commutativity induces frustration and, starting from a random product state, 
   dynamically stabilizes an ensemble of pure quantum states with a typical entanglement structure.}
   \label{fig:model}
\end{figure}

%%%%%%%%%%%%%%%%%%%%%%%%%%%%%%%%%%%%%%%%%%%%%%%%%%%%%%%%%%%%%%%%%%%%%%%%%%%%
\subsection{Kitaev physics}
%%%%%%%%%%%%%%%%%%%%%%%%%%%%%%%%%%%%%%%%%%%%%%%%%%%%%%%%%%%%%%%%%%%%%%%%%%%%

%%%%%%%%%%%%%%%%%%%%%%%%%%%%%%%%%%%%%%%%%%%%%%%%%%%%%%%%%%%%%%%%%%%%%%%%%%%%
\subsubsection*{Hamiltonian model}
%%%%%%%%%%%%%%%%%%%%%%%%%%%%%%%%%%%%%%%%%%%%%%%%%%%%%%%%%%%%%%%%%%%%%%%%%%%%
The Kitaev Hamiltonian can be defined on any lattice geometry that allows for a tricolorization of its bonds. This is naturally the case for \emph{trivalent} lattice geometries such as the honeycomb lattice, but can also be expanded to more complicated lattice geometries with higher coordination number 
\cite{Kimchi2014,OBrien2016,Hermanns2018}. 
On each bond, an Ising-like interaction couples two neighboring spin-1/2 degrees of freedom along the $X, Y, Z$ directions (in spin space) 
associated with the bond's color
\begin{equation}
    H = \sum_{\langle jk\rangle_{\mu}} K_{\mu} \sigma_j^\mu\sigma_k^\mu \,,
    \label{eq:Hamiltonian}
\end{equation}
where $\mu =x,y,z$ indicates the easy-axis of the bond-directional exchange.

For any tricoordinated lattice geometry, this model can be \emph{exactly solved} via rewriting the model in a Majorana basis~\cite{Kitaev2006}.
To this end, each spin-1/2 is mapped onto four Majorana fermions $\{b^x, b^y, b^z, c\}$ satisfying a gauge constraint $b^x b^y b^z c = 1$, see also the illustration in Fig.~\ref{fig:model} above and Fig.~\ref{fig:mapping} below. This new basis allows us to reexpress the Pauli operator as Majorana bilinears $\sigma_j^\mu=ib_j^\mu c_j^{\phantom \mu}$. The bond-dependent Ising interaction is then equivalent to a gauged Majorana hopping bilinear $\sigma_j^\mu\sigma_k^\mu =  i u_{kj}c_jc_k $, where $u_{kj}\equiv i b_k^\mu b_j^\mu=\pm 1$ denotes the \bbZ gauge connection on the bond.
Note that the gauge field $u$ is not a gauge-invariant physical observable, but instead one needs to consider the Wilson loop plaquette operator, as a product of the Ising interactions surrounding a plaquette, which evaluates the gauge flux penetrating the plaquette. Importantly, these \bbZ gauge fields remain entirely static and assume a fixed low-energy configuration (favored by a Majorana fermion mediated interaction between the gauge fluxes)
that can typically be predicted via Lieb's theorem \cite{Lieb_1994}.
We will discuss this gauge physics in more detail below for the two principal lattice geometries of our manuscript.

The virtue of this reformulation, which at first sight might be perceived as simple operator algebra, is that it faithfully captures the low-energy physics
of the Hamiltonian model: The spin-1/2 degrees of freedom fractionalize into an itinerant Majorana fermion coupled to a static \bbZ lattice gauge theory in the background --  key characteristics of quantum spin liquid ground states \cite{Savary2017}. The precise nature of these spin liquids depend on the underlying lattice geometry and the relative strength of couplings, but generally include gapless, gapped, and chiral spin liquids which are accompanied by long-range entanglement structures \cite{Hermanns2018}. In fact, it is this formation of long-range entanglement in a quantum magnet that is the defining characteristic of a quantum spin liquid state and allows to unambiguously distinguish them from ordered ground states or finite-temperature paramagnetic states \cite{zeng2019quantum}.

%%%%%%%%%%%%%%%%%%%%%%%%%%%%%%%%%%%%%%%%%%%%%%%%%%%%%%%%%%%%%%%%%%%%%%%%%%%%
\subsubsection*{Quantum circuits}
\begin{figure}[t]    
	\centering
    \includegraphics{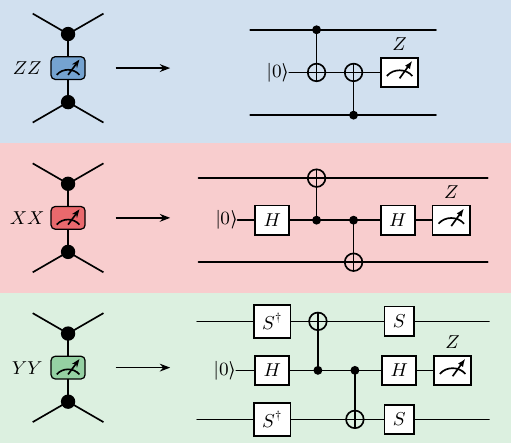}
   \caption{{\bf Building blocks of two-qubit parity measurements.} 
   	The three panels correspond to parity measurements of $ZZ$ (blue), $XX$ (red), and $YY$ (green), respectively. 
	The most important ingredient is an ancilla qubit $\ket{0}$ which,
	entangled via two CNOT gates to the two qubits involved in the parity check,
	is the only qubit that is actually measured (and collapsed). Note that for all three parity checks, the ancilla qubit is initialized in a 
	$Z$ eigenstate and measured in the $Z$ basis.
	Depending on the parity check, the two-body entangling CNOT gates are sandwiched 
	by one-body Hadamard gates $H$ or a phase gate $S^\dagger=\sqrt{Z}$ that rotates the Pauli basis. 
	Note that any circuit built from these building blocks only relies on Clifford gates.
	}   
\label{fig:circuit_buildingblocks}
\end{figure}

%%%%%%%%%%%%%%%%%%%%%%%%%%%%%%%%%%%%%%%%%%%%%%%%%%%%%%%%%%%%%%%%%%%%%%%%%%%%
In a new twist on Kitaev physics, several groups have recently started to look into monitored Kitaev models \cite{Vijay23Kitaev,Ippoliti23kitaev,Zhu23structuredVolumeLaw} or ``Kitaev circuits". 
In these circuits one replaces the bond-directional exchange terms of the Hamiltonian by bond-directional joint two-qubit measurements, also known as parity checks. These are illustrated in Fig.~\ref{fig:circuit_buildingblocks}, along with their circuit implementation using an ancilla qubit.
On a theoretical level, these bond-directional parity checks implement a non-unitary Kraus operator
\begin{equation}
	 M_{jk} \equiv \exp\left(-\frac{\tau}{2}s^{\phantom\mu}_{jk}\sigma_j^\mu\sigma_k^\mu\right) / \sqrt{2 \cosh(\tau)}\ ,
	\label{eq:KrausOp}
\end{equation}
where $\mu = x, y, z$ indicates the type of bond-directional parity check, $s=\pm 1$ is the measurement outcome of the ancilla qubit, 
and $\tau\in[0,+\infty)$ characterizes the strength of measurement that is controlled by a unitary entangling gate parameter $t\in[0,\pi/4]$: 
$\tanh(\tau/2)=\tan(t)$. 
The entangling CNOT gates of Fig.~\ref{fig:circuit_buildingblocks} correspond to $t = \pi/4$ rotations, resulting in strong, projective measurements.
Here we focus on this Clifford regime of projective measurements and {\it randomly sample} the two-qubit parity checks in every microstep without evoking any spatio-temporal ordering (see Fig.~\ref{fig:circuit_introduction} above).
On a global scale one thereby transitions from (continuous) Hamiltonian dynamics where each exchange term is a projector onto a single, energetically favored two-spin state to a (discrete) \emph{measurement-only dynamics} \cite{Lang2020,Vedika2021measure,Barkeshli2021measuretoriccode,Buchold22measurespt} where each projective measurement results in two possible, random outcomes.

%%%%%%%%%%%%%%%%%%%%%%%%%%%%%%%%%%%%%%%%%%%%%%%%%%%%%%%%%%%%%%%%%%%%%%%%%%%%
%%%%%%%%%%%%%%%%%%%%%%%%%%%%%%%%%%%%%%%%%%%%%%%%%%%%%%%%%%%%%%%%%%%%%%%%%%%%

%%%%%%%%%%%%%%%%%%%%%%%%%%%%%%%%%%%%%%%%%%%%%%%%%%%%%%%%%%%%%%%%%%%%%%%%%%%%
\begin{figure*}[t]
    \centering
    \includegraphics{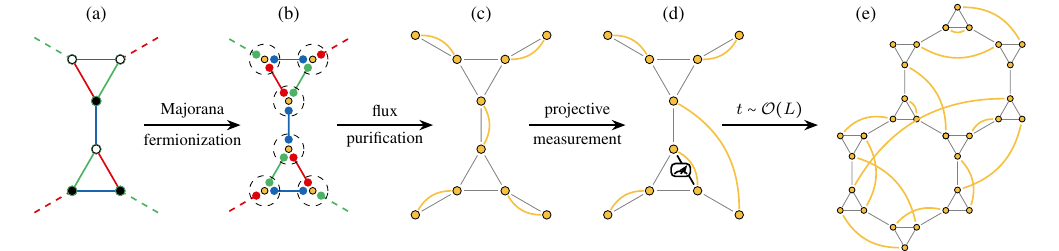}
    \caption{\textbf{Majorana fermions, measurement-induced teleportation, and long-range entanglement in Kitaev circuits.}
    (a,b) Each elementary qubit is represented by a four Majorana fermions. 
    (c) The pairings of the $b_i^\alpha$ Majoranas are fixed after a short purification time. 
    Plaquette operators are stabilized by the random protocol and commute with all following measurements, effectively allowing a decomposition $\rho_f(t) = \rho_c(t) \otimes \rho_b$ (up to a global gauge constraint), where $\rho_b = \Ket{\psi_b}\Bra{\psi_b}$ 
    with $ib_{i_1}b_{i_2} \Ket{\psi_b} = \Ket{\psi_b}$ -- thus only the $c$ Majoranas (orange circule) are involved in the subsequent dynamics.
    (d) A projective measurement (here: $(1+X_i X_j)/2$ on the lower triangular bond) fixes the parity $ic_i c_j = 1$ of the Majorana fermions involved; 
    global parity conservation forces $ic'_i c'_j = 1$ for the now unpaired Majoranas 
    thereby effectively inducing Majorana teleportation \cite{Nayak2008measurement}.
    (e) After long times, $t \sim \bigO(L)$, the state can exhibit long-range entanglement, 
    which is captured by the fixed parities of Majorana pairs far from each other (arising from measurement-induced long-distance teleportation).}
    \label{fig:mapping}
\end{figure*}
%%%%%%%%%%%%%%%%%%%%%%%%%%%%%%%%%%%%%%%%%%%%%%%%%%%%%%%%%%%%%%%%%%%%%%%%%%%%

While this might sound like a somewhat obscure idea, one should be reminded that such a connection between Hamiltonian and measurement-only circuit is also deeply embedded in the concept of the toric code~\cite{Kitaev2003} -- it points to the implementation of a topological quantum memory via 
stabilizer measurements, i.e.\ a measurement-only quantum circuit that implements rounds of four-qubit measurements 
(whose outcome is interpreted as a syndrome and fed into a decoder to perform quantum error correction). 
Of course, the toric code is also very well known in its Hamiltonian version which serves as an exactly solvable spin model that gives rise to a topological spin liquid ground state. 
A key distinction between the toric code and the monitored Kitaev models of interest here is the commutativity of operators. 
While all Hamiltonian terms/measurement operators commute with one another for the former, they do not for the latter. 
This elevated level of measurement-induced frustration has important consequences on the entanglement structures 
that can be stabilized in these circuits as we have discussed in the introduction and which, in contrast to the toric code, 
are also distinct from the entanglement in the ground states of the corresponding Hamiltonians. 

Returning to the monitored Kitaev models at hand, it is important to note that running such a Kitaev circuit on a given initial state 
will give rise to a multitude of possible trajectories, depending on the sequence of measurement outcomes. 
It is the ensuing \emph{ensemble of pure states},
created by running a Kitaev circuit multiple times, that is the object of interest in the following. It allows to
define (in a statistical sense) typical states with a characteristic entanglement structure which we wish to classify
and compare to the ground-state entanglement stabilized by Hamiltonian dynamics 
(see Appendix \ref{app:hamiltonian_phase_diag} for some Hamiltonian phase diagrams).

To understand the formation of long-range entanglement in the circuit setting, one can again rely on the language of fractionalization
which we have introduced above in the context of the exact solution of the Hamiltonian model in a Majorana basis. To do so, let us
perform a completely analogous rewriting of the 2-level qubit states in terms of four Majorana fermions and a subsequent regrouping into one
$c$-Majorana fermion per site and \bbZ gauge connections (arising from Majorana bilinears on every bond) as illustrated in Fig.~\ref{fig:mapping} below.
In this language, every two-qubit parity check has the effect of measuring the local Majorana fermion parity, 
while effectively teleporting single Majorana fermions~\cite{Nayak2008measurement} -- for a schematic illustration see panels (c) and (d) in Fig.~\ref{fig:mapping}. Repeated local measurements can then quickly lead to teleportation over large distances and the formation
of long-range entanglement, see Fig.~\ref{fig:mapping}(e).

%%%%%%%%%%%%%%%%%%%%%%%%%%%%%%%%%%%%%%%%%%%%%%%%%%%%%%%%%%%%%%%%%%%%%%%%%%%%
\subsection{Time-reversal symmetry and gauge physics}
%%%%%%%%%%%%%%%%%%%%%%%%%%%%%%%%%%%%%%%%%%%%%%%%%%%%%%%%%%%%%%%%%%%%%%%%%%%%

A crucial ingredient in the classification of entanglement emerging either in the ground states of Kitaev Hamiltonians or in typical states stabilized by Kitaev circuits is the role of time-reversal symmetry and gauge physics. 
To set the stage, let us recall that, for a bipartite lattice/circuit geometry, the time reversal transformation can be defined as $\mathcal{T}=i\sigma^y \mathcal{K}$ where $\mathcal{K}$ is the complex conjugate, which leads to $\sigma^{x(y)(z)}\to -\sigma^{x(y)(z)}$. The combination of time-reversal symmetry $\mathcal{T}^2=+1$ with the intrinsic particle-hole symmetry of Majorana fermions puts both the Hamiltonian model and the quantum circuit in the BDI symmetry class.
The situation is fundamentally different when the symmetry class is changed from BDI to class D, which can be realized by moving to a non-bipartite geometry as first noted by Kitaev~\cite{Kitaev2006} and later explored in the context of a lattice model by Yao and Kivelson~\cite{YaoKivelson2007}. 
Even though on the microscopic level of the Hamiltonian or quantum circuit description the system remains time-reversal invariant, 
the low-temperature ground state or dynamically stabilized typical state is \emph{not}.
Instead, time-reversal symmetry is spontaneously broken in this non-bipartite setting as we will discuss in the following.

%%%%%%%%%%%%%%%%%%%%%%%%%%%%%%%%%%%%%%%%%%%%%%%%%%%%%%%%%%%%%%%%%%%%%%%%%%%%
\subsubsection*{Hamiltonian model}
%%%%%%%%%%%%%%%%%%%%%%%%%%%%%%%%%%%%%%%%%%%%%%%%%%%%%%%%%%%%%%%%%%%%%%%%%%%%

Let us first inspect time-reversal symmetry in the context of Hamiltonian models, where its role in stabilizing ground-state flux patterns 
and its spontaneous breaking for non-bipartite lattice geometry is well established.
As a first example, consider the seminal honeycomb model with its bipartite lattice geometry. 
Here the Wilson loop, indicating the flux through one elementary hexagonal plaquette, can be expressed as 
$W_{p} = \sigma_1^x\sigma_2^y\sigma_3^z\sigma_4^x\sigma_5^y\sigma_6^z=\pm 1$, 
with a clock-wise ordering of the six spin-1/2s around the plaquette. 
This flux is invariant under time reversal symmetry, i.e., $\mathcal{T}^{-1} W_p \mathcal{T} = W_p$. 
In addition, the Wilson loop operator commutes with the Hamiltonian, resulting in a static flux pattern, which by virtue of Lieb's theorem \cite{Lieb_1994}
is the completely flux-free state in the ground state. As a consequence, the ensuing spin liquid is primarily characterized by the band
structure of the free Majorana fermions. For isotropic coupling, $K_x = K_y = K_z$ in Eq.~\eqref{eq:Hamiltonian}, this is the celebrated
quasi-relativistic band structure with two gapless Dirac cones. 
In terms of entanglement, this is an area-law entangled state albeit one with a subleading $O(\ln L)$ correction~\cite{Qi10kitaevent, Chen_2015}
\[
	S(A) \sim L + \ln L + \ldots \,.
\]

Now, if we move to the Yao-Kivelson model on its non-bipartite lattice, note that while the flux through the large, dodecagonal plaquette (of length 12)
is invariant under time reversal, the triangle flux is not -- $W_{\triangle} = \sigma_1^x\sigma_2^y\sigma_3^z$ changes sign under time reversal. In the low-temperature regime of the Hamiltonian (i.e.\ below the thermal crossover at which the spin degrees fractionalize \cite{Nasu14mc}), the Majorana fermions mediate an effective Ising interaction between the triangle fluxes~\cite{vidal08}. This leads to a flux-ordering transition at a finite transition temperature \cite{Nasu2015} at which the system {\it spontaneously} breaks time reversal symmetry. Simultaneously, the Majorana fermions respond to the chiral triangle flux and form a $p+ip$ superconductor which falls into two topologically distinct phases~\cite{readgreen00}: a topological phase at weak pairing and a trivial phase at strong pairing. The transition between the two states can be driven by the relative strength of the triangle versus hexagonal coupling ($J$ and $K$ in the overview of Fig.~\ref{fig:overview_circuit_diagram}).
In the language of quantum magnetism, the time-reversal symmetry breaking, topological state is called a chiral spin liquid \cite{KalmeyerLaughlin},
with an entanglement structure 
\[
	S(A) \sim L  - \ln 2 \,,
\]
where the topological nature of this phase is reflected by a non-vanishing, subleading topological correction $\gamma_{\rm topo} = \ln(2)$ to the area-law entanglement~\cite{KitaevPreskill2006,LevinWen2006}.

%%%%%%%%%%%%%%%%%%%%%%%%%%%%%%%%%%%%%%%%%%%%%%%%%%%%%%%%%%%%%%%%%%%%%%%%%%%%
\subsubsection*{Quantum circuits}
%%%%%%%%%%%%%%%%%%%%%%%%%%%%%%%%%%%%%%%%%%%%%%%%%%%%%%%%%%%%%%%%%%%%%%%%%%%%

Let us now turn to a discussion of the role of time-reversal symmetry and gauge physics in the measurement-only circuit analogs of the Kitaev model. 
If starting from a featureless, maximally mixed state, the purification dynamics due to measurements can be separated into two contributions~\cite{Vijay23Kitaev,Ippoliti23kitaev,Zhu23structuredVolumeLaw} -- a gauge flux purification and a Majorana purification. 
Since the gauge flux commutes with the measurement operators, it is purified with a constant rate leading to an exponential decay of its entropy in time.
The Majorana purification, in contrast, depends on the degree of frustration -- it is exponentially fast for the small frustration phase, but algebraically slow for the large frustration phase. Its behavior is tightly related to the pure state entanglement entropy that will be discussed in the following. 
To summarize the main results, the purification time for the gauge flux always scales logarithmically with the system size $\bigO(\ln N)$, while the purification time of the Majorana fermion takes $\bigO(\ln N)$ in the less frustrated and less entangled phase, but $\bigO(N^{p})$ power-law scaling with the system size in the highly frustrated and entangled liquid-like phase. 
Such two-stage purification dynamics was found to be a rather generic feature for measurement-only circuits built from subsystem codes~\cite{Placke2024}.

\begin{figure}[t]
    \centering
    \includegraphics[width=\columnwidth]{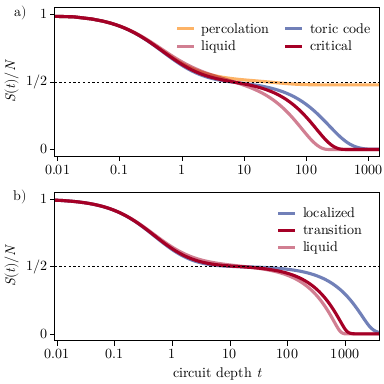}
    \caption{{\bf Purification dynamics.}
        Time evolution of the state entropy (not to be confused with the pure state entanglement entropy) under the action of Kitaev (a) and Yao-Kivelson (b) circuits at a system size of $L=24$. Color coding is in accordance with Fig.~\ref{fig:overview_circuit_diagram}. The initial state is an empty generating set for which $S=N$ bits. The dynamics are averaged over 3840 disorder realizations. As before, a single timestep in the honeycomb (dodecagonal) geometry corresponds to $N=2L^2$ ($N=6L^2$) random measurements. 
        After the initial drop from adding (any) mutually commuting stabilizers to the generating set, the entropy decays exponentially from the constant rate at which the gauge flux is purified. This rate depends on the degree of anisotropy in the measurement probabilities.
        The percolation point (yellow) corresponds to the decoupled stacking limit, say, a stack of horizontal chains with $XX$ and $YY$ parity checks. 
        It leaves a single logical operator as the global flux pumped horizontally. When infinitesimal weak $ZZ$ measurement is turned on, 
        the inter-chain Majorana hopping measures the $\prod_{j=1}^{L_y} \sigma_j^y$ operator, 
        that purifies the remaining horizontal flux (manifesting in the curve changing from the orange to the red). 
    }
    \label{fig:HoneycombPurification}
\end{figure}

%%%%%%%%%%%%%%%%%%%%%%%%%%%%%%%%%%%%%%%%%%%%%%%%%%%%%%%%%%%%%%%%%%%%%%%%%%%%
\paragraph*{Gauge purification.---}
%%%%%%%%%%%%%%%%%%%%%%%%%%%%%%%%%%%%%%%%%%%%%%%%%%%%%%%%%%%%%%%%%%%%%%%%%%%%
Let us first discuss the gauge flux purification dynamics in more detail. Note that the flux is not directly measured in our protocol but instead indirectly via the Majorana fermion parity checks. 
Every single Majorana parity measurement enforces a Majorana hopping (see Fig.~\ref{fig:mapping}), and when a set of measurements encircles a plaquette the moving Majorana detects and collapses the gauge flux to one of its eigenstates. 
As a consequence, a particular block in the block-diagonal matrix representing the Gaussian Majorana fermion state is picked up
\begin{equation*}
\left(
\begin{matrix}
\rho(W) & 0 & 0 & \cdots\\
0 & \rho(W') & 0 & \cdots\\
0 & 0 & \rho(W'') & \cdots\\
\cdots & \cdots & \cdots & \cdots\\
\end{matrix}
\right) \ ,
\end{equation*}
where each block corresponds to a fixed gauge flux configuration $W$, $W'$, $W''$. 
Over time, this block matrix is being purified towards a pure state, akin to lowering the temperature of a free fermion system. 
Due to the Clifford nature of the Kitaev circuits discussed here, the probability distribution for any possible flux configuration is equal, such that the state is purified into a \emph{random flux configuration}, distinct from the flux-ordered Hamiltonian ground state~\cite{Kitaev2006}. Nevertheless, the states associated with any gauge flux configuration share exactly the same entanglement entropy. This is due to the generic property of Clifford stabilizer states: the entanglement of a Clifford stabilizer state $\rho =  \prod_{j=1}^{N}\frac{1\pm O_j}{2}$ where $\{O_j\}$ are the stabilizer generators, does not depend on the eigenvalues being positive or negative. A minimal example is that all four possible Bell pairs $\ket{\uparrow\uparrow}+\ket{\downarrow\downarrow}$, $\ket{\uparrow\downarrow}+\ket{\downarrow\uparrow}$, $\ket{\uparrow\uparrow}-\ket{\downarrow\downarrow}$, $\ket{\uparrow\downarrow}-\ket{\downarrow\uparrow}$ share one bit of entanglement entropy. 
Since the Majorana parity checks are performed stochastically, there is a constant rate of purifying the gauge flux (denoted as $r$ in the following), 
which leads to an exponential decay of the flux entropy in time ~\cite{Vijay23Kitaev,Ippoliti23kitaev,Zhu23structuredVolumeLaw, Zhu23qubit}
\[
	S_{\text{flux}}(t) \sim N_p e^{-rt} \,,
\]
where $N_p$ is the number of the plaquettes of the same type. 
This decay is illustrated in Fig.~\ref{fig:HoneycombPurification} 
for various points in the phase diagram of the monitored honeycomb Kitaev model -- the initial decay of the state entropy towards a plateau at $S(t)/N=\tfrac12$ signals the flux purification, which is independent of the circuit phase.
In general, the purification rate $r$ depends on microscopic details, such as the number of edges of the plaquette 
-- the larger the plaquettes, the slower the gauge purification. 
Apart from these microscopic details, the time scale to purify the flux to a pure configuration can be estimated as $t \propto \ln N_p / r $
-- the purification evolution here is akin to lowering the temperature to approach the ground state flux configuration in the Hamiltonian~\cite{Nasu14mc}. 

For plaquettes with an \emph{odd} number of edges~\cite{Placke2024}  there is an {\it obstruction} to purification (that renders $r=0$) due to the time-reversal symmetry of the protocol, as proved in the following: Any stabilizer of the time-evolved state $O$ is composed of a subset of the local quantum measurement operators $K_j$
\begin{equation*}
O = \prod_{j} K_j \ .
\end{equation*}
Thus, when every local measurement respects a global symmetry such as time reversal, so does any stabilizer operator of the evolved state
\begin{equation*}
\mathcal{T}^{-1} O \mathcal{T} =\prod_j ( \mathcal{T}^{-1} K_j\mathcal{T}) = O \ .
\end{equation*}
For example, a hexagon or dodecagon plaquette can be decomposed into its surrounding six bond operators, while a triangle plaquette operator is time reversal odd and cannot be generated by any two-body time-reversal-even operator measurements. Nevertheless, an even product of triangle plaquettes are time reversal even and can be purified, for example considering the sequence of stabilizer generators depicted in Fig.~\ref{fig:yk_measurement_sequence} below.
\begin{figure}[h!]
	\begin{center}
	\includegraphics{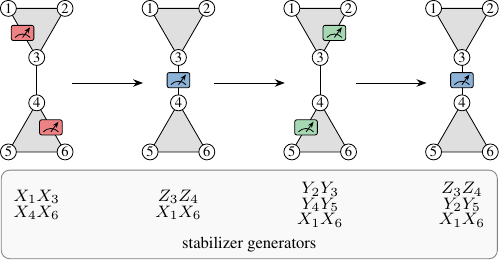}
	\end{center}
    \caption{{\bf Purification of paired triangle fluxes.}
    		Shown is a measurement sequence that can purify the fluxes in two gray-shaded triangles.
    		In the last step, the product of the three stabilizer generators is equal to the product of the two triangle fluxes.}
    \label{fig:yk_measurement_sequence}
\end{figure}\\
This implies that a product of an even number of triangle plaquettes can still be purified exponentially fast, while a product of an odd number of triangle plaquettes remains always undetermined, respecting the global time-reversal symmetry. Thus, one global bit, corresponding to the action of a chiral Ising time-reversal transformation, remains conserved, leaving one bit of entropy, the same as the time reversal symmetric mixed state of the Yao-Kivelson Hamiltonian at low temperature. 

Crucially, for the quantum circuits, the entanglement dynamics does {\it not} depend on the individual flux assignments of the gauge configurations. 
Thus we have two strategies to break time-reversal symmetry: 
(i) we directly measure the 3-body next-nearest neighbor (n.n.n.) Kitaev interaction on the honeycomb lattice, 
which {\it explicitly} breaks the time reversal symmetry by fixing the $\pm\pi/2$ fluxes on this triangle spanned by the n.n.n.\ spins; 
(ii) we move to the decorated honeycomb lattice with triangle plaquettes (a la Yao and Kivelson) and fix a single triangle flux from the initial state. 
Both strategies yield, as we will show in the following, the same entanglement phases. 

%%%%%%%%%%%%%%%%%%%%%%%%%%%%%%%%%%%%%%%%%%%%%%%%%%%%%%%%%%%%%%%%%%%%%%%%%%%%
\paragraph*{Majorana purification.---}
%%%%%%%%%%%%%%%%%%%%%%%%%%%%%%%%%%%%%%%%%%%%%%%%%%%%%%%%%%%%%%%%%%%%%%%%%%%%
The purification dynamics of the Majorana fermions (the second stage of purification in Fig.~\ref{fig:HoneycombPurification}), in contrast, primarily depends on the amount of frustration, given by the relative weight
of non-commuting operators, and not the specifics of the underlying lattice geometries.
For small frustration, i.e.\ a large bias towards certain commuting measurement patterns (such as the toric code stabilizers), 
the Majorana fermion purifies into the preferred dimer states exponentially fast, thereby stabilizing only a few, exponentially long-lived, 
global topological states (potentially serving as logical qubits). 
For large frustration, i.e.\ closer to the point of isotropic measurement strengths, in the liquid-like phases of the phase diagram, 
the Majorana fermion generally purifies in a power-law, governed by the physics of the statistical loop model to be introduced in Section \ref{sec:loop_models}. 

%%%%%%%%%%%%%%%%%%%%%%%%%%%%%%%%%%%%%%%%%%%%%%%%%%%%%%%%%%%%%%%%%%%%%%%%%%%%
\subsection{Relation to Hastings-Haah code}
%%%%%%%%%%%%%%%%%%%%%%%%%%%%%%%%%%%%%%%%%%%%%%%%%%%%%%%%%%%%%%%%%%%%%%%%%%%%

This section has introduced and discussed Kitaev circuits as measurement-only quantum circuit analogs of the Kitaev model with an eye on the underlying lattice/circuit geometry. Before closing we also want to draw a few connections to the Hastings-Haah code~\cite{Haah21honeycomb,Hastings22honeycomb}. 
The latter has attracted considerable interest from the quantum information community for its ability to {\it dynamically} stabilize a topological qubit,
while also being intimately connected to the Kitaev physics discussed above. 
The Hastings-Haah code imposes two-qubit parity checks with a strong spatio-temporal ordering. In space, it assigns a Kekul\'e pattern 
of couplings to the honeycomb geometry (see Fig.~\ref{fig:overview_circuit_diagram}) and then performs time-periodic measurements of 
all $XX$, $YY$, $ZZ$ parity checks. This Floquet dynamics periodically swaps the instantaneously stabilized state among the three corners of the
Kekul\'e Kitaev circuit phase diagram (see Fig.~\ref{fig:overview_circuit_diagram}), 
which like their conventional honeycomb counterpart, are toric code phases. This is what allows it to create a topological qubit state. 

Here we take the broad interest in the Hastings-Haah code as motivation to study the measurement-only circuit variant of the 
Kekul\'e Kitaev model~\cite{Schmidt10honeycomb} along with its conventional honeycomb siblings.

%%%%%%%%%%%%%%%%%%%%%%%%%%%%%%%%%%%%%%%%%%%%%%%%%%%%%%%%%%%%%%%%%%%%%%%%%%%%

%%%%%%%%%%%%%%%%%%%%%%%%%%%%%%%%%%%%%%%%%%%%%%%%%%%%%%%%%%%%%%%%%%%%%%%%%%%%
\section{Majorana loop models in \ensuremath{\mathbf{(2+1)}} dimensions}
\label{sec:loop_models}
%%%%%%%%%%%%%%%%%%%%%%%%%%%%%%%%%%%%%%%%%%%%%%%%%%%%%%%%%%%%%%%%%%%%%%%%%%%%

Loop models arise quite generically in systems of quadratic Majorana fermions~\cite{Nahum_Annihilation,Klocke_2023,Merritt2023,Chalker2002,Chalker_1988}. 
Here we describe how the loop picture emerges, both in Hamiltonian dynamics and in quantum circuits.
We briefly review the statistical mechanics of loops in 3D and then discuss the universality classes of loop models, 
based on symmetry and loop fugacity, and name representative models therein.
Finally, we define important observables and entanglement quantities for loops and Majorana circuits, providing a dictionary between the two.

%%%%%%%%%%%%%%%%%%%%%%%%%%%%%%%%%%%%%%%%%%%%%%%%%%%%%%%%%%%%%%%%%%%%%%%%%%%%
\begin{figure}[t]
    \centering
    \includegraphics{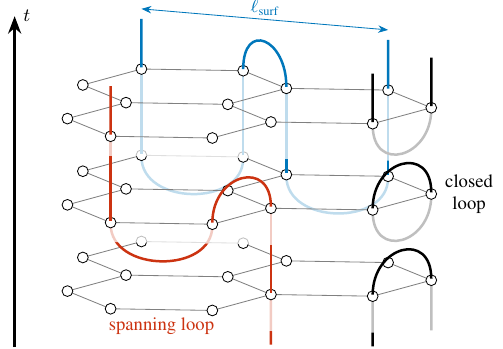}
    \caption{{\bf Majorana loop models in 3D space-time.}
        The loop model description emerges from the dynamics of the Majorana fermions in space-time. 
        The final state produced by the corresponding circuit is insensitive to closed loops contained in the space-time bulk (such as the black loop).
        Dynamical purification is sensitive to the presence of spanning loops (red) which connect the two temporal boundaries.
        Open arcs with both ends on a single temporal boundary (blue) have both a bulk path length $\ell_\textrm{bulk}$ and a projected surface length $\ell_\textrm{surf}$.
    }
    \label{fig:3dloops}
\end{figure}
%%%%%%%%%%%%%%%%%%%%%%%%%%%%%%%%%%%%%%%%%%%%%%%%%%%%%%%%%%%%%%%%%%%%%%%%%%%%

In a quadratic Majorana model all information about the state is encoded in the two-point Majorana correlation functions $\langle i \gamma_l \gamma_m \rangle$. 
If one asks, in addition, that the quadratic Majorana state is also a stabilizer state, then all Majoranas have definite pairings $\langle i \gamma_l \gamma_m \rangle = 0, \pm 1$. 
Such a {\it stabilizer state} can be graphically represented by a pairing diagram wherein nodes $l$ and $m$ are connected by an open arc if and only if 
$\abs{i\langle \gamma_l \gamma_m \rangle} = 1$. 
Any generic Gaussian state can then be written as a linear combination of stabilizer states or, alternatively, of pairing diagrams. 
Evolving the state over time leads to rearrangement of the Majorana pairings, with fermion world lines tracing out loops in space-time as schematically illustrated in Fig.~\ref{fig:3dloops}.
The entanglement and mutual information between two distinct subregions in space $A,B$ are given by the number of arcs connecting $A$ and $B$~\cite{Klocke_2023} (e.g., the blue loop in Fig.~\ref{fig:3dloops}),
independent of the sign of the Majorana parities~\footnote{Diagrammatic representations including the parity sign are possible but follow a more complicated loop algebra than required here~\cite{Klocke_2023}.}.
When the number of Majoranas is conserved at all times (e.g., as for a spin-chain after a Jordan-Wigner transformation), the space-time is \emph{fully packed} with loops.

\begin{figure}[t]
    \centering
    \includegraphics{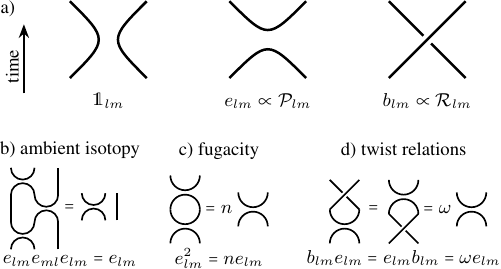}
    \caption{
    \textbf{Generators and relations of the loop algebra.}
    (a) Local generators of the loop algebra and the corresponding quadratic Majorana operation.
    The non-trivial generators $e_{lm}$ generate a Temperley-Lieb subalgebra and are proportional to the fermion parity projector $\calP_{lm}$.
    The $b_{lm}$ are proportional to the Majorana swap operators and generate a braid group.
    (b-d) Elementary relations between the generators.
    (b) Ambient isotopy of loops allows smooth deformation.
    (c) Closed loops can be replaced by a scalar factor $n$, the loop fugacity.
    (d) Untwisting loops after a braid operation introduces a scalar factor $\omega$.
    }
    \label{fig:loop_model}
\end{figure}

Generic models of fully-packed loops may be described by a braid-monoid algebra~\cite{Wadati_1989, Grimm_1993}.
Loop configurations and transformations between them are then generated by a sequence of local operations (i.e., generators of the algebra) which rearrange the world lines.
The allowed local operations acting on Majoranas $\gamma_l$ and $\gamma_m$ are depicted in Fig.~\ref{fig:loop_model}(a).
The identity $\mathds{1}_{lm}$ leaves the pairings unmodified, amounting to uninterrupted propagation forward in time.
By contrast, the Temperley-Lieb (TL) generators $e_{lm}$ result in spatial propagation of loops.
On the Majoranas, this generator acts as a projector onto the local fermion parity, 
\[
	e_{lm} \propto \calP_{lm} = \tfrac12(1 + i\gamma_l\gamma_m) \,.
\]
In particular, for initial pairings $(k,l)$ and $(m,n)$, $e_{lm}$ implements a \emph{loop surgery} to yield new pairings $(l,m)$ and $(k,n)$.
Finally, the braid operator $b_{lm}$ acts as the Majorana swap operator 
\[
	\calR_{lm} = \tfrac{1}{\sqrt{2}}\left(1 + \gamma_l \gamma_m\right) \,, 
\]
exchanging the two Majoranas and causing the world lines to cross over one another.

Loop models corresponding to a braid-monoid algebra have three key properties -- ambient isotopy, scalar fugacity ($d$-isotopy), and twist relations -- which are depicted in Fig.~\ref{fig:loop_model}(b-d).
Ambient isotopy reflects the ability to smoothly deform loops.
The scalar fugacity endows closed loops with a weight $n$.
Lastly, the twist relations allow braids $b_{lm}$ ($b_{lm}^{-1}$) to be undone by incurring a factor $\omega$ ($\omega^{-1}$) related to spin statistics.
All three leave the connectivity of the loop endpoints unchanged and may thus be viewed as equivalence relations (up to a scalar) between Majorana trajectories with identical entanglement in the initial and final states.

%%%%%%%%%%%%%%%%%%%%%%%%%%%%%%%%%%%%%%%%%%%%%%%%%%%%%%%%%%%%%%%%%%%%%%%%%%%%
%%%%%%%%%%%%%%%%%%%%%%%%%%%%%%%%%%%%%%%%%%%%%%%%%%%%%%%%%%%%%%%%%%%%%%%%%%%%

Throughout this work, we are interested exclusively in the dynamics of entanglement and mutual information in Majorana models, such that only the connectivity $\abs{\langle i\gamma_l\gamma_m\rangle}$ matters.
Braiding phases can thus be neglected and we fix $\omega = 1$.
This yields the equivalence relation $b_{lm} \sim b_{lm}^{-1}$ on the space of loop configurations, reducing the braid-monoid algebra to the Brauer algebra $\calB(n)$~\cite{Candu_2009}.

\begin{figure}[t]
    \centering
    \includegraphics{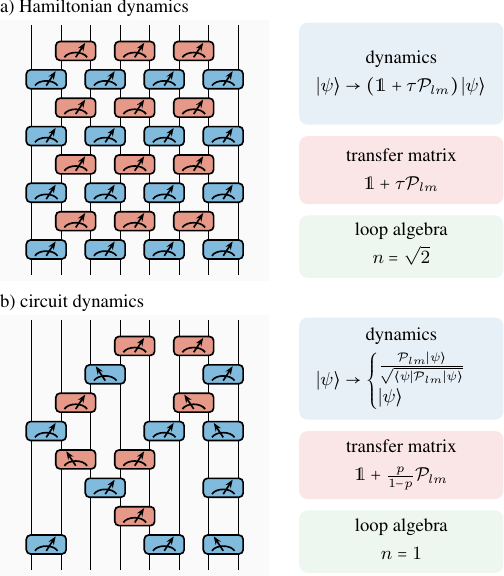}
    \caption{{\bf Quantum dynamics.}
        (a)
        Trotterizing the imaginary time evolution of a quadratic Majorana Hamiltonian $H$ leads to a product of transfer matrices 
        involving projectors $\calP_{lm}$ onto fixed parity for pairs of Majoranas corresponding to edges in the lattice. This can be viewed as a ``post-selected" measurement dynamics, where normalization only appears at the end of the evolution, hence yielding a loop algebra with fugacity $n=\sqrt{2}$. 
        (b)
        For a measurement-only quantum circuit, random spatio-temporal concatination of individual projective measurements, each with random outcome and subsequent normalization, 
        yield a loop model with fugacity $n=1$.    
    }
    \label{fig:overview_quantum_dynamics}
\end{figure}

By decomposing the Majorana dynamics into a sequence of operations $\{\mathds{1}_{lm}, e_{lm}, b_{lm}\}$, one obtains a transfer matrix representation of the evolution, i.e.\ a brickwall quantum circuit as illustrated on the left-hand side of Fig.~\ref{fig:overview_quantum_dynamics}. 
We may now distinguish two principal scenarios where Majorana loop models arise: Hamiltonian and circuit dynamics.

%%%%%%%%%%%%%%%%%%%%%%%%%%%%%%%%%%%%%%%%%%%%%%%%%%%%%%%%%%%%%%%%%%%%%%%%%%%%
%%%%%%%%%%%%%%%%%%%%%%%%%%%%%%%%%%%%%%%%%%%%%%%%%%%%%%%%%%%%%%%%%%%%%%%%%%%%

\subsubsection*{Hamiltonian dynamics}
Suppose that the local transfer matrix takes the form 
\[
	T \sim \mathds{1} + \tau \calP_{lm} \,.
\]
Such a transfer matrix arises naturally from Trotterizing the imaginary time evolution $e^{-\beta H}$ of a quadratic Majorana Hamiltonian $H = i\sum_{\langle l m \rangle} A_{lm} \gamma_l \gamma_m$.
In order to satisfy the loop isotopy condition $e_{lm} e_{mn} e_{lm} = e_{lm}$, we take normalized projectors $e_{lm} = \sqrt{2}\calP_{lm}$ akin to defining the generators of the TL algebra~\cite{TemperleyLieb}.
This implies the relation $e_{lm}^2 = \sqrt{2} e_{lm}$ and identifies the loop fugacity 
\[ 
    n = \sqrt{2} \,
\]
for the imaginary time evolution of a Gaussian Majorana Hamiltonian.
The details of the Majorana adjacency matrix $A_{lm} \neq 0$ fix the geometry and the symmetries for the model.

\subsubsection*{Circuit dynamics}
Alternatively, the transfer matrix may arise in a measurement-only circuit consisting of projective measurements of parities $i\gamma_l\gamma_m$.
Here the sign of the measurement outcome poses a challenge to specifying a loop representation.
In particular, the two possible outcomes correspond to $\calP_{lm}$ and $\calP_{ml}$.
Distinguishing such phase information in the graphical language of loops would require non-trivial twisting relations ($\omega \neq 1$), thereby complicating the relative weights in the trajectory-averaged loop ensemble.
If, however, we are interested only in entanglement and other quantities which depend exclusively on the modulus $\abs{\langle i \gamma_l \gamma_m \rangle}$, then the sign of the measurement outcome may be neglected.
This amounts to imposing a twist relation $\omega = 1$ and thus an equivalence relation $\calP_{lm} \sim \calP_{ml}$, with a loop configuration now representing an equivalence class of circuits.
Unlike the Trotterized imaginary-time evolution from before, here the projector $\calP_{lm}$ is always accompanied by a normalization of the state (i.e. $\ket{\psi} \rightarrow \tfrac{\calP_{lm}}{\sqrt{\langle \calP_{lm} \rangle}}\ket{\psi}$) yielding a transfer matrix of the form
\[ 
	T \sim \mathds{1} + \tfrac{p}{1-p} \calP_{lm} \,.
\]
Due to the normalization, projective measurements automatically satisfy the loop isotopy condition.
We may thus identify the generator $e_{lm}$ as the operator which acts on any state $\ket{\psi}$ as the normalized projective measurement $\calP_{lm}$.
In addition, the idempotence of projectors $\calP_{lm}^2 = \calP_{lm}$ fixes the loop fugacity 
\[ 
    n=1 \,
\]
for each individual measurement in the circuit.

%%%%%%%%%%%%%%%%%%%%%%%%%%%%%%%%%%%%%%%%%%%%%%%%%%%%%%%%%%%%%%%%%%%%%%%%%%%%
\subsubsection*{Universality classes of loop models}
\label{ss:symmetry}
%%%%%%%%%%%%%%%%%%%%%%%%%%%%%%%%%%%%%%%%%%%%%%%%%%%%%%%%%%%%%%%%%%%%%%%%%%%%
The behavior of correlation functions and observables at large distances in loop models is determined by two ingredients: the loop fugacity $n$ and the underlying symmetry of the loop model. Despite realizing a different loop fugacity, the transfer matrix of both the Hamiltonian and measurement-only circuit are generated by the operators $\mathcal{P}_{lm}$, which are quadratic in Majorana fermions. Thus each Hamiltonian symmetry class  has a corresponding symmetry class counterpart in the measurement-only circuit. The loop fugacity then yields a fine structure of universality classes for each symmetry class.

For quadratic Majorana theories, particle-hole (PH) symmetry is always present and we thus distinguish two different symmetry classes based on whether time-reversal symmetry is present or not. In class BDI time-reversal is present, which enables a bipartition of the Majorana lattice into sublattices $A$ and $B$, such that each operator $\mathcal{P}_{lm}$ acts exactly on one Majorana fermion on sublattice $A$ and on one Majorana fermion on sublattice $B$. The paradigmatic example of such a model is Kitaev's honeycomb model in two spatial dimensions~\cite{Kitaev2006}. In the loop model framework, this symmetry is known as ``orientability''. It allows to assign a unique orientation to each lattice site, either forward or backward in time, such that each loop carries this orientation through space-time.

In the absence of time-reversal, no bipartition of the lattice can be found, translating to the absence of orientable loop configurations. In the loop framework, the latter is known as non-orientable loops models or completely packed loop models with crossings (CPLC). This is symmetry class D and paradigmatic Hamiltonian models include the Yao-Kivelson model~\cite{YaoKivelson2007}, the Kitaev honeycomb model with next-nearest neighbor interactions as well as in the presence of a magnetic field. In the following, we will synonymously apply the language of Hamiltonian symmetry classes and loop model symmetry classes, i.e., we will refer to symmetry class BDI for orientable loop models and symmetry class D for non-orientable ones. 

%%%%%%%%%%%%%%%%%%%%%%%%%%%%%%%%%%%%%%%%%%%%%%%%%%%%%%%%%%%%%%%%%%%%%%%%%%%%
\subsection{Statistical mechanics of loops in 3D}\label{ss:loop_stat_mech}
%%%%%%%%%%%%%%%%%%%%%%%%%%%%%%%%%%%%%%%%%%%%%%%%%%%%%%%%%%%%%%%%%%%%%%%%%%%%

A $d$-dimensional Majorana circuit ($d=2$ in this work) corresponds to a $d+1$-dimensional loop model in space-time. 
Each lattice site $l$ with Majorana fermion $\gamma_l$ at a given time $t$ when measurements are performed represents a vertex in a regular space-time lattice.
When all Majorana fermions are included in at least one measurement $\mathcal{P}_{lm}$ during the circuit evolution, the loop model is fully packed, meaning that every vertex in the lattice has coordination number $z=4$.
Then the space of possible loop configurations is generated by routing loops through each vertex in one of the three ways depicted in Fig.~\ref{fig:loop_model}(a).
Such a model is described by a partition function 
\[
	Z = \sum_\mathcal{C} W(\mathcal{C}) n^{N(\mathcal{C})} \,,
\] 
where the sum is over all possible loop configurations $\mathcal{C}$.
Each term in the sum involves two components: (i) a ``local'' part $W(\mathcal{C})$ and (ii) a ``non-local'' part $n^{N(\mathcal{C})}$.
The local term $W(\mathcal{C})$ is a product of Boltzmann weights associated to the choice of loop connections at each vertex, giving the probability of performing the unique series of measurements (or Hamiltonian evolution steps) that yields the loop configuration $\mathcal{C}$.
This is the product of local measurement probabilities (or Hamiltonian matrix elements $A_{lm}$). 
The non-local term $n^{N(\mathcal{C})}$ accounts for the loop fugacity $n$ associated to each of the $N(\mathcal{C})$ closed loops in configuration $\mathcal{C}$.

As a side remark, we note that the weight $n^{N(\mathcal{C})}$ poses a challenge to Monte Carlo simulations of loop models for any fugacity $n\neq1$ since it cannot be represented by a local update rule for evolving a single time slice.
This singles out measurement-only circuits with $n=1$ as a particularly attractive example for simulating novel entanglement phases.
For larger integer fugacity $n \in \mathbb{Z}_+$, simulation is possible by more complicated Monte-Carlo methods involving the whole space-time lattice~\cite{Nahum_2011_3D, Nahum_2013_3D, Nahum_2013_soup, Nahum_2015_DQC, Serna_2021} or additional ancilla degrees of freedom~\cite{Klocke_2024}, but this is beyond the scope of this work.

%%%%%%%%%%%%%%%%%%%%%%%%%%%%%%%%%%%%%%%%%%%%%%%%%%%%%%%%%%%%%%%%%%%%%%%%%%%%
\subsubsection*{Non-linear sigma models}
%%%%%%%%%%%%%%%%%%%%%%%%%%%%%%%%%%%%%%%%%%%%%%%%%%%%%%%%%%%%%%%%%%%%%%%%%%%%
A field theory formulation of the loop model partition function $Z$ is provided by $Z=\int D[Q]\exp(-S[Q])$ with the non-linear sigma model (\nlsm) action
\begin{align}\label{eq:NonLinS}
    S[Q] = \frac{1}{2g} \int \dif^d x \, \Tr\left[(\nabla Q)^2 \right] + (\textrm{topological terms})
\end{align}
Here, $Q$ takes values in \cpn for symmetry class BDI and \rpn for symmetry class D. 
The field $Q$ is parametrized by a vector $z$, which is complex for class BDI and real for class D, such that $Q^{\alpha\beta} = z^\alpha \bar{z}^\beta - \delta_{\alpha\beta}$ is a traceless $n \times n$ Hermitian (orthogonal) matrix with normalization condition $z^\dagger z = n$. 
At short distances, the coupling constant $g$ scales with the microscopic diffusion constant $\mathcal{D}_{\text{mic}}\sim g^{-2}$, which is derived in Sec.~\ref{ss:crit_phase}.
The measurement-only limit $n=1$ is obtained by taking $n\rightarrow 1^+$ as a replica limit or alternatively by considering a supersymmetric formulation~\cite{Jacobsen_2003, Candu_2009, Martins_1998}.

Both the \cpn and the \rpn non-linear sigma models appear in a variety of different applications. 
Prominent examples are disordered fermion systems and quantum Hall layers in two dimensions. 
Although the loop models considered here are in $(2+1)$ dimensions, the relation to disordered fermions remains prominent: independent of the loop fugacity and the symmetry class, but depending on the parameter $g$ and the topological terms, the \nlsm support: (i) a long-loop phase with $\mathcal{D}\neq0$ at the largest distances, yielding long-range entangled Majorana pairs akin to a disordered metal for which the topological terms are irrelevant,
(ii) a short-loop phase with a diffusion constant $\mathcal{D}$ flowing to zero under renormalization group transformations, yielding area-law entanglement akin to localized fermions states but with distinct topological properties~\footnote{We note that this is different from the $(1+1)$-dimensional case, for which class BDI is always localized, except for at the critical points separating different topological phases, while class D may support a metallic state akin to weak anti-localization, known as Goldstone phase.}.

While for $n=2$ the metallic phases of the action $S[Q]$ in $(2+1)$ dimensions are known from the Hamiltonian case, the $n=1$ limit describes loops undergoing Brownian motion with diffusion constant $\mathcal{D}\sim g^{-2}$. For $n=1$, the intermediate distance behavior, i.e., on distances $\ell\le L_x L_y$ of the \emph{spatial} volume, of Brownian random walkers is the same for both symmetry classes. This, however, manifestly changes on larger distances $\ell\ge L_x L_y$: here, the Brownian random walkers have non-zero probability of forming space-time volume filling loops which occupy a non-zero fraction of the \emph{space-time }lattice. These macroscopic loops are sensitive to the symmetry class and yield a means to distinguish symmetry classes BDI and D in a measurement-only quantum circuit, see Sec.~\ref{ss:crit_phase}.

At the transition from the metallic to the localized regime, a phase transition occurs with universal scaling behavior depending on the symmetry class and the fugacity $n$~\cite{Nahum_2013_3D, Serna_2021}, which we explore below.

%%%%%%%%%%%%%%%%%%%%%%%%%%%%%%%%%%%%%%%%%%%%%%%%%%%%%%%%%%%%%%%%%%%%%%%%%%%%
\subsection{Loop-Circuit Dictionary}\label{ss:dictionary}
%%%%%%%%%%%%%%%%%%%%%%%%%%%%%%%%%%%%%%%%%%%%%%%%%%%%%%%%%%%%%%%%%%%%%%%%%%%%
Here, we connect the key observables of the loop model framework to the Majorana circuits, providing a mapping between the two pictures.
In passing from loops in $(2+1)$-dimensional space-time to the dynamics of Majorana fermions in two dimensions, one dimension in the loop picture is designated as time.
The state of the Majorana circuit at a given time corresponds to the surface of the loop model along one direction.
There are two such temporal surfaces: (i) that at $t=0$ where boundary conditions on the loops correspond to a choice of initial state for the Majoranas, and (ii) that at time $t=T$ corresponding to the final state produced by the circuit.
Correlations in this output state amount to \emph{boundary} correlations in the loop model.
By contrast, loop correlations in the space-time bulk require the ability to peek at the state midway through the circuit evolution.

To track information regarding both the surface and the bulk of the loop model, we keep track of not only the loop connectivity, but also their integrated path lengths.
Any open arc is labeled by a tuple $(l, m, \ell)$, where $l$ and $m$ represent the endpoints and $\ell$ is the path length. 
In a numerical simulation this information is readily incorporated into the update scheme:
Consider a state with initial pairings $(k, l, \ell_1)$ and $(m, n, \ell_2)$ where $l$ and $m$ are nearest neighbors.
Measuring the parity $i\gamma_l\gamma_m$ yields a state with new pairings and loop lengths $(k, n, \ell_1 + \ell_2 + 1)$ and $(l, m, 1)$.
When $k = m$, such a measurement closes the loop with total length $\ell_1 + 1$ which is then recorded in a histogram for the trajectory.

%%%%%%%%%%%%%%%%%%%%%%%%%%%%%%%%%%%%%%%%%%%%%%%%%%%%%%%%%%%%%%%%%%%%%%%%%%%%
\subsubsection*{Loop length distributions}
%%%%%%%%%%%%%%%%%%%%%%%%%%%%%%%%%%%%%%%%%%%%%%%%%%%%%%%%%%%%%%%%%%%%%%%%%%%%

The central observable for the loop models we consider here is the probability distribution for the length of loops, both along the (temporal) surface and through the space-time bulk. 
The surface loop length between two points $l,m$ at fixed time $T$ describes the distances between two entangled Majorana fermions $\gamma_l,\gamma_m$ in the circuit. 
The distribution of surface loops thus yields complete information on the quantum state at time $T$ with regard to its entanglement and mutual information. 
The bulk loop length yields information on out-of-time ordered correlators (OTOCs), explored below.

For an open arc connecting Majoranas $\gamma_l$ and $\gamma_m$ on the $t=T$ boundary, we define the surface length $\ell_\textrm{surf}^\alpha$ along the $\alpha$-direction as follows.
Let $\ell_x \hat{x} + \ell_y \hat{y}$ be the vector connecting sites $l$ and $m$ in the two-dimensional plane. 
Using periodic boundary conditions (PBC), the surface lengths of the loop are defined as $\ell_\textrm{surf}^\alpha \equiv \min\left(\abs{\ell_\alpha}, L_\alpha - \abs{\ell_\alpha}\right)$.
We define the corresponding length distributions $P_\textrm{surf}^\alpha(\ell)$ as that obtained in the stationary state of the circuit when taking a pure initial state with only local pairings.

Characteristics of the entanglement dynamics during the evolution are encoded in the bulk loop length distribution $P_\textrm{bulk}(\ell)$. 
It is the probability distribution for the total path length of a randomly selected loop.
Here we need to make a distinction between open arcs and closed loops, as these will generically exhibit distinct length distributions.
During the circuit evolution, many closed loops are formed, but the number of open arcs is fixed by boundary conditions and remains constant.
If we impose PBC in time, all loops are closed and there is no need to make this distinction.
In practice, temporal PBC can be implemented with the assistance of ancilla measurements, which is discussed in Sec.~\ref{ss:crit_phase} together with the subtle but important role of boundary conditions for the bulk loop statistics.

%%%%%%%%%%%%%%%%%%%%%%%%%%%%%%%%%%%%%%%%%%%%%%%%%%%%%%%%%%%%%%%%%%%%%%%%%%%%
\subsubsection*{Surface observables: Entanglement entropy and mutual information}
%%%%%%%%%%%%%%%%%%%%%%%%%%%%%%%%%%%%%%%%%%%%%%%%%%%%%%%%%%%%%%%%%%%%%%%%%%%%

Observables described by the loop distribution of a temporal surface at time $t=T$ correspond to the entanglement content of the Majorana fermion state $\rho(T)$. 
We consider the von Neumann entanglement entropy $S_A \equiv -\Tr\left[\rho_A \log_2 \rho_A \right]$ for a subsystem $A$, where the reduced density matrix $\rho_A \equiv \Tr_{\bar{A}}[\rho]$ is obtained by tracing out the complementary system $\bar{A}$,
and the mutual information $I_2(A,B) \equiv S_A + S_B - S_{AB}$ between subsystems $A$ and $B$.
For Majorana stabilizer states, these measures of entanglement are determined wholly by the non-vanishing two-point Majorana correlations $\abs{\langle i\gamma_l \gamma_m \rangle} = 1$.
Thus the connectivity encoded by the loop representation contains all necessary information to specify $S_A$ and $I_2(A,B)$.
The mutual information $I_2(A,B)$ therefore corresponds precisely to the number of loops connecting subsystems $A$ and $B$, while the von Neumann entanglement entropy for a pure state is $S_A = \tfrac12 I_2(A, \bar{A})$, reflecting that each Majorana arc carries a half qubit of entanglement.
Multipartite entanglement measures such as the tripartite mutual information $I_3(A,B,C)$ vanish due to the additivity of mutual information in the loop framework.

Throughout this work we will frequently consider the entanglement entropy of a cylindrical subsystem of length $\ell$, with boundaries encircling the system in the $\hat{x}$ or $\hat{y}$ directions.
For such a partition, the entanglement entropy can be computed from the surface loop length distribution~\cite{Klocke_2023}.
In particular, the average entanglement entropy for subsystem $A$ in a cylindrical region of length $\ell$  along the $\hat{x}$ direction is given by
\begin{align}\label{eq:entanglement_integral}
    \llangle S^x_\ell \rrangle = L_y \sum_{\ell'} \min(\ell, \ell') P_\textrm{surf}^x(\ell') \,.
\end{align}
The mutual information between such cylindrical subsystems can be computed in a similar manner.

%%%%%%%%%%%%%%%%%%%%%%%%%%%%%%%%%%%%%%%%%%%%%%%%%%%%%%%%%%%%%%%%%%%%%%%%%%%%
\subsubsection*{Spanning loops and purification}
%%%%%%%%%%%%%%%%%%%%%%%%%%%%%%%%%%%%%%%%%%%%%%%%%%%%%%%%%%%%%%%%%%%%%%%%%%%%

A single loop connecting the two temporal surfaces at $t=0$ and $t=T$ correspond to Majorana fermions, which remain unpaired during the entire circuit evolution. Such loops are called \emph{spanning} and reflect the dynamical purification under measurement: Consider a maximally mixed initial state $\rho(t=0) \propto \mathds{1}$, corresponding to each Majorana being unpaired at $t=0$. The purity of the Majorana state $\rho$ at time $t=T$ is given by the total entropy $S_L(T) = -\Tr\left[\rho(T)\log_2\rho(T)\right]$. Each unpurified qubit at time $T$ corresponds to two unpaired Majorana fermions and thus to half the number of spanning loops $n_s(T)$, i.e., $S_L(T) = \tfrac12 n_s(T)$.

In the loop model framework, the spanning number, i.e., the dynamical purification, serves as an accurate quantifier for the different loop model phases and the critical point separating them~\cite{Ortuno_2009, Nahum_2011_3D, Nahum_2013_3D, Nahum_2013_crossing}.
To do so, we assume fixed aspect ratios $T / L_y$ and $L_x / L_y$.
Then for small distance $\delta$ from the critical point, we employ a scaling ansatz,
\be
\begin{aligned}
    n_s(\delta, L) & = f(x)\left[1 + \beta L^{\yirr} \right] \,,           \\
    x              & = L^{1/\nu} \delta \left[1 + \alpha \delta \right] \,,
\end{aligned}
\label{eq:FSS_ansatz}
\ee
where \yirr is the dimension of the leading irrelevant operator and $\alpha \neq 0$ gives a polynomial correction.

%%%%%%%%%%%%%%%%%%%%%%%%%%%%%%%%%%%%%%%%%%%%%%%%%%%%%%%%%%%%%%%%%%%%%%%%%%%%
\subsubsection*{Bulk loops, watermelons, and OTOCs}
\label{sec:Watermelon correlators}
%%%%%%%%%%%%%%%%%%%%%%%%%%%%%%%%%%%%%%%%%%%%%%%%%%%%%%%%%%%%%%%%%%%%%%%%%%%%

Beyond the distribution of bulk loop lengths $P_\textrm{link}(\ell)$, bulk properties of the loop model are captured by the watermelon correlators $G_k(r)$, the probability that two points in space-time separated by distance $r$ are connected by exactly $k$ distinct arcs.
For example, $G_2(r)$ gives the probability that two points with separation $r$ lie along the same loop.
When the points do not lie along the same fixed-time slice, then $G_k(r)$ takes the form of an out-of-time order correlator (OTOC) for the Majoranas.
As was previously pointed out in Ref.~\onlinecite{Klocke_2023}, watermelon correlators can be accessed in the circuit by an ancilla-based measurement scheme.
For example, the two-leg watermelon correlator $G_2(r)$ can be measured by considering space-time points $(t_1, l)$ and $(t_2, m)$ separated by distance $r$.
Consider ancilla $A$ and $B$, each consisting of a pair of definite parity, e.g. $\abs{\langle i \gamma_{A,1}\gamma_{A,2}\rangle} = 1$.
After evolving the circuit to time $t_1$ ($t_2$), a ``marked'' loop can be effectively inserted into the circuit by coupling the ancilla and bulk via measurement of $i\gamma_{A,2}\gamma_l$ ($i\gamma_{B,2}\gamma_m$).
The history of the world lines which had been passing through these points in space-time is then effectively stored in the remaining ancilla $\gamma_{A,1}$ and $\gamma_{B,1}$.
We then evolve the circuit to late time $t_f \gg t_2$ and impose boundary conditions on the bulk qubits (e.g., by measurement) such that the only remaining open arcs are those through the ancilla.
Had we taken only individual Majoranas for each ancilla, then the watermelon correlator $G_2(r)$ would explicitly take the form of an OTOC, $\langle \gamma_A(t_f) \gamma_A(0) \gamma_B(t_f) \gamma_B(0) \rangle - \langle \gamma_A(t_f) \gamma_A(0) \rangle \langle \gamma_B(t_f) \gamma_B(0) \rangle$, which for stabilizer states coincides with the mutual information between the ancilla $I_2(A,B)$.
By construction, the mutual information $I_2(A, B)$ counts the number of loops connecting points $(t_1, l)$ and $(t_2, m)$ in space-time such that $G_2(r) = \tfrac12 I_2(A, B)$.
For higher order correlators $G_k(r)$, additional ancilla are required, but the general procedure remains the same.

%%%%%%%%%%%%%%%%%%%%%%%%%%%%%%%%%%%%%%%%%%%%%%%%%%%%%%%%%%%%%%%%%%%%%%%%%%%%
\section{Dynamics of Majorana loops}\label{sec:results}
%%%%%%%%%%%%%%%%%%%%%%%%%%%%%%%%%%%%%%%%%%%%%%%%%%%%%%%%%%%%%%%%%%%%%%%%%%%%

In this section we present results from large-scale simulation of the measurement-only circuits in the loop framework. As representatives for the symmetry classes, we consider the honeycomb Kitaev model for class BDI and the next-nearest neighbor honeycomb Kitaev model for class D. Numerical results for the Yao-Kivelson circuit are presented in App.~\ref{ss:YK}.
By varying the relative probability for measuring different bonds, one may tune the circuit between distinct measurement-induced phases, including several topologically distinct area-law phases and an extended metallic phase with a logarithmic area-law violation and long space-time loops.
In symmetry class BDI, i.e., in the measurement-only Kitaev circuit, the area-law phases arise in the toric-code limit where one flavor of bond measurement dominates.
Area-law phases also appear in symmetry class D, i.e., in the next-nearest neighbor Kitaev geometry and the Yao-Kivelson circuit by appropriate choice of parameters, see Fig.~\ref{fig:overview_circuit_diagram}.
Besides the various area-law phases, circuits in both symmetry classes host an entangled liquid phase, reminiscent of a weakly disordered Majorana metal phase, which we characterize in Sec.~\ref{ss:crit_phase}.
Then in Sec.~\ref{ss:universality}, we shed light on the universality of the entanglement transition separating metallic and area-law phases. Finally, we discuss the shape of the phase diagram for class BDI and the properties of the metallic phase based on the picture of random Brownian walkers.

%%%%%%%%%%%%%%%%%%%%%%%%%%%%%%%%%%%%%%%%%%%%%%%%%%%%%%%%%%%%%%%%%%%%%%%%%%%%
\subsection{Long loops and the Majorana liquid phase}\label{ss:crit_phase}
%%%%%%%%%%%%%%%%%%%%%%%%%%%%%%%%%%%%%%%%%%%%%%%%%%%%%%%%%%%%%%%%%%%%%%%%%%%%

When the non-commuting measurements introduce a sufficient degree of frustration, the endpoints of the Majorana world lines no longer remain confined to a finite area but start to undergo a random Brownian motion.
This generates long loops in the circuit and stabilizes a Majorana liquid or metallic phase with characteristic $L\log L$ entanglement. 
The entanglement structure in the stationary state is determined by the surface loop distribution in the loop model. 
In the liquid phase, this distribution does not distinguish between the two symmetry classes BDI and D, which we discuss below. 
A universal distinction between BDI and D, however, is observable in the bulk loop distribution.

%%%%%%%%%%%%%%%%%%%%%%%%%%%%%%%%%%%%%%%%%%%%%%%%%%%%%%%%%%%%%%%%%%%%%%%%%%%%
\begin{figure*}[t]
    \centering
    \includegraphics[width=\textwidth]{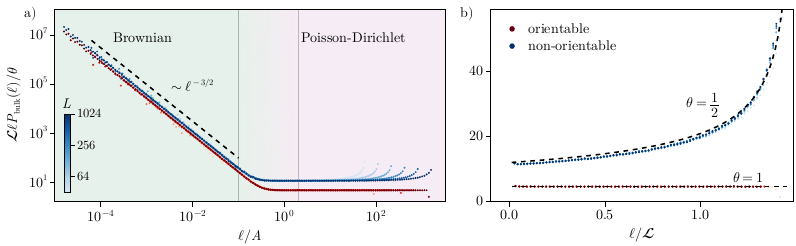}
    \caption{{\bf Bulk loop statistics.}
        Distribution of bulk loop lengths $P_{\textrm{bulk}}(\ell)$ in the measurement-only Kitaev circuit with PBC in both time and space.
        The isotropic point of the non-orientable circuit $K_x = K_y = K_z = J = 1/4$ (blue) is contrasted with the orientable circuit $K_x = K_y = K_z = 1/3$ (red).
        (a) Both the orientable and non-orientable cases exhibit the expected Brownian scaling regime $\ell P_{\textrm{bulk}}(\ell) \propto \ell^{-3/2}$ for loop lengths $\ell \lesssim L_x L_y$.
        (b) Breaking orientability leads to a distinct scaling behavior for macroscopic loops $\ell \gg L_x L_y$, following a Poisson-Dirichlet (PD) distribution with universal parameter $\theta$ as in Eq.~\eqref{eq:PD_distrib}.
        We find $\theta = 1$ and $\theta = 1/2$ in the orientable and non-orientable circuits, respectively, with the corresponding fit shown as a black dashed line.
    }
    \label{fig:HC_BulkLoops}
\end{figure*}
%%%%%%%%%%%%%%%%%%%%%%%%%%%%%%%%%%%%%%%%%%%%%%%%%%%%%%%%%%%%%%%%%%%%%%%%%%%%

%%%%%%%%%%%%%%%%%%%%%%%%%%%%%%%%%%%%%%%%%%%%%%%%%%%%%%%%%%%%%%%%%%%%%%%%%%%%
\subsubsection*{Surface loops and entanglement}\label{sss:surface_loops}
%%%%%%%%%%%%%%%%%%%%%%%%%%%%%%%%%%%%%%%%%%%%%%%%%%%%%%%%%%%%%%%%%%%%%%%%%%%%

The entanglement structure of a Majorana state $\rho(T)$ at time $t=T$ is captured by the probability distribution $P^{x,y}_\textrm{surf}(\ell)$ of open loop arcs along the temporal boundary of the loop model.
In the liquid phase, the loop length distribution exhibits power-law scaling 
\[
	P^\alpha_\textrm{surf}(\ell) \sim \ell^{-2} \,.
\]
For a cylindrical subsystem of length $L$ and circumference $L_y$, Eq.~\eqref{eq:entanglement_integral} yields an entanglement entropy $S(L, L_y) \sim L_y \log(L)$ (see Appendix~\ref{app:supplemental_data} for a numerical confirmation). 
Such a logarithmic violation of the area-law of fermions in 2D is often associated with the presence of a Fermi surface~\cite{Wolf2006,Klich2006,Li2006,Swingle2010}, but it is also expected for metallic states with non-vanishing conductivity in the presence of weak disorder~\cite{Burmistrov_2017}. 

The scaling of $P^{\alpha}_\textrm{surf}(\ell)$ and of the entanglement entropy is a general feature of the liquid phase for both class BDI and D.
Let us denote $\tilde{c}_\alpha$ for $\alpha = x,y$ as the coefficient for the entanglement $S=\tilde{c}_\alpha L_\alpha\log(L)$ for a cut along the $\alpha$-direction. Away from the critical point, this coefficient is generally not universal but rather varies continuously with microscopic details of the model (i.e.\ the lattice geometry and the bond weights). This may be seen analogously to a smooth deformation of the Fermi surface yielding a continuous modification of the logarithmic entanglement scaling in the Hamiltonian setting, an analogy, which 
we develop further below. 

The effect of orientability here is a rather trivial one.
Suppose that we fix the length $\ell_x$.
In the orientable case, there are fewer allowed Majorana pairings for any $\ell_x$, as correlations between sites on the same sublattice must necessarily vanish.
As a result, the typical total loop length $\ell = \sqrt{\ell_x^2 + \ell_y^2}$ is longer, and the probability density $P_\textrm{surf}(\ell_x)$ is reduced by a constant factor.

%%%%%%%%%%%%%%%%%%%%%%%%%%%%%%%%%%%%%%%%%%%%%%%%%%%%%%%%%%%%%%%%%%%%%%%%%%%%
\subsubsection*{Bulk loops and universal circuit correlations}\label{sss:bulk_loops}
%%%%%%%%%%%%%%%%%%%%%%%%%%%%%%%%%%%%%%%%%%%%%%%%%%%%%%%%%%%%%%%%%%%%%%%%%%%%

The two symmetry classes BDI and D are distinguishable by their bulk loop statistics. On short distances $\ell\le L_x L_y$ and away from the critical point, Brownian motion in three space-time dimensions yields a mean-field decay 
\[
	P_\textrm{bulk}(\ell) \propto \ell^{-\tau}
\]
with $\tau=5/2$ and a fractal dimension $d_f=3/(\tau-1)=2$ for both symmetry classes,  depicted in Fig.~\ref{fig:HC_BulkLoops}(b). However, Brownian walkers in 3D have a non-zero probability of never returning to their starting point, and to form ``macroscopic'' loops which occupy a non-zero fraction of the space-time volume.
The statistics of such extensive loops provide an unambiguous means for distinguishing between the two symmetry classes.

For finite system sizes, macroscopic loops are sensitive to the spatiotemporal boundary conditions. 
Here, one distinguishes \emph{absorbing} boundary conditions, yielding open Majorana world lines, such as, e.g.\ a mixed initial state, from \emph{reflecting} boundary conditions in space-time~\footnote{Strictly speaking, macroscopic loops would emerge also for periodic boundary conditions in time, which are, however, challenging to implement in a circuit.}. 
For the latter, all world lines are closed including at the final and initial time. 
This is realized by starting with a pure initial state at $t=0$ and by terminating the circuit at $t=T$ with a fixed set of measurements that closes all world lines.
With reflecting boundaries all loops are closed, and macroscopic loops manifest in the distribution $P_\textrm{bulk}(\ell)$ giving rise to a distinct scaling regime at distances $\ell\ge L_x L_y$, shown in Fig.~\ref{fig:HC_BulkLoops}(b).
Here, macroscopic loops follow a Poisson-Dirichlet (PD) distribution~\cite{Grosskinsky_2012, Nahum_2013_soup, Barp_2015}
\be
\ell \cdot P_\textrm{bulk}(\ell) = \frac{\theta}{\mathcal{L}}\left(1 - \frac{\ell}{f\mathcal{L}}\right)^{\theta - 1}, \quad (\ell \gg L_x L_y)
\label{eq:PD_distrib}
\ee
where $\mathcal{L}$ is the total number of links in space-time (i.e. the volume). Whereas $f$ is a non-universal quantity, 
the parameter $\theta$ is a \emph{universal} parameter which depends only on the loop fugacity $n$ and the symmetry class.
In the most general case, one finds $\theta = n$ for symmetry class BDI, while for symmetry class D, $\theta = n / 2$.
For the Majorana circuits at hand with fugacity $n=1$, this leads the normalized bulk loop length distribution to approach a constant 
\begin{align}
	&\ell\cdot P_\textrm{bulk}(\ell)  \sim \mathcal{L}^{-1} & {\rm(orientable)}
\end{align}
for $\ell\to\mathcal{L}$ in class BDI, 
while class D displays a square-root divergence 
\begin{align}
	&\ell\cdot P_\textrm{bulk}(\ell)  \sim \mathcal{L}^{-1/2}(\mathcal{L}-\ell/f)^{-1/2} & {\rm(non-orientable)} \,. 
\end{align}
Such a divergence reflects the tendency for a concentration of probability density in larger macroscopic clusters, due to the greater mobility of loops without the BDI constraint of time-reversal.
The results of numerical simulations, shown in Fig.~\ref{fig:HC_BulkLoops}(b), confirm the PD scaling regime for macroscopic loops, with the parameter $\theta$ consistent with loop fugacity $n=1$.

%%%%%%%%%%%%%%%%%%%%%%%%%%%%%%%%%%%%%%%%%%%%%%%%%%%%%%%%%%%%%%%%%%%%%%%%%%%%
\begin{figure*}[t]
    \centering
    \includegraphics{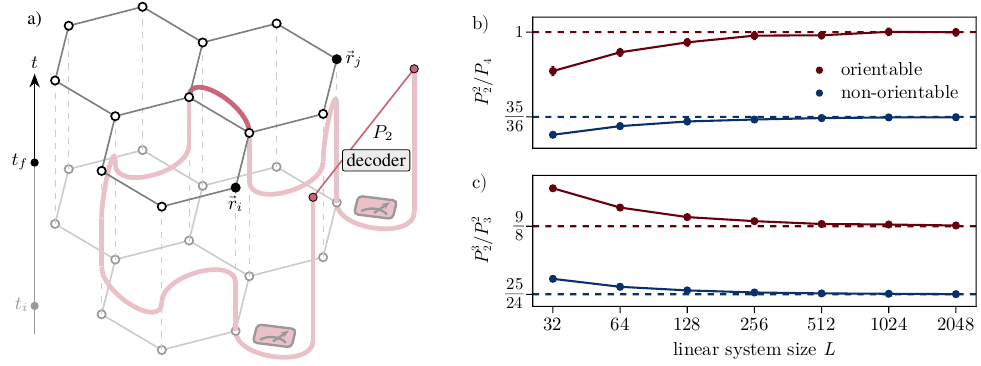}
    \caption{{\bf Loop statistics reveal universal ratios of long-distance loop connectivities.}
        (a) Illustration of an ancilla scheme for probing the statistics of macroscopic loops, shown here for the correlator $P_2$ that requires two ancilla probes. The two ancillae have a well-defined parity iff they are connected by a long string through the bulk. The definite parity can be  probed by a {\it decoder} taking the Majorana parity measurement outcomes in the bulk as input.
        This scheme can be generalized to probe  $P_n$ as defined in Eq.~\eqref{eq:PD_nPt}, i.e., the probability that $n$ points $\vec{r}_1, \dots, \vec{r}_n$ in space-time lie along the same loop.
        The distance $r_{ij} \equiv \abs{\vec{r}_i - \vec{r}_j}$ is taken to be macroscopically large such that non-zero $P_n$ 
        necessarily implies that the points lie along an infinite loop.
        At initial time $t_i$, a marked loop is injected into the circuit at each $\vec{r}_i$ by preparing a pair of ancilla Majoranas 
        and then measuring the parity between one ancilla and the corresponding bulk Majorana.
        In the qubit language, this amounts to measuring the parity between a system qubit and an ancilla qubit.
        After running the circuit to late times ($t_f$), all system Majoranas are projected out.
        Information about $P_n$ is then encoded in the mutual information between different ancillae.
        The ratios
        (b) $P_2^2 / P_4$ and (c) $P_2^3 / P_3^2$ in the entangled phase approach universal values (dashed lines) which differ for the two symmetry classes.
        Data shown here are taken at the isotropic point, but the ratios are universal throughout the entire phase. Data averaged over 100 disorder realizations with 10,000 samples each.
    }
    \label{fig:ancilla_scheme}
    \label{fig:watermelon_ratios}
\end{figure*}
%%%%%%%%%%%%%%%%%%%%%%%%%%%%%%%%%%%%%%%%%%%%%%%%%%%%%%%%%%%%%%%%%%%%%%%%%%%%

In order to access the universal part of the PD distribution, i.e., the parameter $\theta$, one faces two challenges: 
(i) implementing measurements of loop quantities in space-time and on distances $\ell\gg L_x L_y$ much larger than the spatial extent and
(ii) eliminating the dependence on the non-universal parameter $f$ in Eq.~\eqref{eq:PD_distrib}. 
Let us start by solving the latter.
Consider $m$ points $\vec{r}_i$ with $i=1,\dots,m$ in space-time which are separated by macroscopic distances $\abs{\vec{r}_i - \vec{r}_j} \gtrsim L_x,L_y$.
In the thermodynamic limit, if any two points $\vec{r}_i$ and $\vec{r}_j$ are on the same loop, then that loop is macroscopic.
The probability that all $m$ points lie along the same loop is~\cite{Caci_2023}
\be
    P_m \equiv P(\vec{r}_1, \dots, \vec{r}_m) = f^m \frac{\Gamma(1 + \theta)\Gamma(m)}{\Gamma(m + \theta)}.
    \label{eq:PD_nPt}
\ee
Here, $\Gamma$ is the $\Gamma$-function and the non-universal quantity $f$ appears as a prefactor.
Thus, $f$ can be eliminated by taking appropriate ratios $P_{m_1}^{k_1} / P_{m_2}^{k_2}$ for $m_1 k_1 = m_2 k_2$, leading to universal quantities which distinguish the two symmetry classes~\cite{Nahum_2013_soup}.

We propose a way to measure the probabilities $P_m$ directly by generalizing existing ancilla schemes~\cite{Gullans_2020_probe, Klocke_2023}. 
Since $P_2$ is simply the large-separation limit $r\rightarrow\infty$ of the two-point watermelon correlator $G_2(r)$, it may be measured via the protocol described earlier in Sec.~\ref{sec:Watermelon correlators}.
For higher-order quantities $P_m$, we use $m$ ancillae to insert marked loops at all $m$ points in space-time, as depicted in Fig.~\ref{fig:ancilla_scheme}.
Once again we evolve the circuit to late times and then measure all bulk Majoranas, leaving open loops only terminating on the ancillae.
The mutual information between groups of ancillae counts precisely the probability that the space-time points where we measured the bulk and ancilla had been along the same loop, so a quantity like $P_3$ amounts to requiring $I_2(A,B) = I_2(B, C) = I_2(C,A) = 1$.
In practice this can be obtained by comparing the magnitude of $I_2(A,B)$ before and after tracing out ancilla $C$, where the latter case simply reduces to $P_2$.
Here we consider such a scheme with four points $\vec{r}_i$ at separations on the order of $L_y$ and measure $P_2$, $P_3$, and $P_4$ to obtain the following ratios,
\[
    \begin{aligned}
        \frac{P_2^2}{P_4}   & = \frac{\Gamma(1+\theta) \Gamma(4+\theta)}{6\Gamma(2+\theta)^2}   & = \begin{cases}1, & \theta = 1 \\ \frac{35}{36}, & \theta=\frac12 \end{cases},       \\
        \frac{P_2^3}{P_3^2} & = \frac{\Gamma(1+\theta)\Gamma(3+\theta)^2}{4 \Gamma(2+\theta)^3} & = \begin{cases} \frac98, & \theta = 1 \\ \frac{25}{24}, & \theta=\frac12\end{cases}.
    \end{aligned}
\]
Importantly, both give values which differ between the two symmetry classes.
In Fig.~\ref{fig:watermelon_ratios}, we show that the ratios indeed converge toward the universal value as system size is increased.
This then offers a practical scheme for distinguishing between the long-loop phases which would otherwise not be possible from the steady-state entanglement.

We note that when implementing (open) absorbing boundaries in space-time, any macroscopic loop will be cut into an extensive number of individual \emph{open loops}.  
Then the loop length distribution matches the first passage time distribution for a random walk in one spatial dimension with absorbing boundaries 
\[
	P_\textrm{bulk}(\ell) \sim P_\textrm{FP}(\ell) \sim \ell^{-5/2} e^{-\alpha \ell} \,
\] 
with $\alpha^{-1} \sim L_x L_y$.
In Fig.~\ref{fig:open_loops_FBC} (see Appendix), we verify this expected scaling for open loops in the metallic phase of both symmetry classes.

%%%%%%%%%%%%%%%%%%%%%%%%%%%%%%%%%%%%%%%%%%%%%%%%%%%%%%%%%%%%%%%%%%%%%%%%%%%%
\subsubsection*{Analytical entanglement scaling and phase boundaries \texorpdfstring{\\}{} from the diffusion picture}\label{sss:diffusion_coeff}
%%%%%%%%%%%%%%%%%%%%%%%%%%%%%%%%%%%%%%%%%%%%%%%%%%%%%%%%%%%%%%%%%%%%%%%%%%%%

Viewing the dynamics in the metallic phase as Majorana world lines undergoing Brownian motion in space-time allows us to establish a direct link between (i) the non-linear sigma model in Eq.~\eqref{eq:NonLinS}, (ii) the entanglement entropy and, (iii) for class BDI, the microscopic measurement probabilities. 
Note that the latter enables \emph{accurate analytical predictions}, for symmetry class BDI, for the entanglement scaling and the location of the transition line between the metallic and the localized phases for arbitrary geometries. 

We start by considering loop endpoints undergoing Brownian motion in space-time with an effective diffusion constant $\mathcal{D}$ at large wavelengths. 
The entanglement entropy at a fixed time $t=T$ is then determined from open arcs terminating at the temporal boundary. 
Designating one of the arc endpoints as the ``start'', we may view the arc as the path of a random walker which starts near and eventually terminates upon an absorbing boundary at $t=T$.
Then the total bulk length of the arc is equivalent to the first passage time $\tau$ of the random walker
\footnote{More precisely, the movement of the loop endpoint along the temporal direction should be viewed as a persistent random walk, where the direction of propagation is reflected whenever the Majorana is involved in a measurement. This yields a mild correction to the timescales, but does not alter the universality.}.
The distribution of first passage times for one absorbing boundary in three dimensions is known to be $P_\textrm{FP}(\tau) \approx \sqrt{\frac{2}{\pi}} \tau^{-3/2}$. 
During this time, both endpoints undergo Brownian motion in the two-dimensional spatial plane, yielding a distribution for the distances between endpoints $P_\textrm{2D}(r,\tau) \approx \tfrac{2r}{\mathcal{D}\tau}e^{-r^2/\mathcal{D}\tau}$.
Averaging with respect to the time $\tau$, yields the expected distribution of spatial displacements
\[
    P_\textrm{surf}(r) = \int \dif \tau \, P_\textrm{FP}(\tau) P_\textrm{2D}(r,\tau) = \sqrt{2\mathcal{D}}/r^2
\]
with $r^2=\ell_x^2+\ell_y^2$. 
The entanglement entropy along a cut depends on the distribution of loop lengths projected along a fixed axis.
For isotropic Brownian motion in space, this gives
\[
    P_\textrm{surf}^x(\ell_x) = 4\int_0^\infty \dif \ell_y \, \frac{P_\textrm{surf}(r)}{2\pi r} = \frac{2\sqrt{2\mathcal{D}}}{\pi \ell_x^2}.
\]
From this distribution we infer that the logarithmic entanglement scaling has coefficient $\tilde{c}_x = 6\sqrt{2\mathcal{D}}{\pi}$.
Varying the relative measurement probabilities of different bonds will generally introduce some anisotropy to the Brownian motion such that $\mathcal{D}^x \neq \mathcal{D}^y$ and $\tilde{c}_\alpha \propto \sqrt{D^\alpha}$. This scaling of the logarithmic entanglement proportional to the square root of the diffusion constant appears also in monitored fermion systems with unitary dynamics~\cite{Chahine_2023, Poboiko2023}, and is seemingly generic for measurement-induced fermion liquid-type phases. 

The effective diffusion constant $\mathcal{D}$ emerges at long wavelengths from the renormalization group (RG) flow of the \nlsm in Eq.~\eqref{eq:NonLinS}. The \nlsm is a single-parameter theory, initialized with the microscopic short-distance diffusion constant $\mathcal{D}_\text{mic}$, which depends on the lattice geometry. The RG flow only depends on the fugacity $n$, the symmetry class and the dimensionality $d$. For $n=1$ and $d=3$, each symmetry class thus has a one-to-one correspondence $\mathcal{D}_{\text{mic}}\leftrightarrow \mathcal{D}$. For bipartite lattices, i.e., for class BDI, the random walker is always on the same sublattice after two steps and the microscopic diffusion constant is readily inferred from the measurement probabilities.
Consider the measurement-only Kitaev circuit, where bonds are measured with probabilities $K_x + K_y + K_z = 1$. The diffusion constants may be estimated by considering the mean-squared displacement of a random walker with jump rates set by $K_\alpha$ and bonds of unit length. Setting $\mathcal{D}^z_{\text{mic}}$ as the diffusion constant for the direction parallel to the $ZZ$-bonds and $\mathcal{D}^{\perp}_{\text{mic}}$ as the diffusion constant perpendicular to it yields
\[
    \begin{aligned}
        \mathcal{D}_{\text{mic}}^z & = \frac98 K_z(K_x + K_y) \,, \ \mathcal{D}_{\text{mic}}^{\perp}  = \frac16 (2\mathcal{D}_{\text{mic}}^z + 9K_x K_y) \,,                                                       \\
        \mathcal{D}_{\text{mic}}  &= \frac12(\mathcal{D}_{\text{mic}}^z + \mathcal{D}_{\text{mic}}^{\perp}) =\frac34 (K_x K_y + K_x K_z + K_yK_z)  \,.
    \end{aligned}
\]
The total diffusion constant $\mathcal{D}_{\text{mic}}$ is \emph{invariant under continuous rotation} of $K_{x/y/z}$ with respect to the isotropic point 
-- this explains the circular symmetry of both the metallic phase and its phase boundary, as numerically found in Refs.~\cite{Vijay23Kitaev,Ippoliti23kitaev,Zhu23structuredVolumeLaw}.

\begin{figure}[t]
    \centering
    \includegraphics[width=\columnwidth]{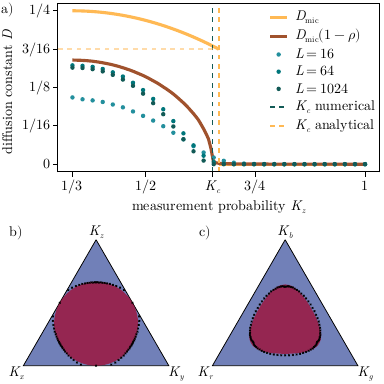}
    \caption{
        {\bf Effective diffusion constant} for the critical phase of the measurement-only Kitaev model.
        (a) A comparison of the effective diffusion constant $\mathcal{D}$ extracted from the distribution $P_\textrm{surf}(\ell)$ (blue) and the microscopic diffusion constant $\mathcal{D}_\textrm{mic}$ (orange), computed along the line $K_x = K_y = \tfrac12(1 - K_z)$.
        At the transition, $\mathcal{D}$ vanishes, while the microscopic $\mathcal{D}_\textrm{mic}$ crosses the threshold value $\mathcal{D}_\textrm{mic,c} = \tfrac{3}{16}$.
        The vertical dashed lines mark the critical measurement probability $K_c$ found by finite-size scaling in Fig.~\ref{fig:HC_span_FSS} (blue) and by $D_\textrm{mic}$ (orange).
        Rescaling by the numerically obtained fraction $1-\rho$ of volume available to the Brownian walkers brings $\mathcal{D}_\textrm{mic}(1-\rho)$ (brown) and $\mathcal{D}$ closely in line with one another.
        Phase diagrams for the measurement-only (b) Kitaev and (c) Kekul\'e-Kitaev models.
        We show a direct comparison of the phase boundary determined via the microscopic diffusion constant $\mathcal{D}_\textrm{mic} = \mathcal{D}_\textrm{mic,c}$ (solid colors) and via scaling collapse of the spanning number (black crosses).
       }
    \label{fig:diffusion_vs_c}
\end{figure}

We emphasize two particular regimes: (i) In the vicinity of the isotropic point $K_\alpha=1/3$, i.e., deep in the metallic phase, we expect the RG flow corrections to $\mathcal{D}$ to be weak. Thus setting $\mathcal{D}=\mathcal{D}_{\text{mic}}$ proves to be a good approximation in the vicinity of the isotropic point, see Fig.~\ref{fig:diffusion_vs_c}(a). Since under RG, the diffusion constant will generally flow towards a smaller value, $\mathcal{D}_{\text{mic}}$ at the isotropic point serves as an upper bound for the average entanglement growth in each direction $\mathcal{D}\le 1/4 $. (ii) In the vicinity of the critical line marking the transition between the metallic and the localized phase, the renormalization of $\mathcal{D}_{\text{mic}}$ is strong and $\mathcal{D}_{\text{mic}}\gg\mathcal{D}$. Nevertheless, the one-to-one correspondence $\mathcal{D}_{\text{mic}}\leftrightarrow\mathcal{D}$~\footnote{Neglecting the effect of anisotropy in the microscopic values.} allows us to extract the position of the critical line for any bipartite geometry. At $K_x = K_y = \tfrac12$ and $K_z = 0$, the location of the critical point is exactly known: the system decouples into disconnected one-dimensional strings each of which displays percolation critical behavior~\cite{Klocke_2023}. Using this point as an estimate for the critical value $\mathcal{D}_{\text{mic},c}$ provides
\begin{align}\label{eq:critical_D}
    \text{critical diffusion constant:} \ \mathcal{D}_{\text{mic},c}=\frac{3}{16}.
\end{align}
For the Kitaev honeycomb lattice, this yields a circular contour in parameter space (see Fig.~\ref{fig:diffusion_vs_c}(b)) which coincides exceptionally well with numerical estimates for the entanglement transition. 
To further underpin the strength of the diffusion picture, we provide the numerically obtained phase diagram for the measurement-only Kekulé-Kitaev geometry illustrated in Fig.~\ref{fig:overview_circuit_diagram}(b). This system also belongs to symmetry class BDI but its phase diagram exhibits a different shape of the critical line as displayed in Fig.~\ref{fig:diffusion_vs_c}(c). Using Eq.~\eqref{eq:critical_D} again provides a rather accurate estimate for its phase boundary based on a calculation of the diffusion constant as
 $\mathcal{D}_\textrm{mic}  = \frac{3}{8}\bigl [K_g K_r (2-3K_gK_r) + K_b(K_g + K_r)(2 + K_g K_r) - 3 K_b^2 (K_g^2 - K_g K_r + K_r^2)\bigr ] $, 
 where again we have considered the mean-squared displacement of a random walker after two steps along lattice edges, averaging over starting points in the unit cell.

%%%%%%%%%%%%%%%%%%%%%%%%%%%%%%%%%%%%%%%%%%%%%%%%%%%%%%%%%%%%%%%%%%%%%%%%%%%%
%%%%%%%%%%%%%%%%%%%%%%%%%%%%%%%%%%%%%%%%%%%%%%%%%%%%%%%%%%%%%%%%%%%%%%%%%%%%

Lastly, let us discuss a more rigorous approximation scheme for the entanglement entropy. We note that even deep in the metallic phase, the microscopic diffusion constant $\mathcal{D}_\textrm{mic}$ is larger than $\mathcal{D}$ determined from the entanglement scaling by a factor of approximately $3/2$.
This difference reflects the fact that the microscopic calculation treated the loop endpoints as \emph{non-interacting} Brownian walkers.
In actuality, due to the Pauli principle, the loops are constrained to not overlap with themselves or one another, corresponding to an effective excluded volume.
To account for this, consider the role of a nonzero density $\rho$ of closed loops in the bulk of space-time.
These act as obstructions to the Brownian motion of open arc endpoints, suppressing diffusion.
At leading order, this may be treated as a uniform probability $\rho$ that at any step the random walker remains in place, sending $\mathcal{D} \rightarrow \mathcal{D}(1-\rho)$. The density $\rho$ is hard to extract analytically but easily accessible numerically. 
We find that the fraction $\rho$ varies continuously from $\rho\approx1/3$ at the isotropic point to $\rho=1$ in the area-law phase.
In Fig.~\ref{fig:diffusion_vs_c}(a), we show that the rescaled microscopic diffusion coefficient $\mathcal{D}_\textrm{mic}(1-\rho)$ better matches the fitted $\mathcal{D}$, reproducing the magnitude deep in the metallic phase and going to zero continuously at the transition.
Thus, accounting for the excluded volume effect accurately captures the interactions neglected in $\mathcal{D}_\textrm{mic}$.
Alternatively we might arrive at the same correction by considering a self-avoiding walk on a lattice with coordination number $z$ and connectivity constant $\mu$.
Here the quantity $1-\rho$ corresponds to the ratio $\mu/z$, which falls in the range $\mu/z \in [1.2, 1.5]$ for various 3D lattices~\cite{Hughes_1995}.

The direct relation between the entanglement scaling, i.e., the prefactor $\tilde{c}$ of the metallic $L\log(L)$ term, and the effective diffusion constant $\mathcal{D}$ at distance $L$ allows us to numerically track the RG flow of $\mathcal{D}$ from its microscopic starting point $\mathcal{D}_{\text{mic}}$ towards its asymptotic value as illustrated in Fig.~\ref{fig:diffusion_vs_c}(a) for linear system sizes $L=16,64,1024$. This is remarkable, since in $d=3$ dimensions, the \nlsm is no longer perturbatively controlled and the RG flow of $\mathcal{D}$ is not accessible analytically. 
We emphasize the connection between $\tilde{c}$, the diffusion constant for Brownian motion of loops, and their relation to the logarithmic entanglement scaling found in weakly disordered Fermi liquids.
In the ground state of a disordered metal with disorder strength $\gamma$, the diffusion constant is related to the Fermi velocity $v_F$ and the disorder by $\mathcal{D} \sim v_F^2 / \gamma$, implying $\tilde{c} \sim v_F / \sqrt{\gamma}$.
Indeed, such proportionality is expected from calculations of the entanglement entropy from a Fermi surface~\cite{Wolf2006,Klich2006,Li2006,Swingle2010}. 

%%%%%%%%%%%%%%%%%%%%%%%%%%%%%%%%%%%%%%%%%%%%%%%%%%%%%%%%%%%%%%%%%%%%%%%%%%%%
\subsection{Quantum Lifshitz criticality and universal scaling behavior at the entanglement transition}
\label{ss:universality}
%%%%%%%%%%%%%%%%%%%%%%%%%%%%%%%%%%%%%%%%%%%%%%%%%%%%%%%%%%%%%%%%%%%%%%%%%%%%

In both symmetry classes, the Majorana liquid and localized phases are separated by a critical line exhibiting quantum Lifshitz scaling of the entanglement entropy. At this critical line, the effective diffusion constant vanishes, resulting in an area law for the entanglement entropy. The distribution of Majorana loop lengths, however, is not yet localized but displays universal algebraic scaling behavior that clearly differentiates the two symmetry classes. This unique combination of universal algebraic scaling and the area law leads to the distinctive subleading entanglement scaling characteristic for quantum Lifshitz criticality.

In the Kitaev honeycomb model for symmetry class BDI, the critical line is reached by tuning the probabilities $K_\alpha$. Quantum Lifshitz scaling is observed, except at the three points where one of the $K_\alpha = 0$, reducing the model to an array of one-dimensional chains. For class D, we set the next-nearest neighbor bond measurement probabilities to $J_x = J_y = J_z = J$ and tune ${J, K_\alpha}$ through the critical regions. For $J > 0$, quantum Lifshitz scaling is generally present.

%%%%%%%%%%%%%%%%%%%%%%%%%%%%%%%%%%%%%%%%%%%%%%%%%%%%%%%%%%%%%%%%%%%%%%%%%%%%
\subsubsection*{Quantum Lifshitz entanglement and surface loops}
%%%%%%%%%%%%%%%%%%%%%%%%%%%%%%%%%%%%%%%%%%%%%%%%%%%%%%%%%%%%%%%%%%%%%%%%%%%%
\begin{figure}[t]
    \centering
    \includegraphics[width=\columnwidth]{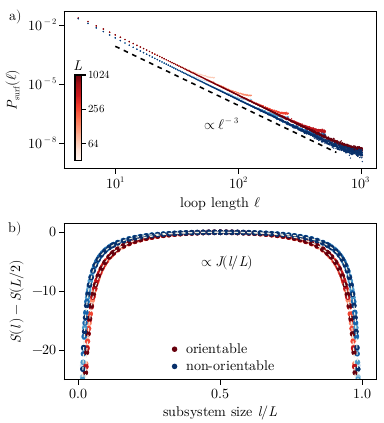}
    \caption{
    \textbf{Quantum Lifshitz criticality from surface loops}
    (a) Surface loop length distribution at the critical point along the line $p_y = p_z = (1-p_x)/2$ for the orientable (red) and non-orientable (blue) loop models.
    The dashed black line reflects an asymptotic scaling $P_\textrm{surf}(\ell) \sim \ell^{-3}$.
    (b) Data collapse of the subsystem entanglement entropy $S(\ell)$ onto a single curve described by the Lifshitz scaling function $J(l/L)$ of Eq.~\eqref{eq:lifshitz_func}. 
    In both panels we show data for linear system sizes $L=64,128,256,512,1024$.
    }
    \label{fig:critical_EE_scaling}
\end{figure}

The entanglement entropy in the loop framework is computed from Eq.~\eqref{eq:entanglement_integral}. 
At a genuine two-dimensional transition we observe an algebraic decay of surface loop lengths, $P_\textrm{surf}(\ell) \propto \ell^{-\gamma_{\text{surf}}}$. 
Away from the percolation points with one $K_\alpha=0$, we find that $\gamma_{\text{surf}}\approx3$, shown in Fig.~\ref{fig:critical_EE_scaling}(a). 
This is faster than in the Majorana liquid phase, where $\gamma_{\text{surf}} = 2$ leads to an $L\log(L)-$growth.
As a result, the transition exhibits a subsystem dependence of the entanglement entropy $S(\ell, L) = aL + b J(\ell / L)$. 
Here $J(u)$ is the ``Lifshitz'' scaling function~\footnote{For unit aspect ratio of the spatio-temporal volume.}
\be
    \begin{aligned}
        J(u) &= \log\left(\frac{\theta_3(i\lambda u)\theta_3(i\lambda(1-u))}{\eta(2iu)\eta(2i(1-u))}\right) \,,
    \end{aligned}
    \label{eq:lifshitz_func}
\ee
where $\theta_3$ is the Jacobi theta function, $\eta$ is the Dedekind eta function, and $\lambda$ is a fitting parameter~\cite{Fradkin04RK, Stephan_2013}.
In Fig.~\ref{fig:critical_EE_scaling}(b) we show the collapse of the entanglement data onto a curve accurately described by this Lifshitz scaling function.
For both symmetry classes, we find $\lambda = 3.4 \pm 0.1$ for cuts along either direction of the system.

Although derived initially for a quantum Lifshitz model~\cite{Fradkin04RK}, this torus entanglement scaling has been observed in a variety of $(2+1)$-dimensional CFTs~\cite{Inglis_2013,Chen_2015}.
Notably, the R\'enyi entropy takes precisely the form of Eq.~\eqref{eq:lifshitz_func} also for quantum dimer models and resonating valence bond (RVB) states~\cite{Stephan_2012, Stephan_2013}.
Given the close link between models of dimers, loops, and height fields~\cite{Sutherland_1988, Kohmoto_1988, Rokhsar_1988, Affleck_1991, Fradkin04RK, Fendley_2006}, one may expect that this entanglement scaling ought to extend also to the case of loops with fugacity $n=1$ (relevant to the quantum circuits at hand).
Indeed, higher fugacity loop models have been employed in numerical studies to verify the Lifshitz scaling of entanglement in quantum magnets~\cite{Inglis_2013, Kallin_2011}. 
Similarly, the bond-length distribution in the antiferromagnetic Heisenberg model on the square lattice exhibits the same scaling exponent as $P_\textrm{surf}(\ell)$ in our circuit model~\cite{Sandvik_2005}.
We note here that quantum Lifshitz scaling is compatible with the simultaneous presence of conformal symmetry and an area law of the entanglement in $(2+1)$ dimensions. 
It is remarkable that this entanglement scaling, though apparently quite universal, has been reported previously only in the context of \emph{interacting} circuits~\cite{Zhu23structuredVolumeLaw, Barkeshli2021measuretoriccode}.
For fermion-parity symmetric circuits, the mapping to an interacting dimer model suggests that Lifshitz entanglement scaling is quite generic.
More broadly, requiring conformal invariance at the entanglement transition imposes strict constraints such that one might generically expect scaling of the form Eq.~\eqref{eq:lifshitz_func} except at fine-tuned points.

%%%%%%%%%%%%%%%%%%%%%%%%%%%%%%%%%%%%%%%%%%%%%%%%%%%%%%%%%%%%%%%%%%%%%%%%%%%%
\subsubsection*{Correlation length exponent \texorpdfstring{$\nu$}{ν}}
%%%%%%%%%%%%%%%%%%%%%%%%%%%%%%%%%%%%%%%%%%%%%%%%%%%%%%%%%%%%%%%%%%%%%%%%%%%%

%%%%%%%%%%%%%%%%%%%%%%%%%%%%%%%%%%%%%%%%%%%%%%%%%%%%%%%%%%%%%%%%%%%%%%%%%%%%
\begin{figure}[b]
    \centering
    \includegraphics[width=\columnwidth]{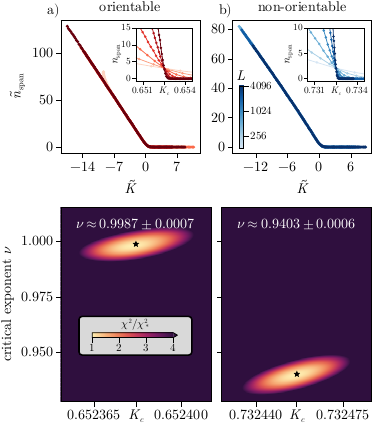}
    \caption{{\bf Correlation length exponent from spanning number.}
    	Data shown is for the honeycomb Kitaev circuit. Row (a) corresponds to symmetry class BDI with orientable loops and (b) to class D with broken orientability.
         We take $K=K_x$ in both cases, while $K_y = K_z = (1-K)/2$ in (a) and $K_y = K_z = J = (1-K)/3$ in (b).
        Top panels: Scaling collapse of the spanning number in the vicinity of the transition, with scaling variable $\Tilde{K}$ as defined in Eq.~\eqref{eq:FSS_ansatz}.
        Data shown is for linear system sizes $L = 128, 256, 512, 1024, 2048, 4096$
        with up to $N= 2L^2 = 2^{25} \approx 3.35 \cdot 10^7$ qubits. The inset shows the unscaled data close to the critical point.
        Bottom panels: Error estimation of critical exponent $\nu$ for the non-orientable and orientable loop model. Keeping all other scaling coefficients fixed we vary the estimate for the critical point $K_c$ and correlation length exponent $\nu$ and evaluate the cost function. The heatmap then shows the normalized cost function $\chi^2/\chi^2_*$ for the FSS analysis of the spanning number shown in the top panels, stars mark the minima $(K_c^*, \nu^*)$. The colormap has been truncated at four times the minimum cost to give a visual estimate for the uncertainty in $K_c^*$ and $\nu^*$. The error value has been estimated from the $\chi^2_* + 4$ contour.
        The optimization was performed on system sizes $L \geq 256$ for $\tilde{K} \in (-6,6)$,  $\tilde{K} \in (-7,7)$ for the orientable and non-orientable case, respectively. In both cases, the final cost function divided by the number of degrees of freedom is sufficiently close to one $|\chi^2_r -1 | < 0.025$, indicating a good fit.
    }
    \label{fig:HC_span_FSS}
\end{figure}
%%%%%%%%%%%%%%%%%%%%%%%%%%%%%%%%%%%%%%%%%%%%%%%%%%%%%%%%%%%%%%%%%%%%%%%%%%%%

To accurately determine the critical point $K_c$ and the correlation length exponent $\nu$, we consider the spanning number $n_s$ for fixed aspect ratios $L_x = 2L_y$ (i.e. $L_x \times L_x$ plaquettes) and circuit depth $t=L_y$.
We then perform a finite-size scaling (FSS) analysis using an ansatz generically of the form in Eq.~\eqref{eq:FSS_ansatz} and a cubic B-spline to fit the scaling function.
Moreover, by taking sufficiently large system sizes and a narrow window around the critical point, we find that any irrelevant scaling variables may be dropped from the ansatz.
In Fig.~\ref{fig:HC_span_FSS} we show the resulting data collapse for the transition in (i) the (orientable) honeycomb Kitaev circuit along the line $K_x = K_y = (1 - K_z)/2$
and (ii) the (non-orientable) honeycomb Kitaev circuit with next-nearest neighbor parity checks along $K_x = K_y = J = (1 - K_z)/3$.

The results are summarized in Tab.~\ref{tab:nu}. We find consistent values of $\nu$ in both symmetry classes for several other lattice geometries (see Appendix~\ref{app:nonorientable_geometries}).
These results thereby highlight that the entanglement transitions in (2+1)-dimensional measurement-only Majorana circuits have \emph{distinct universality} depending on the presence or absence of orientability, i.e., depending on the symmetry class. We fix the critical points $K_c$ at the values reported here and use them in all further analysis.
\begin{table}[h]
	\begin{tabular}{|c||c|c|}\hline
	symmetry 	& BDI (orientable)		& D (non-orientable) \\
 \hline
	$K_c$	& $0.6523817 \pm 0.0000021$ & $0.7324564 \pm 0.0000015$ \\
	\hline
	$\nu$ 	& $0.9987 \pm 0.0007$ 		& $0.9403 \pm 0.0006$ \\
	\hline
	\end{tabular}
	\caption{{\bf Critical point and correlation length exponent} from the spanning number in Fig.~\ref{fig:HC_span_FSS}.
		}\label{tab:nu}
\end{table}

%%%%%%%%%%%%%%%%%%%%%%%%%%%%%%%%%%%%%%%%%%%%%%%%%%%%%%%%%%%%%%%%%%%%%%%%%%%%
% \subsubsection*{Fractal dimension \texorpdfstring{$d_f$}{d_f}}
\subsubsection*{Fisher exponent \texorpdfstring{$\tau$}{τ}}
%%%%%%%%%%%%%%%%%%%%%%%%%%%%%%%%%%%%%%%%%%%%%%%%%%%%%%%%%%%%%%%%%%%%%%%%%%%%

At the transition and throughout the extended critical phase, the bulk loop length distribution is expected to follow a power-law $P_\textrm{bulk}(\ell) \propto \ell^{-\tau}$, where the Fisher exponent $\tau$ is related to the fractal dimension $d_f$ via the hyperscaling relation $\tau = (d / d_f) + 1$.
In the extended critical phase, the Brownian nature of the loops yields $d_f = 2$ and thus $\tau = 5/2$.
By contrast, a mean-field analysis~\cite{Amit_1976, McKane_1976, Nahum_2012_vortex, Nahum_2013_3D} predicts $\eta = 0$, corresponding to $d_f = 5/2$ and $\tau = 11/5$ at the transition.
Tuning to the critical point determined from FSS of the spanning number, we obtain the exponent $\tau$ by fitting a power-law to the bulk loop length distribution $P_\textrm{bulk}(\ell)$. 
The result is provided in Tab.~\ref{tab:Fisher_exp}.

These results deviate very slightly from previous numerical studies on 3D classical loop models, where hyperscaling gave $\tau = 2.184 \pm 0.003$ for the orientable case~\cite{Ortuno_2009} and $\tau=2.178 \pm 0.002$ for the non-orientable case.
In both symmetry classes, the decay of $P_\textrm{bulk}(\ell)$ is \emph{slower} than expected for $\tau = 11/5$, indicative of small but finite renormalization due to fluctuations beyond the mean-field prediction.
However, the difference in $\tau$ between the two symmetry classes is rather subtle, making it challenging to distinguish them even at the large system sizes accessible in the loop framework.

%%%%%%%%%%%%%%%%%%%%%%%%%%%%%%%%%%%%%%%%%%%%%%%%%%%%%%%%%%%%%%%%%%%%%%%%%%%%
\begin{figure}[t]
    \centering
    \includegraphics[width=\columnwidth]{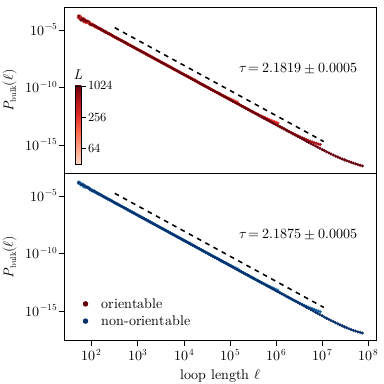}
    \caption{{\bf Fisher exponent at the critical point.}
        Shown is the power-law scaling of the bulk length distribution for closed loops $P_\textrm{bulk}(\ell)$ at the transition with circuit depth $t=4L_y$.
        Data are shown for linear dimension $L_y = 32, 64, 128, 256, 512, 1024$.
        We fix $K=K_c$ known from the spanning number collapse in Fig.~\ref{fig:HC_span_FSS}.
        The black dashed line marks the power-law found by fitting $P_\textrm{bulk}(\ell) \sim \ell^{-\tau}$ to all data in the scaling regime $10^{5/2} \leq \ell \leq 10^{-2}L^3$.
        In both symmetry classes, the resulting Fisher exponent $\tau$ deviates clearly from the mean-field value $\tau = 11/5$.
    }
    \label{fig:HC_fractal_dim}
\end{figure}
%%%%%%%%%%%%%%%%%%%%%%%%%%%%%%%%%%%%%%%%%%%%%%%%%%%%%%%%%%%%%%%%%%%%%%%%%%%%

Finally, we note that a direct observation of the Fisher exponent $\tau$ might be practically infeasible as it would require access to the bulk lengths of loops in space-time.
As an alternative, it might be far more practical to employ ancilla probes~\cite{Gullans_2020_probe, Klocke_2023} to instead extract the anomalous dimension $\eta$, from which we may determine $\tau$ via hyperscaling relations, see below.

\begin{table}[h]
	\begin{tabular}{|c||c|c|}\hline
	symmetry 	& BDI (orientable)		& D (non-orientable) \\
 \hline
	$\tau$	& $2.1819 \pm 0.0005$ & $2.1875 \pm 0.0005$ \\
	\hline
	$d_f$ 	& $2.5383 \pm 0.0011$ 		& $2.5263 \pm 0.0011$ \\
	\hline
	\end{tabular}
	\caption{{\bf Fisher exponent and fractal dimension} from Fig.~\ref{fig:HC_fractal_dim}.
		}
  \label{tab:Fisher_exp}
\end{table}

%%%%%%%%%%%%%%%%%%%%%%%%%%%%%%%%%%%%%%%%%%%%%%%%%%%%%%%%%%%%%%%%%%%%%%%%%%%%
\subsubsection*{Anomalous dimension \texorpdfstring{$\eta$}{η}}
%%%%%%%%%%%%%%%%%%%%%%%%%%%%%%%%%%%%%%%%%%%%%%%%%%%%%%%%%%%%%%%%%%%%%%%%%%%%

\begin{figure}[t]
    \centering
    \includegraphics[width=\columnwidth]{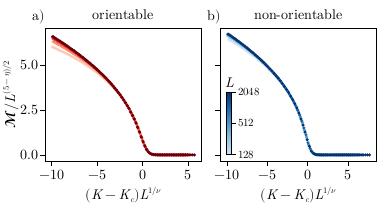}
    \caption{{\bf Anomalous dimension.}
        Scaling collapse of the quantitiy $\mathcal{M}/L^{(5-\eta)/2}$ as a function of $x = (K-K_c)L^{1/\nu}$ 
        for the orientable (a) and non-orientable (b) circuit. 
        Data shown is for linear system sizes $L=128,256,512,1024,2048$.
        To avoid the need for higher order terms in the finite-size scaling ansatz, optimization was restricted to $x \in (-1,2)$ and $x \in (-2,2)$ for the orientable and non-orientable case respectively, $L \geq 256$. Also here the cost function per degree of freedom is close to one, $|\chi^2_r -1 | < 0.05$.
    }
    \label{fig:HC_span_length}
\end{figure}

The anomalous dimension $\eta$ can be determined by examining the finite-size scaling behavior of the total \emph{length} of spanning loops $\mathcal{M}$,
\[
    \mathcal{M} L^{-(5-\eta)/2} = f(L^{1/\nu} (K - K_c)).
\]
Moreover, this provides a second, independent estimate of the correlation length exponent $\nu$.
In Fig.~\ref{fig:HC_span_length}, we show the scaling collapse of $\mathcal{M}$ in both symmetry classes. The results are displayed in Tab.~\ref{tab:anomalous_dim}. We note that the estimates for $\nu$ obtained in this finite-size scaling analysis are independent of the values extracted from the spanning number collapse, Tab.~\ref{tab:nu}.
Both approaches yield almost perfect agreement, which is a strong internal consistency check of our numerical approach.

\begin{table}[h]
	\begin{tabular}{|c||c|c|}\hline
	symmetry 	& BDI (orientable)		& D (non-orientable) \\
 \hline
	$\nu$ 	& $\phantom{-}1.002 \pm 0.002$ 		& $\phantom{-}0.942 \pm 0.004$ \\
	\hline
 $\eta$	& $ -0.084 \pm 0.004$ & $-0.066 \pm 0.007$ \\
	\hline\hline
$\eta $ (alt)	& $ -0.073 \pm 0.007$ & $-0.066 \pm 0.009$ \\
 \hline
	\end{tabular}
	\caption{{\bf Anomalous dimension}. Top rows: anomalous dimension $\eta$ and correlation length exponent $\nu$ from the spanning lengths in Fig.~\ref{fig:HC_span_length}. Bottom row: Anomalous dimension $\eta$, obtained in an alternative way from the watermelon correlators in Fig.~\ref{fig:critical_watermelon}.
		}\label{tab:anomalous_dim}
\end{table}

An alternative way to obtain an estimate for the anomalous dimension $\eta$ is to probe the two-leg watermelon correlator $G_2(r)$.
At the transition, this correlator decays algebraically with the distance $r$, $G_2(r) \propto r^{-(1 + \eta)}$.
Analogously to the correlators $P_n$ discussed in Fig.~\ref{fig:ancilla_scheme} which we considered earlier, $G_2(r)$ can be measured in an experiment by using an ancilla scheme: $G_2(r)$ is the average mutual information between two ancillae that were entangled with the circuit at space-time points separated by a distance $r$.

%%%%%%%%%%%%%%%%%%%%%%%%%%%%%%%%%%%%%%%%%%%%%%%%%%%%%%%%%%%%%%%%%%%%%%%%%%%%
\begin{figure}[t]
    \centering
    \includegraphics[width=\columnwidth]{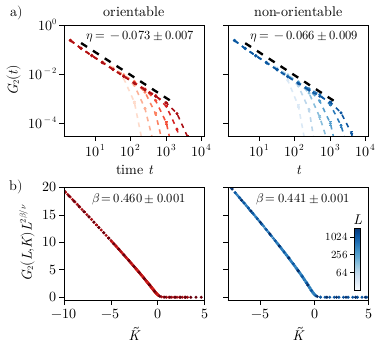}
    \caption{{\bf Watermelon correlator.}
        (a) Scaling of the two-leg watermelon correlator $G_2(r)$ at the critical point for the orientable (red) and non-orientable (blue) circuit.
        The separation $r$ is varied by increasing the circuit depth $t$ between the two points where ancillae couple to the bulk.
        Data here and in (b) are shown for system sizes $L=32,64,128,256,512,1024$.
        The black dashed lines show a regime of power-law scaling $t^{-(1+\eta)}$, from which we find $\eta = -0.073 \pm 0.007$ for the orientable class and $\eta = -0.066 \pm 0.009$ for the non-orientable class.
        (b) Scaling collapse of the two-leg watermelon correlator $G_2(K,r)$ at macroscopic distances $r \sim L$.
        The rescaled correlator $G_2(K, L) L^{2\beta/\nu}$ is a smooth function of the scaling variable $\tilde{K}$ defined in Eq.~\eqref{eq:beta_ansatz}.
        The critical point $K_c$ and critical exponent $\nu$ are fixed using the results from FSS of the spanning number in Fig.~\ref{fig:HC_span_FSS}.
        We find order parameter exponent $\beta = 0.460 \pm 0.001$ in the orientable circuit and $\beta = 0.441 \pm 0.001$ in the non-orientable circuit.
    }
    \label{fig:critical_watermelon}
\end{figure}
%%%%%%%%%%%%%%%%%%%%%%%%%%%%%%%%%%%%%%%%%%%%%%%%%%%%%%%%%%%%%%%%%%%%%%%%%%%%

Here, one would entangle ancillae at identical spatial positions but at times $t_1, t_2$ such that $r = \abs{t_1 - t_2}$.
Then, for a fixed system size, the scaling of $G_2(r)$ can be studied by simply varying the temporal separation (i.e.\ the circuit depth).
As shown in Fig.~\ref{fig:critical_watermelon}(a), we observe a power-law decay of the watermelon correlator with respect to the space-time separation that allows one to extract the anomalous dimension for both symmetry classes, displayed in Tab.~\ref{tab:anomalous_dim}.
We note that this approach is highly sensitive to subleading corrections and small deviations from the exact local of the critical point. We consider it thus generally less accurate to extract $\eta$ in this manner.

%%%%%%%%%%%%%%%%%%%%%%%%%%%%%%%%%%%%%%%%%%%%%%%%%%%%%%%%%%%%%%%%%%%%%%%%%%%%
\subsubsection*{Order parameter exponent \texorpdfstring{$\beta$}{ß}}
%%%%%%%%%%%%%%%%%%%%%%%%%%%%%%%%%%%%%%%%%%%%%%%%%%%%%%%%%%%%%%%%%%%%%%%%%%%%

In the loop framework, the Majorana liquid phase acquires a non-zero ferromagnetic-type order parameter, represented by the probability that two infinitely far separated space-time coordinates are connected by a single closed loop, i.e., $P_2\sim |K-K_c|^{2\beta}$. 
The critical exponent $\beta$ associated to this order parameter can be detected in an ancilla scheme similar to Fig.~\ref{fig:ancilla_scheme}. 
In practice, for systems of linear dimension $L$, one measures $G_2(r)$ for large space-time separations $r\sim L$. 
This introduces a subleading correction $G_2(r) = A r^{-x_2} + P_2$, where $x_2 = 2$ for Brownian loops~\cite{Nahum_2013_3D, Nahum_2012_vortex}.
For even moderate system sizes $L$, this subleading correction becomes negligible.
We may then treat finite-size effects for $G_2$ similarly to those for the order parameter in percolation,
\be
	\begin{aligned}
	G_2(r\sim L, K) &= L^{-2\beta/\nu}f(\tilde{K}),\\
	\tilde{K} &= (K - K_c)L^{1/\nu}\left[1 + A(K-K_c)\right].
	\end{aligned}
    \label{eq:beta_ansatz}
\ee
In the liquid phase, this scaling function reduces to the simpler form $G_2(r\sim L, K) \propto \abs{K_c - K}^{2\beta}$ on distances $L$ much larger than the correlation length $\xi$.

We determine $\beta$ by fitting the ansatz in Eq.~\eqref{eq:beta_ansatz} to the data in a narrow window enclosing the critical point.
In order to avoid systematic drift in the fitted exponents, we fix $\nu$ and $K_c$ at the values found previously from examining the spanning number.
The resulting scaling collapse is shown in Fig.~\ref{fig:critical_watermelon}(b). 
This yields accurate estimates of the critical exponent $\beta$ from the ancilla scheme which clearly distinguish both symmetry classes and which are summarized in Tab.~\ref{tab:beta_exp}.

\begin{table}[t]
	\begin{tabular}{|c||c|c|}\hline
	symmetry 	& BDI (orientable)		& D (non-orientable) \\
 \hline
    $\beta$ & $0.460 \pm 0.001$ & $0.441 \pm 0.001$\\
	\hline
	\end{tabular}
	\caption{{\bf Order parameter exponent} from the watermelon correlator of Fig.~\ref{fig:critical_watermelon}.
    }\label{tab:beta_exp}
\end{table}

%%%%%%%%%%%%%%%%%%%%%%%%%%%%%%%%%%%%%%%%%%%%%%%%%%%%%%%%%%%%%%%%%%%%%%%%%%%%
\subsubsection*{Hyperscaling and Discussion}
%%%%%%%%%%%%%%%%%%%%%%%%%%%%%%%%%%%%%%%%%%%%%%%%%%%%%%%%%%%%%%%%%%%%%%%%%%%%

Having obtained several independent estimates of various critical exponents, we now examine their consistency with one another via hyperscaling relations.
We further comment on the reliability of the different approaches for determining critical exponents and justify the choice of values reported in Table~\ref{tab:crit_exp}.

Given the Fisher exponent $\tau$, the anomalous dimension is given by the scaling relation $\eta = 5 - \tfrac{6}{\tau-1}$.
The resulting values are given in Tab.~\ref{tab:hyperscaling_exp}.
For class BDI, scaling gives a value of $\eta$ which lies between the two estimates determined earlier in Tab.~\ref{tab:anomalous_dim}.
In symmetry class D, scaling suggests a slightly smaller but not inconsistent magnitude of $\eta$ than previously found.
The anomalous dimension $\eta$ is notoriously difficult to determine directly, even for the large system sizes accessible with our loop model simulations.
Computing $\eta$ via the hyperscaling relation is highly susceptible to small deviations in the Fisher exponent $\tau$ and thus even smallest uncertainties in the critical point $K_c$.
By contrast, determining $\eta$ directly via FSS of $\mathcal{M}$ leaves $K_c$ as a fitting parameter and is generally found to be more robust.
This is supported not only by the near perfect scaling collapse shown in Fig.~\ref{fig:HC_span_length}, but also by the close agreement with the value of $\nu$ found by FSS of the spanning number.
As such we opted to report the corresponding value of $\eta$ in Tab.~\ref{tab:crit_exp}.

Hyperscaling allows for two alternative ways of estimating the order parameter exponent $\beta$: 
(i) given exponents $\eta$ and $\nu$ from the FSS of $\mathcal{M}$, one finds $\beta = \nu(\eta + 1)/2$, and
(ii) using the correlation length exponent $\nu$ and Fisher exponent $\tau$ from the FSS of $\mathcal{M}$ and $P_\textrm{bulk}(\ell)$, one finds  $\beta = 3\nu(\tau - 2)/(\tau-1)$.
We find that both approaches give values consistent with one another and clearly distinct in the two symmetry classes.
Moreover, hyperscaling gives values of $\beta$ which are in very good agreement with those found from the FSS analysis of the watermelon correlator $G_2(r,K)$ (see Tab.~\ref{tab:beta_exp}).
For consistency we also report $\beta$ as determined by the hyperscaling relation $\beta = \nu(\eta+1)/2$ in Tab.~\ref{tab:crit_exp}.

Lastly, let us comment on how the exponents we determine here compare with previous numerical studies.
The orientable loop model was studied previously in Ref.~\onlinecite{Ortuno_2009}.
Our results for the measurement-only Kitaev model yields critical exponents perfectly consistent with these earlier results 
(see Tab.~\ref{tab:crit_exp}), confirming the universality of the entanglement transition.
With our loop model simulations alloing us to reach appreciably larger system sizes (by two orders of magnitude) 
we are able to refine the estimated critical exponents to greater precision. 

\begin{table}[t]
	\begin{tabular}{|c||c|c|}\hline
	hyperscaling relation 	& BDI (orientable)		& D (non-orientable) \\
 \hline
	$\eta=5-\frac{6}{\tau-1}\phantom{\Big]}$	& $-0.0766 \pm 0.0021$ & $-0.0526 \pm 0.0021$ \\
	\hline
	$\beta = \nu(\eta + 1)/2 \phantom{\Big]}$ 	& $\phantom{-}0.4590 \pm 0.0020$ 		& $\phantom{-}0.4400 \pm 0.0040$ \\
	\hline
 	$\beta = 3\nu(\tau - 2)/(\tau-1) \phantom{\Big]}$ 	& $\phantom{-}0.4611 \pm 0.0013$ 		& $\phantom{-}0.4454 \pm 0.0012$ \\
	\hline
	\end{tabular}
	\caption{{\bf Hyperscaling relations} and corresponding alternative estimates of critical exponents.
		}\label{tab:hyperscaling_exp}
\end{table}

By contrast, the non-orientable loop model was studied only relatively recently in Ref.~\onlinecite{Serna_2021}.
Unlike the orientable case, the critical exponents we report for this symmetry class seem to differ appreciably from 
these previous estimates.
In particular, whereas we find $\nu=0.9403\pm 0.0006$, Ref.~\onlinecite{Serna_2021} suggested a smaller value $\nu=0.918 \pm 0.005$.
This in turn leads to a systematic difference in all other critical exponents found by hyperscaling relations or FSS which involves $\nu$ \footnote{
For instance, consider the anomalous dimension $\eta$ as found by FSS of $\mathcal{M}$.
From the RG flow~\cite{Amit_1976}, one expects that $\nu^{-1} - 2 + \eta$ is a constant.
Thus, a small error in the fitted value of $\nu$ is compensated by a corresponding shift in $\eta$.
For $\Delta(\nu^{-1}) = (0.9403^{-1} - 0.918^{-1})$, one finds $\Delta\eta = 0.026$, almost precisely matching the difference in reported values of $\eta$.
Since the fractal dimension was determined in Ref.~\onlinecite{Serna_2021} via the scaling relation $d_f = (5-\eta)/2$, we find that the discrepancy in $d_f$ can also be accounted for using a similar reasoning.
Finally, the assumption that $\nu^{-1} + \eta$ is constant in tandem with the scaling relation $\beta=\nu(1 + \eta)$ allows one to show that $\Delta\beta=\Delta\nu = 0.0223$, consistent with the values reported in Tab.~\ref{tab:crit_exp}.
}.
To reconcile this apparent discrepancy, we have simulated two alternative circuit/lattice geometries, including the L-lattice studied in Ref.~\onlinecite{Serna_2021}, which has allowed us to track down its source: In our approach we have expanded the FSS ansatz by {\it non-linear} terms, which we find to give superior data collapses over wider parameter ranges and system sizes. As discussed in Appendix \ref{app:nonorientable_geometries}, this also allows us to smoothly interpolate between our estimates and the ones found in Ref.~\cite{Serna_2021} as a function of (i) the range of included system sizes (ii) the choice of scaling variable, and (iii) the density of spline points. This leads to conclude that an improved FSS ansatz for the numerical finite-size data of Ref.~\onlinecite{Serna_2021} would give a somewhat higher estimate of the critical exponent $\nu$, perfectly consistent with what we report here.

Despite the discrepancy with earlier works, there is strong reason to believe that the critical exponent values we report here represent the most accurate and precise determination for this symmetry class to date.
Our numerical simulations reach system sizes which are much larger than previously studied, enabling greater control over subleading corrections in the FSS analysis.
This is further bolstered by the internal consistency between the two independent approaches for estimating $\nu$.
In Appendix.~\ref{app:nonorientable_geometries} we further demonstrate the universality of $\nu \approx 0.94$ by examining two additional lattice geometries, namely the measurement-only Yao-Kivelson model and Cardy's 3D L-lattice.

%%%%%%%%%%%%%%%%%%%%%%%%%%%%%%%%%%%%%%%%%%%%%%%%%%%%%%%%%%%%%%%%%%%%%%%%%%%%
\subsection{Simulation setup}
\label{sec:loop_simulation_setup}
%%%%%%%%%%%%%%%%%%%%%%%%%%%%%%%%%%%%%%%%%%%%%%%%%%%%%%%%%%%%%%%%%%%%%%%%%%%%

Let us close this Section by highlighting the benefits of the loop model representation in numerical simulations. 
In short, the loop model representation allows for a substantially more efficient numerical simulation of Majorana circuits~\cite{Klocke_2023, Nahum_2013_crossing} reaching system sizes of up to $10^8$ qubits far exceeding those accessible via standard Clifford circuit methods (employed and discussed in the next Section). 
To understand this efficiency gain of about two orders of magnitude in number of simulated qubits remember that, in the tableau representation of conventional Clifford simulations, 
an $N$ qubit stabilizer state requires $\mathcal{O}(N^2)$ memory, and each Clifford operation (e.g.\ a measurement) requires $\mathcal{O}(N^2)$ time.
By contrast, in the loop model representation,  the loop connectivity compactly encodes the state in only $\mathcal{O}(N)$ space and can be updated after a measurement in $\mathcal{O}(1)$ time.

This substantial computational advantage of the loop model representation is further enhanced by noting that a circuit of depth $t$ acts as a transfer matrix on the loop connectivity.
Two such depth $t$ circuits can be efficiently composed in time $\mathcal{O}(N)$, yielding a new circuit of depth $2t$.
The loop model simulations are then carried out as follows:
We begin by preparing a ``pool'' of $N_p$ shallow circuits of depth $t=1$.
In particular, we measure $N$ bonds from the Majorana honeycomb lattice chosen at random with relative probabilities $K_\alpha$ for nearest-neighbor bonds and $J$ for next-nearest-neighbor bonds.
Then all trajectories are evolved to depth $T$ by $\log_2(T)$ rounds of concatenation.
When concatenating, the circuits are randomly translated with respect to one another in order to enlarge the accessible configuration space as well as to reduce spurious correlations arising from finite $N_p$.
From this pool, $N_s$ independent sample trajectories are drawn by choosing two random elements, translating them with respect to one another, and concatenating.
Due to self-averaging in large systems, random translations before concatenating allows one to take $N_s \gg N_p$.

For each such sample, quantities such as the entanglement entropy and the spanning number may be directly computed from the loop connectivity after imposing appropriate boundary conditions at the temporal boundaries.
Moreover, for PBC, the entanglement entropy averaged over all cuts can be efficiently computed in one $\mathcal{O}(N)$ shot.
By contrast, computing entanglement entropy in the tableau (Clifford) representation requires $\mathcal{O}(N^3)$ time for \emph{each} subsystem, making it far more costly.
To probe bulk loop statistics, we track not only the loop connectivity, but also the total length of each loop, storing the lengths of closed loops in a histogram during the evolution.

Our simulations are completely based on self-implemented loop model codes (as opposed to codes using highly optimized open-source community packages as in the case of the Clifford simulations, see next Section). These codes are run on national high-performance computing resources, specifically the AMD EPYC (v3 Milan)-based Noctua2 cluster at the Paderborn Center for Parallel Computing (PC$^2$), the AMD EPYC (v4 Genoa)-based RAMSES cluster at RRZK/University of Cologne, and the Intel XEON Platinum 8168-based JUWELS cluster at FZ Julich.
Limited by the amount of system memory available, we initialize pool sizes between 120 ($L=128$) and 20 ($L=4096$) and generate $10^5$ samples from a given configuration pool. Depending on the system size, we generate between 50 and 100 independent pools.
Let us provide an example of the resulting computational cost: at each set of parameters in Fig.~\ref{fig:HC_span_FSS}, generating $10^6$ samples consumes 3 core-hours at $L=128$ and 1000 core-hours at $L=4096$. For both systems, across all system sizes and 100 different parameters, a budget of less than 200,000 core-hours was sufficient to generate the data shown in Fig.~\ref{fig:HC_span_FSS}. Using a budget of about 2000 core-hours per data point, this has allowed us to simulate system sizes up to $N \sim 10^8$ qubits in the loop representation (versus $N \sim 10^4$ in the Clifford tableau representation). An important technical detail is that our loop model code is highly parallelizable and limited in performance primarily by memory bandwidth, inviting GPU acceleration in future simulations.

%%%%%%%%%%%%%%%%%%%%%%%%%%%%%%%%%%%%%%%%%%%%%%%%%%%%%%%%%%%%%%%%%%%%%%%%%%%%
\section{Clifford simulations}\label{sec:clifford_simulations}
%%%%%%%%%%%%%%%%%%%%%%%%%%%%%%%%%%%%%%%%%%%%%%%%%%%%%%%%%%%%%%%%%%%%%%%%%%%%
In this final section, we provide numerical results from conventional Clifford simulations relying on stabilizer tableau algebra. 
Such simulations often serve as a starting point for a quick exploration of entanglement phase diagrams before indulging into
more complex numerical simulations (such as the loop model numerics discussed in the previous Section) or analytical 
considerations. It is the purpose of this Section to demonstrate what kind of qualitative and quantitative insight can already be
afforded when performing such Clifford simulations using state-of-the-art codes \cite{Gidney2021stimfaststabilizer,QuantumClifford}, 
with which one can readily simulate systems of the order of $10^4$ qubits.

%%%%%%%%%%%%%%%%%%%%%%%%%%%%%%%%%%%%%%%%%%%%%%%%%%%%%%%%%%%%%%%%%%%%%%%%%%%%
\subsection{Phase diagrams from tripartite mutual information}
\label{sec:Clifford_tripartite}
\label{sec:Clifford_universality}
%%%%%%%%%%%%%%%%%%%%%%%%%%%%%%%%%%%%%%%%%%%%%%%%%%%%%%%%%%%%%%%%%%%%%%%%%%%%

\begin{figure}[t]
    \centering
    \includegraphics[width=\columnwidth]{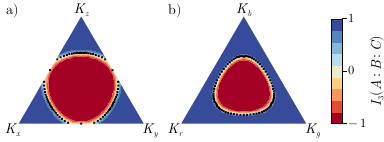}
    \caption{{\bf Tripartite mutual information heat maps.} The tripartite mutual information $I_3$ for (a) the Kitaev and (b) Kekule circuits 
    at linear system size $L=48$ is shown as a contour plot. 
    The location of critical points extracted from scaling collapses overlay as black crosses (same as Fig.~\ref{fig:diffusion_vs_c}). 
    The data is averaged over 96 disorder realizations with 100 samples each and was obtained for 211 points in parameter space 
    $K_x \geq K_y \geq K_z$, with the rest obtained via $\pi/6$ rotations. 
    The critical points align well with the visually apparent phase boundaries, establishing this kind of visual analysis as a viable method 
    when exploring new circuits.
    }
    \label{fig:CliffordHeatmaps}
\end{figure}

One particularly powerful observable to quickly map out the entanglement phase diagram of a given Clifford circuit is the {\it tripartite} mutual information
\begin{gather*}
    I_3(A:B:C) = I_2(A:B) + I_2(A:C)-I_2(A:BC)                        \\
               =S_A + S_B + S_C - S_{AB} - S_{BC} - S_{AC} + S_{ABC} \,.
\end{gather*} 
where the subregions $A,\ldots, D$ are equal partitions of the torus on which the lattice is defined. 
Already for a fixed, finite linear system size, this observable allows to distinguish the entanglement structure of the localized and liquid phases,
for which is assumes values \cite{Vijay23Kitaev,Ippoliti23kitaev,Zhu23structuredVolumeLaw} of $I_3(A:B:C) = +1$ (localized) and $I_3(A:B:C) = -1$ (liquid), respectively.
As we show in Fig.~\ref{fig:CliffordHeatmaps} for the original honeycomb Kitaev circuit and its Kekul\'e variant, a heat map-style visualization of $I_3$ readily maps out the entanglement phase diagram already with reasonable accuracy (even though the critical point does not necessarily coincide with $I_3 = 0$) -- compare, e.g., with the phase diagrams in the overview of Fig.~\ref{fig:overview_circuit_diagram}.

\begin{figure}[t]
    \centering
    \includegraphics[width=\columnwidth]{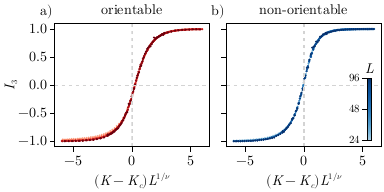}
    \caption{{\bf Correlation length exponents. }
        Scaling collapse of the tripartite mutual information $I_3$ calculated with the Clifford framework
        for (a)  the orientable and (b) non-orientable circuit 
        showing data for linear system sizes $L=24,32,40,48,56,64,72,80,96$ in the vicinity of the critical point at $K_c$. 
        The fitted exponents are 
        a) $\nu = 0.96 \pm 0.01$ with $\chi^2_r = 1.39$, $x \in (0,1)$ and 
        b) $\nu = 0.91 \pm 0.02$ with $\chi^2_r = 1.46$, $x \in (0,1)$, where $x = (K-K_c)L^{1/\nu}$.
        Dashed gray lines mark $x=0$ and $y=0$, showing that $I_3 \approx 0$ at or very near to the critical point, corroborating Fig.~\ref{fig:CliffordHeatmaps}.
    }
    \label{fig:CliffordExponents}
\end{figure}

A more precise quantitative estimate of the phase boundaries can be obtained via a finite-size scaling analysis of the tripartite mutual information, which exhibits an easy-to-locate crossing point for different system sizes at the phase transition $K=K_c$.
Performing a finite-size data collapse as shown in Fig.~\ref{fig:CliffordExponents} one can further extract the correlation length exponent, though with far less accuracy as what is achievable with the loop model framework discussed in the previous section.

The values of the correlation length exponent $\nu$ are clearly distinct between the two symmetry classes, even at these small system sizes. 
However, the specific values do not agree with our previous results from the loop simulations, even within margin of error. In the Clifford simulations, 
there are several limitations at play which affect the accuracy of the extracted correlation length exponents and which are likely to lead to this disagreement. This includes 
(i) strong finite-size effects arising from the limited system sizes available in the Clifford simulations (which are hard to overcome with the current algorithms/numerical platforms), and
(ii) a necessity to extend the first-order scaling ansatz $x = (K - K_c) L^{1/\nu}$ to higher orders (which is not straight-forward).
One manifestation of these limitations is a system-size dependent drift of the tripartite mutual information within the liquid phase, 
making it impossible to properly collapse the data. The origin of this drift can be rationalized as follows: even deep in the liquid phase, there is always a non-zero probability that the system will not be long-range entangled at a given time. In essence, a series of measurements on the $K$ bonds can localize the Majorana pairs and temporarily destroy long-range entanglement. Alternatively, this can be phrased as a random walk in configuration space with a small fraction of area-law states, which vanishes in the thermodynamic limit. In taking the average over a large number of samples, this small fraction of localized (area-law) states then manifests itself as a small, finite offset in the tripartite information and is most pronounced for small system sizes and more so for the orientable circuit (due to the reduced number of paths for long-range strings). This is indeed visible in Fig.~\ref{fig:CliffordExponents}, where $I_3$ never actually saturates to $-1$ (in the liquid phase). 
In our finite-size scaling analysis we have worked around this by performing the optimization of the fitting cost function only for data points at $K > K_c$.

\begin{table}[t]
	\begin{tabular}{|c||c|c|}\hline
	symmetry 	& BDI (orientable)		& D (non-orientable) \\
 \hline
	$K_c$	& $0.6520 \pm 0.0002$ & $0.7315 \pm 0.0002$ \\
	\hline
	$\nu$ 	& $0.96 \pm 0.01$ 		& $0.91 \pm 0.02$ \\
	\hline
	\end{tabular}
	\caption{{\bf Critical point and correlation length exponent} from the tripartite mutual information in Fig.~\ref{fig:CliffordExponents}.
		}\label{tab:nu_Clifford}
\end{table}

\begin{figure}[t]
    \centering
    \includegraphics[width=\linewidth]{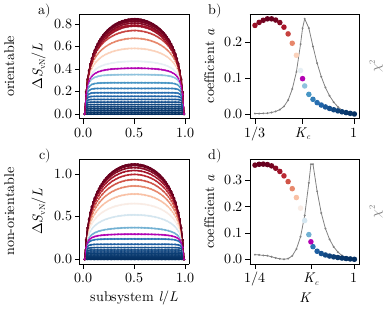}
    \caption{{\bf Subsystem entanglement entropy.}
       For $L=72$, we take cuts through the phase diagrams of the orientable (top panels) and non-orientable circuit models (bottom panels). 
       In the former case, we take $K=K_x$, $K_y = K_z = (1-K/2)$, while in the latter case we take $K=K_x$, $K_y = K_z = J = (1-K/3)$. 
       Figures a) and c) show the resulting bipartite entanglement entropy $S(l/L)/L$, excluding constant contributions. 
       The color denotes the log-law contribution, as indicated on the right side, with the critical point highlighted in purple. 
       For each curve, we fit the data to the entanglement arc of Eq.~\eqref{eq:CFT-Entanglement-Arc} 
       and show the resulting fit coefficient $a$ in panels b) and d). 
       Gray line shows the fit residual (not to scale), with the peak indicating the qualitatively different entanglement scaling 
       close to the critical point. 
       The data shown is averaged over 192 disorder realizations with 150 samples each.
    }
    \label{fig:CliffordArches}
\end{figure}

%%%%%%%%%%%%%%%%%%%%%%%%%%%%%%%%%%%%%%%%%%%%%%%%%%%%%%%%%%%%%%%%%%%%%%%%%%%%
\subsection{Entanglement entropy and Lifshitz criticality}
%%%%%%%%%%%%%%%%%%%%%%%%%%%%%%%%%%%%%%%%%%%%%%%%%%%%%%%%%%%%%%%%%%%%%%%%%%%%
Another way to explore the entanglement phase diagram of a Kitaev circuit and to zoom in on the phase transitions is, 
of course, to directly explore the entanglement structure of the ensemble of stationary states. 
To this end, one might want to look at ``entanglement arcs",
i.e.\ the bipartite entanglement entropy for subsystem cuts of varying size $0 \leq l/L \leq 1$. This is shown in Fig.~\ref{fig:CliffordArches}
for parameter scans through the phase diagram of the honeycomb Kitaev circuit as an example of an orientable circuit as well as the non-orientable version with next-nearest neighbor parity checks. One can clearly distinguish the liquid region where the entanglement entropy $S/L$ spans an arc akin to the well-known Calabrese-Cardy result \cite{CalabreseCardy2004, Turkeshi20}  
\begin{equation}
	S(l/L)/L = a \, \log \sin(\pi l/L) + b \,,
	\label{eq:CFT-Entanglement-Arc}
\end{equation}
for conformally invariant systems (red traces), versus the localized phase whose area-law scaling reveals itself as a wide plateau of constant entanglement entropy (blue traces).

\begin{figure}
    \includegraphics[width=\linewidth]{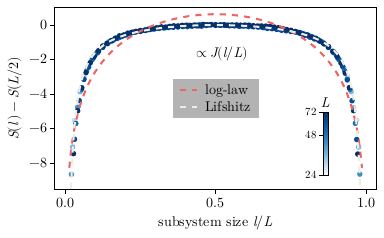}
    \caption{{\bf Lifshitz scaling at the phase transition. }
        Collapse of the entanglement entropy $S(l/L)$ at the critical point calculated with the Clifford framework 
        for the non-orientable circuit model for varying subsystem sizes $l$ and overall linear system sizes $L=24,32,40,48,56,64,72$. 
        The dashed lines correspond to fits of the data to the CFT scaling form $S(l/L) = a \, \log \sin(\pi l/L) + b$ (red) 
        and the Lifshitz scaling form Eq.~\eqref{eq:lifshitz_func} (white), with the latter clearly being the better fit.
        The data is averaged over 384 disorder realizations with 100 samples each. 
        This figure is the Clifford analogue to what is shown (for much larger system sizes) in Fig.~\ref{fig:critical_EE_scaling} 
        for data obtained from  loop model simulations. 
    }
    \label{fig:CliffordLifshitz}
\end{figure}

When moving towards the phase boundary within the entanglement phase diagram, a close inspection of the scaling of the bipartite entanglement entropy reveals the characteristic shape of Lifshitz critical scaling, marked by the pink trace in Fig.~\ref{fig:CliffordArches}, which shows a clearly distinct behavior from the ``entanglement arcs" in the liquid regime and the almost constant behavior in the localized phase. This deviation from the liquid-like scaling is also reflected in an enlarged $\chi^2$ value of the respective fit, as shown in the right panels of Fig.~\ref{fig:CliffordArches}. 
Zooming in on the critical behavior and plotting the entanglement entropy for different system sizes in Fig.~\ref{fig:CliffordLifshitz}, we find already for the system sizes achievable in our Clifford simulations 
an indisputable signature of the Lifshitz scaling, consistent with our discussion in the context of the loop model simulations in Section \ref{ss:universality} above.

These findings support the usefulness of Clifford simulations to not only map out the entanglement phase diagram 
of a given monitored circuit with relatively modest numerical resources (for instance via scans of the tripartite mutual information), but also to extract qualitative features of the phase transitions (such as a distinction of $\sim L\log L$ versus 
Lifshitz scaling) and even get a rough estimate of critical exponents. At the same time, high-quality fits of critical 
exponents require substantially larger system sizes than attainable in Clifford simulations ($\sim$20,000 qubits), leaving 
an opening for more advanced numerical schemes (such as the loop model simulations) that allow to go beyond 
these limitations.

%%%%%%%%%%%%%%%%%%%%%%%%%%%%%%%%%%%%%%%%%%%%%%%%%%%%%%%%%%%%%%%%%%%%%%%%%%%%
\subsection{Simulation setup}
\label{sec:Clifford_simulation_setup}
%%%%%%%%%%%%%%%%%%%%%%%%%%%%%%%%%%%%%%%%%%%%%%%%%%%%%%%%%%%%%%%%%%%%%%%%%%%%

Let us close also this Section on Clifford simulations with a detailed review of our numerical approach. 
To obtain the stationary state as efficiently as possible, we stabilize all conserved quantities at initial time. Depending on system size, we take between $5 \times 10^2$ and $4\times 10^4$ samples for each set of parameters $K_{x,y,z}, J$, which determine the probability to projectively measure a bond (see Fig.~\ref{fig:overview_circuit_diagram}). We use a highly efficient implementation \cite{QuantumClifford} of stabilizer tableau algebra. While it is not possible to improve upon either the $\mathcal{O}(N^2)$ scaling for performing a single measurement (and thus $\mathcal{O}(N^3)$ per circuit layer) or the $\mathcal{O}(N^3)$ scaling for canonicalization, we perform several HPC optimizations to reduce memory access, since tableau-based simulations are always bound by memory bandwidth. This includes a hybrid parallelization strategy based on the physical layout of the hardware used, which in this case is a $50 \times 2 \times$ AMD EPYC 9654 cluster.
Furthermore, we disregard phase information and measurement outcomes -- ensuring that measurements always scale $\sim \mathcal{O}(N^2)$ -- and make use of the fact that there are $\frac{3L}{4}$ possible partitions of the honeycomb lattice, allowing to perform self-averaging. Even though computing the entanglement entropy of a stabilizer state \emph{always} requires to canonicalize the tableau ($ \mathcal{O}(N^3)$), it is possible to get a slight advantage when the only quantity of interest is the scaling of $S(l/L)$ by converting to the clipped gauge. While this also requires an $ \mathcal{O}(N^3)$ operation, it allows one to extract the entanglement entropy for all subsystem sizes in one shot. This trick, however, is of little use for the tripartite mutual information, since it requires tracing out non-contiguous regions -- which is not possible in the clipped gauge representation. In the end, we are able to generate high quality data for up to $2\times 96^2=18,432$ qubits, albeit at exploding computational cost. 
As an example, for a sweep of 100 different parameters, generating 192 trajectories with 150 samples each consumes close to 2,000 core hours at $L=32$ while generating only 96 trajectories with 5 samples each takes slightly more than half a million core hours at $L=96$ -- the majority of which is however spent on performing canonicalization.

%%%%%%%%%%%%%%%%%%%%%%%%%%%%%%%%%%%%%%%%%%%%%%%%%%%%%%%%%%%%%%%%%%%%%%%%%%%%

\section{Discussion}\label{sec:discussion}
%%%%%%%%%%%%%%%%%%%%%%%%%%%%%%%%%%%%%%%%%%%%%%%%%%%%%%%%%%%%%%%%%%%%%%%%%%%%

Summarizing our main results in brief, we have investigated measurement-only circuit variants of the Kitaev and Yao-Kivelson model to elucidate that 
(i) monitored quantum circuits give rise to robust entanglement phases, i.e.\ dynamic phases of matter that are stable to local variations of the circuit parameters (measurement probabilities in our case), 
(ii) the notions of symmetry and dimensionality can be used, akin to their Hamiltonian counterparts, to define universal behavior of the entanglement phases and the quantum criticality of phase transitions between them, and
(iii) that the particle-hole symmetric Majorana circuits, relevant to these monitored Kitaev models, can be recast in terms of Majorana loop models.
These Majorana loop models are a powerful resource as they not only allow to recast the entanglement phases in terms of localization physics of loops and map their phase transitions to non-linear sigma models, but they also provide a numerical framework allowing to simulate up to $10^8$ qubits. This has allowed us to clearly discriminate the universality classes of orientable versus non-orientable loop models in $(2+1)$ dimensions, relevant to Majorana circuits in symmetry classes BDI and D, respectively and determine their critical exponents to unprecedented precision.

Furthermore, our work suggests a broader perspective on the tenfold way classification of free fermionic wave functions. The connection between orientable and non-orientable loop models offers a unified framework for embedding both free Majorana circuits (fugacity $n=1$) and ground states of free Majorana Hamiltonians (fugacity $n=\sqrt{2}$) within symmetry classes BDI and D. This raises intriguing possibilities for future research, such as exploring whether other symmetry classes allow similar embeddings or investigating higher loop fugacities, potentially realized in quantum circuits coupled to classical dynamical agents~\cite{Klocke_2024}. This approach could refine the tenfold way classification by distinguishing universality not only by symmetry classes but also by Hamiltonian dynamics versus monitored circuits. Moreover, this offers a tractable handle on broader emerging questions regarding symmetry as an organizing principle in \emph{open} quantum systems~\cite{li2023exact, Lee23criticalityunderdecoherence, You24weaksym, Wang24weaksym}.

One unexpected discovery in our study is the presence of quantum Lifshitz scaling at the phase transition of monitored Majorana circuits—a feature previously associated only with \emph{interacting} circuits. This finding suggests a new and unexplored connection between loop models in $(2+1)$ dimensions, monitored free fermions, and $ (2+1)$-dimensional conformal field theories, warranting further investigation.

There are many future directions one might pursue in studying universal behavior in monitored quantum circuits and measurement-only Kitaev models. 
This includes variations of dimensionality, such as going to higher spatial dimensions and studying three-dimensional monitored Kitaev models~\cite{OBrien2016,Eschmann2020}, which is also closely related to the 3D Floquet code~\cite{Nat24floquetcode3dkitaev}, or exploring quantum systems with arbitrary qudit dimension~\cite{Ludwig2024}. 
One might also think of generalizations of monitored Majorana systems to more general monitored non-Abelian anyon systems, such as a monitored Fibonacci anyons in a measurement-only variant of the golden chain~\cite{GoldenChain}.

This work also raises a number of questions regarding the relation between the universality of entanglement transitions in Majorana circuits and those in more generic (2+1)-dimensional Clifford circuits.
Moving away from the Gaussian limit, we obtain random quantum circuits which can be related to a broad class of interacting loop or dimer models and for which new strong-coupling fixed-points may emerge.
Clifford operations which are even (odd) under fermionic-parity will preserve (break) the orientability of the loop model~\cite{Klocke_2023}.
Exploration of the parity-symmetric case has been initiated already in Ref.~\onlinecite{Zhu23structuredVolumeLaw}, where the one finds volume-law entanglement and a correlation length exponent consistent with the loop model in class BDI.
Far from the Gaussian limit, large-scale numerical simulation of generic (2+1)-dimensional Clifford circuits give compelling evidence that the measurement-induced transition is in the same universality class as 3D percolation~\cite{Sierant_2022}.
However, the exponent $\nu=0.876 \pm 0.004$~\cite{Borinsky_2021, Koza_2016, Wang_2013, Xu_2013} for 3D percolation is markedly smaller than what we find here for non-orientable Majorana circuits via loop simulations.
As is apparent from our own Clifford simulations in class D, finite-size effects lead to an underestimation of $\nu$ even for the largest system sizes.
It would be of interest to better resolve whether this apparent difference is due to finite-size effects or if perturbing away from the Gaussian Majorana limit leads to a genuine renormalization of the critical exponents.

%%%%%%%%%%%%%%%%%%%%%%%%%%%%%%%%%%%%%%%%%%%%%%%%%%%%%%%%%%%%%%%%%%%%%%%%%%%%
%\subsection{Beyond Clifford: Weak measurements}
%%%%%%%%%%%%%%%%%%%%%%%%%%%%%%%%%%%%%%%%%%%%%%%%%%%%%%%%%%%%%%%%%%%%%%%%%%%%

Let us close this outlook by pointing to some of the physics awaiting us when going beyond the (random) Clifford regime 
of strong, project measurements.
To do so, one can consider non-random but weak, non-projective measurements 
--  as realized by moving the rotation $t\in[0,\pi/4]$ in the unitary entangling gates, drawn as CNOT gates in Fig.~\ref{fig:circuit_buildingblocks}, 
away from a perfect $\pi/4$-rotation. 
One crucial implication here is that moving away from a perfect $t=\pi/4$ rotation is akin to reducing $\tau$, 
the effective inverse temperature in the Kraus operator \eqref{eq:KrausOp} above, to a {\it finite} value.
This connection between weak measurements and finite temperature physics has been demonstrated \cite{NishimoriCat} in rigorous terms 
for commuting two-qubit checks, resulting in Nishimori physics that subsequently has been probed in experimental devices \cite{chen2023realizing}. 
In the context of the Hastings-Haah code it allows one to explore the physics of \emph{qubit fractionalization}
 (and intermediate phases) as one goes out of the Clifford limit of projective measurements to weak measurements \cite{Zhu23qubit}. 
 It will be interesting to investigate such weakly monitored Kitaev circuits in more depth in the future. 
 For instance, note that analogous to cooling a Hamiltonian in equilibrium, tuning the measurement strength from weak to strong can purify all the time reversal invariant plaquettes or the product of plaquette operators, leaving a Kramers doublet as a time reversal symmetric mixed state. 
 Applied to the non-orientable circuits of interest in this manuscript this poses an interesting question for future research -- is the {\it spontaneous} breaking of time-reversal symmetry in such non-orientable circuits captured by a hitherto unexplored {\it chiral} Nishimori critical point?\\

%%%%%%%%%%%%%%%%%%%%%%%%%%%%%%%%%%%%%%%%%%%%%%%%%%%%%%%%%%%%%%%%%%%%%%%%%%%%

\begin{acknowledgments}
    KK was supported by an NSF Graduate Fellowship under Grant No.~DGE 2146752 and by the U.S. Department of Energy, Office of Science, National Quantum Information Science Research Centers, Quantum Science Center.
    The Cologne group acknowledges support from the Deutsche Forschungsgemeinschaft (DFG, German Research Foundation) under Germany’s Excellence Strategy Cluster of Excellence Matter and Light for Quantum Computing (ML4Q) EXC 2004/1 390534769, and by the DFG Collaborative Research Center (CRC) 183 Project No.~277101999 - projects B01 and B02.
    GYZ additionally acknowledges support from a start-up fund from HKUST(GZ). 
    The numerical simulations have been performed on the JUWELS cluster at Forschungszentrum J\"ulich,
    the Noctua2 cluster at the Paderborn Center for Parallel Computing (PC$^2$), and the RAMSES cluster at RRZK, University of Cologne.\\
\end{acknowledgments}
%%%%%%%%%%%%%%%%%%%%%%%%%%%%%%%%%%%%%%%%%%%%%%%%%%%%%%%%%%%%%%%%%%%%%%%%%%%%

%%%%%%%%%%%%%%%%%%%%%%%%%%%%%%%%%%%%%%%%%%%%%%%%%%%%%%%%%%%%%%%%%%%%%%%%%%%%
% Data availability
%%%%%%%%%%%%%%%%%%%%%%%%%%%%%%%%%%%%%%%%%%%%%%%%%%%%%%%%%%%%%%%%%%%%%%%%%%%%

{\it Data availability}.-- 
All code for generating the figures shown and the underlying numerical data is available on Zenodo~\cite{zenodo_kitaev}.

\newpage
%%%%%%%%%%%%%%%%%%%%%%%%%%%%%%%%%%%%%%%%%%%%%%%%%%%%%%%%%%%%%%%%%%%%%%%%%%%%
\bibliography{3D_loops}
%%%%%%%%%%%%%%%%%%%%%%%%%%%%%%%%%%%%%%%%%%%%%%%%%%%%%%%%%%%%%%%%%%%%%%%%%%%%

\clearpage
%%%%%%%%%%%%%%%%%%%%%%%%%%%%%%%%%%%%%%%%%%%%%%%%%%%%%%%%%%%%%%%%%%%%%%%%%%%%
\begin{appendix}
%%%%%%%%%%%%%%%%%%%%%%%%%%%%%%%%%%%%%%%%%%%%%%%%%%%%%%%%%%%%%%%%%%%%%%%%%%%%

%%%%%%%%%%%%%%%%%%%%%%%%%%%%%%%%%%%%%%%%%%%%%%%%%%%%%%%%%%%%%%%%%%%%%%%%%%%%
\section{Hamiltonian phase diagrams}
\label{app:hamiltonian_phase_diag}
\begin{figure}[h]
    \centering
    \includegraphics{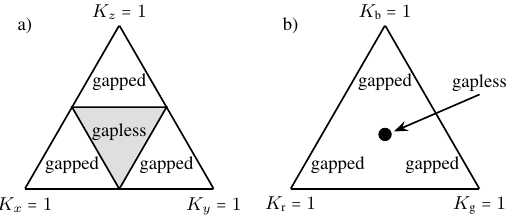}
    \caption{{\bf Hamiltonian phase diagrams.} Schematic phase diagrams for the Kitaev model in its (a) regular and (b) Kekul\'e variants, adapted from \cite{Kitaev2006} and \cite{Schmidt10honeycomb}.}
    \label{fig:HamiltonianCompare}
\end{figure}

%%%%%%%%%%%%%%%%%%%%%%%%%%%%%%%%%%%%%%%%%%%%%%%%%%%%%%%%%%%%%%%%%%%%%%%%%%%%
Here for completeness we show, in Fig.~\ref{fig:HamiltonianCompare}, a comparison of the Hamiltonian phase diagrams of the original honeycomb Kitaev model and the Kekulé-Kitaev model. 
For the isotropic honeycomb Kitaev model, a unit-cell consists of two spin sites, and in the momentum space it hosts two linearly dispersive Majorana Dirac cones at the Brillouin zone corners~\cite{Kitaev2006}. The Dirac cones are stable in the presence of translation and $C_2 \mathcal{T}$ symmetry (a combination of lattice inversion and time reversal transformation). The perturbation away from the isotropic point gradually shifts the location of the two Dirac cones, which merge and annihilate and gap out at the phase boundary between the gapped and gapless phase. 
In contrast, the Kekulé variant of the model tricolors the bonds and triples the size of the unit-cell into 6 spin sites, which induces a periodic potential with large wave vector that can scatter one Dirac cone to the other, which immediately gaps out the Dirac cones when perturbed away from the isotropic point~\cite{Schmidt10honeycomb}. As a result, the gapless phase shrinks to a singular point in the phase diagram. 

The distinct topology of the phase diagram of the Kekulé-Kitaev model allows for an adiabatic path connecting the three gapped phases. This feature is exploited in the Floquet code~\cite{Haah21honeycomb, Hastings22honeycomb}, which measures all the plaquette operators by four rounds of bond measurements without collapsing the global Wilson loops as logical qubits, and inducing an effective Hadamard transformation for the logical qubit due to the winding around the gapless point. \\

%%%%%%%%%%%%%%%%%%%%%%%%%%%%%%%%%%%%%%%%%%%%%%%%%%%%%%%%%%%%%%%%%%%%%%%%%%%%
\section{Frustration graphs}
%%%%%%%%%%%%%%%%%%%%%%%%%%%%%%%%%%%%%%%%%%%%%%%%%%%%%%%%%%%%%%%%%%%%%%%%%%%%

\begin{figure}[h!]
    \centering
    \includegraphics{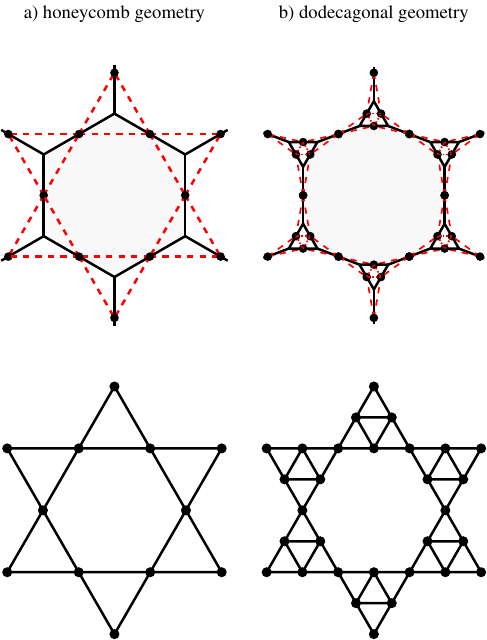}
    \caption{
    \textbf{Frustration graphs.}
    The two columns show the (measurement) frustration graphs associated to the circuits on the (a) honeycomb lattice and (b) dodecagonal lattice.
    In the top row, we overlay the original lattice (black edges) and the frustration graph (red dashed edges and vertices).
    In the bottom row, we show only the frustration graph and deform it better highlight the structure.
    }
    \label{fig:frustration_graphs}
\end{figure}

For generic protocols involving only Majorana parity checks, we can generate the frustration graph by a duality transformation. The original matter Majorana fermion spans a graph where each node hosts a matter Majorana fermion, and each edge corresponds to a parity check. To dualize the graph, we turn every edge to a node, and every Majorana site into a all-to-all connected cluster, see Fig.~\ref{fig:frustration_graphs}. In the dual graph, the node refers to the measured operators, and the edge connects two operators that anti-commutate each other, due to the fact that they overlap with a Majorana fermion. 

If the original graph has uniform coordinate number, say, every matter Majorana fermion has $z$ neighbors, then in the dual graph, each node as an operator has $2(z-1)$ neighbors. For example, in the (decorated) honeycomb lattice $z=3$ and the frustration graph has $4$ neighbors. For the special case of next-nearest neighbor measurement only, the Majorana fermions form two disconnected triangle lattice with $z=6$, and thus the coordinate number of the dual graph is $10$. 

%%%%%%%%%%%%%%%%%%%%%%%%%%%%%%%%%%%%%%%%%%%%%%%%%%%%%%%%%%%%%%%%%%%%%%%%%%%%
\section{Universality and lattice geometry in non-orientable symmetry class D}\label{app:nonorientable_geometries}
%%%%%%%%%%%%%%%%%%%%%%%%%%%%%%%%%%%%%%%%%%%%%%%%%%%%%%%%%%%%%%%%%%%%%%%%%%%%

Here we provide additional evidence for the universality of the critical exponents reported in Tab.~\ref{tab:crit_exp} for the non-orientable symmetry class.
We consider two additional lattice geometries: 
(i) the measurement-only Yao-Kivelson model, and
(ii) Cardy's L-lattice, as studied in Ref.~\onlinecite{Serna_2021}.
Results are summarized in Tab.~\ref{tab:crit_exp_lattice}, showing a striking agreement between the three geometries as expected for universal
critical behavior.

%%%%%%%%%%%%%%%%%%%%%%%%%%%%%%%%%%%%%%%%%%%%%%%%%%%%%%%%%%%%%%%%%%%%%%%%%%%%
\begin{table*}[t]
    \centering
    \begin{tabular}{p{30mm}|p{27mm}|p{27mm}|p{27mm}|p{27mm}|p{30mm}}
        {\bf critical exponent}                       & \multicolumn{1}{c|}{$\nu$}                             &  \multicolumn{1}{c|}{$\eta$ }            &  \multicolumn{1}{c|}{$d_f$} &  \multicolumn{1}{c|}{$\beta$} &  \multicolumn{1}{c}{reference} \\[1mm]
        \hline\hline
        \multicolumn{5}{l}{}                                                                                             \\[-1.7ex]
        \multicolumn{2}{l}{{\it non-orientable loop models}}            & \multicolumn{4}{r}{$10^8$ qubits}                                \\[1mm]
        \hline
        honeycomb	&  $\quad 0.9403 \pm 0.0006$  &  $\quad -0.0660 \pm 0.0070$ & $\quad 2.5263 \pm 0.0011$  & $\quad 0.4400 \pm 0.0040$  & \multicolumn{1}{c}{Figs.~\ref{fig:HC_span_FSS}, \ref{fig:HC_fractal_dim}, \ref{fig:HC_span_length}, \ref{fig:critical_watermelon}} \\
        Yao-Kivelson	&  $\quad 0.9401 \pm 0.0015$  &  $\quad -0.0612 \pm 0.0036$ & $\quad 2.5306 \pm 0.0018$  & $\quad 0.4413 \pm 0.0019$  & \multicolumn{1}{c}{Fig.~\ref{fig:YK_span_FSS}, \ref{fig:YK_fractal_dim}} \\
        Cardy's L-lattice	&  $\quad 0.9390 \pm 0.0013$  &  $\quad -0.0580 \pm 0.0037$ & $\quad 2.5290 \pm 0.0018$  & $\quad 0.4423 \pm 0.0018$  & \multicolumn{1}{c}{Fig.~\ref{fig:L_lattice_FSS}, \ref{fig:L_lattice_fractal_dim}} \\
        \hline\hline
    \end{tabular}
    \caption{{\bf Effect of circuit geometry on critical exponents.} 
    Shown are the critical exponents of the localization transition comparing different circuit geometries. %\\[8mm]
    }
    \label{tab:crit_exp_lattice}
\end{table*}
%%%%%%%%%%%%%%%%%%%%%%%%%%%%%%%%%%%%%%%%%%%%%%%%%%%%%%%%%%%%%%%%%%%%%%%%%%%%

%%%%%%%%%%%%%%%%%%%%%%%%%%%%%%%%%%%%%%%%%%%%%%%%%%%%%%%%%%%%%%%%%%%%%%%%%%%%
\subsection{Measurement-only Yao-Kivelson Model}\label{ss:YK}
%%%%%%%%%%%%%%%%%%%%%%%%%%%%%%%%%%%%%%%%%%%%%%%%%%%%%%%%%%%%%%%%%%%%%%%%%%%%

Let us now consider the decorated honeycomb lattice of the measurement-only Yao-Kivelson model depicted in Fig.~\ref{fig:overview_circuit_diagram}(d).
We take a lattice of $L \times L$ large (dodecagonal) plaquettes such that the corresponding squashed brickwall geometry has dimensions $L_x = 6L$ and $L_y = L$.
Moreover, we fix parameters $K_\alpha = K$ and $J_\alpha = J = (1 - K)$ for simplicity such that for $0 < K < 1$ the loop model is non-orientable.
Then varying $K$ allows one to tune the system through the localization transition for symmetry class D.

%%%%%%%%%%%%%%%%%%%%%%%%%%%%%%%%%%%%%%%%%%%%%%%%%%%%%%%%%%%%%%%%%%%%%%%%%%%%
\subsubsection*{Correlation length exponent \texorpdfstring{$\nu$}{ν}}
%%%%%%%%%%%%%%%%%%%%%%%%%%%%%%%%%%%%%%%%%%%%%%%%%%%%%%%%%%%%%%%%%%%%%%%%%%%%

\begin{figure}[b]
    \centering
    \includegraphics[width=\columnwidth]{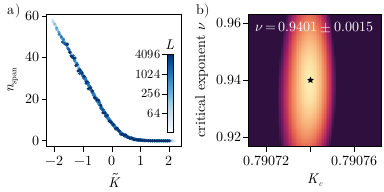}
    \caption{{\bf Correlation length exponent from the Yao-Kivelson circuit.}
    Finite-size scaling analysis of the transition in the Yao-Kivelson circuit, with data shown for linear system sizes $L = 32, 64, 128, 256, 512, 1024, 2048, 4096$.
    (a) Scaling collapse of the data with respect to scaling variable $\tilde{K} = L^{1/\nu}(K-K_c)\left[1 + \alpha (K - K_c)\right]$.
    (b) A heatmap of the normalized cost function $\chi^2 / \chi_*^2$ for the FSS analysis.
    The optimal parameters lie at the minimum $(K_c^*, \nu^*)$ marked by a star.
    The colormap has been truncated to give a visual estimate of the uncertainty in $K_c^*$ and $\nu^*$.
    }
    \label{fig:YK_span_FSS}
\end{figure}

In Fig.~\ref{fig:YK_span_FSS} we show the data collapse for the measurement-only Yao-Kivelson circuit in the vicinity of the critical point.
From this FSS analysis we find correlation length exponent $\nu = 0.9401 \pm 0.0015$, which is in close agreement with the value found both on the honeycomb lattice with NNN couplings and on Cardy's L-lattice.
This result further bolsters the reliability of our reported value of $\nu$ and demonstrates universality of the transition independent of the underlying lattice geometry.

%%%%%%%%%%%%%%%%%%%%%%%%%%%%%%%%%%%%%%%%%%%%%%%%%%%%%%%%%%%%%%%%%%%%%%%%%%%%
\subsubsection*{Fisher exponent \texorpdfstring{$\tau$}{τ}}
%%%%%%%%%%%%%%%%%%%%%%%%%%%%%%%%%%%%%%%%%%%%%%%%%%%%%%%%%%%%%%%%%%%%%%%%%%%%

\begin{figure}[h!]
    \centering
    \includegraphics[width=\columnwidth]{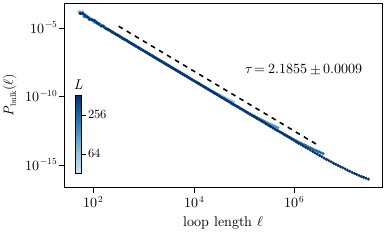}
    \caption{
    \textbf{Fisher exponent for the measurement-only Yao-Kivelson model.}
    We show the bulk length distribution of closed loops $P_\textrm{bulk}(\ell)$ at the critical point of the non-orientable loop model defined by the measurement-only Yao-Kivelson circuit with $K_\alpha = K_c$ and $J_\alpha = J_c = (1 - K_c)$.
    Data are shown for linear dimensions $L = 32, 64, 128, 256, 512$.
    The black dashed line corresponds to the power-law $P_\textrm{bulk}(\ell)\sim \ell^{-\tau}$ fit on the data in the interval $10^{5/2} \leq \ell \leq L^3/50$.
    }
    \label{fig:YK_fractal_dim}
\end{figure}

In order to find the remaining critical exponents, we determine the Fisher exponent $\tau$ from the bulk loop length distribution $P_\textrm{bulk}(\ell)$ at the transition.
Fitting the power-law scaling $P_\textrm{bulk}(\ell) \sim \ell^{-\tau}$ shown in Fig.~\ref{fig:YK_fractal_dim} yields an estimate of $\tau$ which is consistent with the corresponding result on the other two lattice geometries.
As for Cardy's L-lattice, we report the critical exponents $\eta$, $d_f$ and $\beta$ as found by hyperscaling relations in Tab.~\ref{tab:crit_exp_lattice}.

%%%%%%%%%%%%%%%%%%%%%%%%%%%%%%%%%%%%%%%%%%%%%%%%%%%%%%%%%%%%%%%%%%%%%%%%%%%%
\subsubsection*{Bulk loop statistics}
%%%%%%%%%%%%%%%%%%%%%%%%%%%%%%%%%%%%%%%%%%%%%%%%%%%%%%%%%%%%%%%%%%%%%%%%%%%%

%%%%%%%%%%%%%%%%%%%%%%%%%%%%%%%%%%%%%%%%%%%%%%%%%%%%%%%%%%%%%%%%%%%%%%%%%%%%
\begin{figure}[t]
    \centering
    \includegraphics[width=\columnwidth]{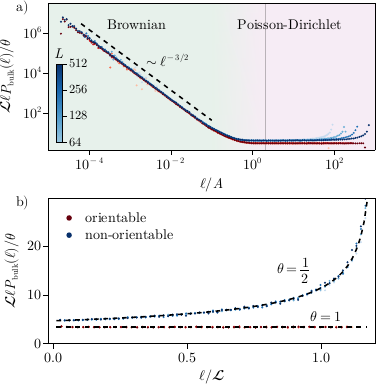}
    \caption{{\bf Bulk loop statistics for the measurement-only Yao-Kivelson model.}
        Analogous to Fig.~\ref{fig:HC_BulkLoops}, we show the distribution of bulk loop lengths $P_{\textrm{bulk}}(\ell)$ in the measurement-only Yao-Kivelson circuit with PBC in both time and space.
        The non-orientable limit $\vec J = \vec K = \{1, 1, 1\}$ (blue) is contrasted with the orientable limit $\vec J = \{1, 1, 0\}, \, \vec K = \{1, 1, 1\}$ (red).
    }
    \label{fig:YK_BulkLoops}
\end{figure}
%%%%%%%%%%%%%%%%%%%%%%%%%%%%%%%%%%%%%%%%%%%%%%%%%%%%%%%%%%%%%%%%%%%%%%%%%%%%

\begin{figure}[h!]
    \centering
    \includegraphics[width=\columnwidth]{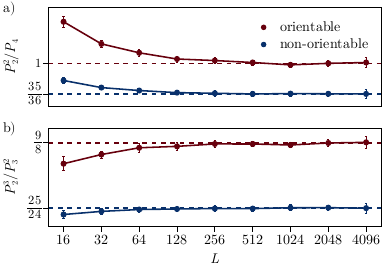}
    \caption{
    \textbf{Universal ratios of long-distance loop connectivities.}
    Analogous to Fig.~\ref{fig:watermelon_ratios} of the main text we show, for the measurement-only Yao-Kivelson model, 
    the ratios (a) $P_2^2 / P_4$ and (b) $P_2^3 / P_3^2$.
    Data shown here are for the isotropic point $K_\alpha = J_\alpha$.
    }
    \label{fig:YK_Watermelon_Ratios}
\end{figure}

Finally, we highlight universal aspects of bulk loop statistics in the metallic phase.
In Fig.~\ref{fig:YK_BulkLoops} we show the distribution of bulk loop lengths $P_\textrm{bulk}(\ell)$ at the isotropic point $K = J$ deep in the critical (metallic) phase.
These results mirror those presented in Fig.~\ref{fig:HC_BulkLoops} for the honeycomb lattice, revealing Brownian and Poisson-Dirichlet scaling regimes.
Moreover, the asymptotic scaling in the Poisson-Dirichlet regime is consistent with Eq.~\eqref{eq:PD_distrib} for both symmetry classes.
We further validate the universality of the metallic phase by examining ratios of the long-distance loop connectivities.
In Fig.~\ref{fig:YK_Watermelon_Ratios}, we show that these ratios rapidly approach the values predicted from the universal quantity $\theta$, consistent with the result in Fig.~\ref{fig:watermelon_ratios} for the honeycomb lattice.

%%%%%%%%%%%%%%%%%%%%%%%%%%%%%%%%%%%%%%%%%%%%%%%%%%%%%%%%%%%%%%%%%%%%%%%%%%%%
%%%%%%%%%%%%%%%%%%%%%%%%%%%%%%%%%%%%%%%%%%%%%%%%%%%%%%%%%%%%%%%%%%%%%%%%%%%%

\newpage
%%%%%%%%%%%%%%%%%%%%%%%%%%%%%%%%%%%%%%%%%%%%%%%%%%%%%%%%%%%%%%%%%%%%%%%%%%%%
\subsection{Cardy's \texorpdfstring{${\bf L}$}{L}-lattice}
%%%%%%%%%%%%%%%%%%%%%%%%%%%%%%%%%%%%%%%%%%%%%%%%%%%%%%%%%%%%%%%%%%%%%%%%%%%%

As noted in the main text, the value of the critical exponent $\nu$ reported here differs notably from that reported previously in Ref.~\onlinecite{Serna_2021}, which employed loop simulations on Cardy's L-lattice~\cite{Cardy_2010}.
Here we re-examine the non-orientable loop model on the L-lattice in order to validate our results from the honeycomb lattice with NNN couplings.

\begin{figure}[h!]
    \centering
    \includegraphics{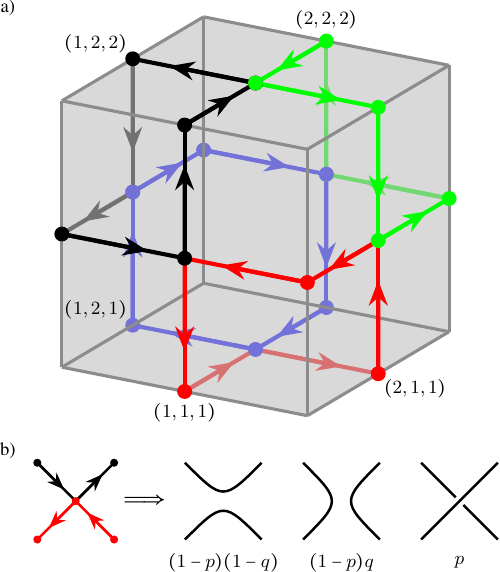}
    \caption{
    \textbf{Loop model on Cardy's ${\bf L}$-lattice.}
    (a) Unit cell of the loop model on Cardy's 3D $L$-lattice, with four-leg vertices living on the faces and edges of a cubic lattice.
    Coordinates are defined for the edge vertices on the top and bottom slices of the unit cell, which may contribute to the spanning number.
    (b) The parameterization of the loop model given in terms of the probabilities of the three ways to resolve the four-leg vertices.
    For $p=0$, the loops are oriented in the manner depicted in (a).
    Coloring of the loops in (a) identifies the elementary loops of length 6 which form for $p=q=0$.
    }
    \label{fig:L_lattice}
\end{figure}

The lattice geometry and loop model parameterization are defined in Fig.~\ref{fig:L_lattice}.
Taking a time slice through the unit cell, we see that only four of the loops may contribute to the spanning number.
We thus define the cross-sectional area of the unit cell to be $2 \times 2$ rather than $4 \times 4$ as was used by Ref.~\onlinecite{Serna_2021}.
For linear dimension $L$, we take the 3D lattice to have unit aspect ratio, corresponding to size $L \times L \times L$.
Here we focus on the line $p=q$, along which the loop model is always non-orientable.\\
%%%%%%%%%%%%%%%%%%%%%%%%%%%%%%%%%%%%%%%%%%%%%%%%%%%%%%%%%%%%%%%%%%%%%%%%%%%%
%%%%%%%%%%%%%%%%%%%%%%%%%%%%%%%%%%%%%%%%%%%%%%%%%%%%%%%%%%%%%%%%%%%%%%%%%%%%

%%%%%%%%%%%%%%%%%%%%%%%%%%%%%%%%%%%%%%%%%%%%%%%%%%%%%%%%%%%%%%%%%%%%%%%%%%%%
\subsubsection*{Correlation length exponent \texorpdfstring{$\nu$}{ν}}
%%%%%%%%%%%%%%%%%%%%%%%%%%%%%%%%%%%%%%%%%%%%%%%%%%%%%%%%%%%%%%%%%%%%%%%%%%%%

%%%%%%%%%%%%%%%%%%%%%%%%%%%%%%%%%%%%%%%%%%%%%%%%%%%%%%%%%%%%%%%%%%%%%%%%%%%%
\begin{figure}[t] 
    \centering
    \includegraphics[width=\columnwidth]{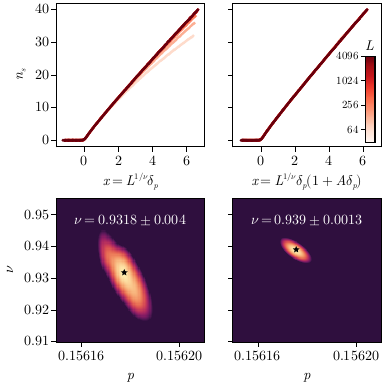}
    \caption{
    \textbf{Scaling collapse of spanning number on the ${\bf L}$-lattice}.
    Scaling collapse of the spanning number (top) near the transition and a heatmap of the corresponding normalized cost function (bottom), with optimal parameters marked with a star.
    Data are shown for linear system sizes $L \in \{64, 128, 256, 512, 1024, 2048, 4096\}$.
    We compare two forms of the scaling function $n_s(p,L) = f(x)$ with (left) $x=L^{1/\nu}\delta_p$ and (right) $x=L^{1/\nu}\delta_p(1+A\delta_p)$, for $\delta_p = p-p_c$.
    }
    \label{fig:L_lattice_FSS}
\end{figure}
%%%%%%%%%%%%%%%%%%%%%%%%%%%%%%%%%%%%%%%%%%%%%%%%%%%%%%%%%%%%%%%%%%%%%%%%%%%%

We compute the spanning number $n_s$ for a set of 100 points in close proximity to the transition for linear dimensions $L = \{64,128,256,512,1024,2048,4096\}$.
For each parameter $p=q$, we average over 100 independent ``pools'', drawing $10^4$ circuit trajectories from each such pool.
In order to fit the scaling function $f(x)$ we again employ a cubic B-spline.
Here we find it useful to take the spline points more densely clustered near $x=0$ where the curvature of $f(x)$ is greatest.
In particular, we follow Ref.~\onlinecite{Ortuno_2009}, taking spline points at spacing $\Delta x = 0.1$ for $x \in [-0.5, 1.5]$ and $\Delta x = 1$ for $x \in [2, 5]$.
While this choice of spline points gives a more robust fit, taking uniformly distributed spline points yields consistent results.
In Fig.~\ref{fig:L_lattice_FSS}, we show the scaling collapse resulting from two different ansatze.
For scaling variable $x = L^{1/\nu}(p-p_c)$, we obtain a good collapse with $\nu=0.9318\pm0.004$ when restricting to a narrow window about $x=0$ and system sizes $L \geq L_\textrm{min} = 256$.
When allowing for a {\it non-linear correction} to the scaling variable $x = L^{1/\nu}(p-p_c)(1 + A(p-p_c))$, we find a nearly perfect collapse with $\nu=0.939 \pm 0.001$ for \emph{all} system sizes and a much wider interval around the transition point.
Note that this value of $\nu$ also nicely matches the one found for the non-orientable circuit on the honeycomb lattice with NNN couplings.
Moreover, this FSS analysis is consistent with the earlier observation that a non-linear term in the scaling variable is necessary for robust scaling collapse.

\begin{figure}[t]
    \centering
    \includegraphics[width=.9\columnwidth]{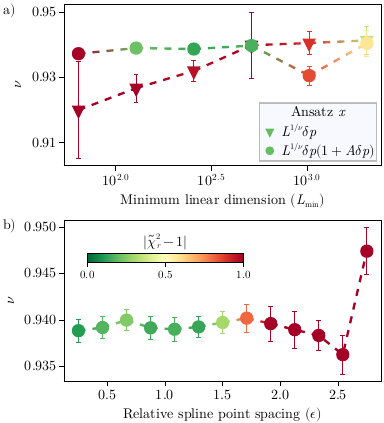}
    \caption{
    \textbf{Quality of scaling collapse for the ${\bf L}$-lattice}.
    We measure the quality of the fit by the reduced $\chi^2$ value, $\tilde{\chi}^2_r$, with $\tilde{\chi}^2_r \approx 1$ indicating good collapse.
    (a) Fitted value of $\nu$ and the fit quality for the two ansatzes shown in Fig.~\ref{fig:L_lattice_FSS} upon varying the minimum linear system size $L_\textrm{min}$ included in the fit.
    The non-linear ansatz shows a stable fitted exponent and high quality fit when excluding $L=64$.
    (b) Quality of fit for the non-linear ansatz when varying number of points used in constructing the spline.
    For very large spacing $\epsilon$ (as defined in the text), the spline fails to capture the scaling function.
    Elsewhere, the fit is robust with respect to the number of spline points, converging toward $\tilde{\chi}^2_r=1$.
    }
    \label{fig:L_lattice_FSS_quality}
\end{figure}

To further validate the reported value of $\nu$, we carry out a careful examination of the fitting procedure.
In Fig.~\ref{fig:L_lattice_FSS_quality}, we show the quality of fit and optimal value of $\nu$ as a function of (i) the minimum system size $L_\textrm{min}$, (ii) the choice of scaling variable, and (iii) the density of spline points.
We find that the simpler (linear) ansatz consistently underestimates the exponent $\nu$ relative to the non-linear ansatz except in the limit of very large system sizes.
Moreover, we see that the non-linear ansatz consistently results in a robust scaling collapse which is largely independent of the density of spline points.
This analysis provides strong support for the universality of critical exponent $\nu = 0.9403 \pm 0.0006$ in non-orientable loop models, independent of the underlying lattice geometry.

%%%%%%%%%%%%%%%%%%%%%%%%%%%%%%%%%%%%%%%%%%%%%%%%%%%%%%%%%%%%%%%%%%%%%%%%%%%%
%%%%%%%%%%%%%%%%%%%%%%%%%%%%%%%%%%%%%%%%%%%%%%%%%%%%%%%%%%%%%%%%%%%%%%%%%%%%

%%%%%%%%%%%%%%%%%%%%%%%%%%%%%%%%%%%%%%%%%%%%%%%%%%%%%%%%%%%%%%%%%%%%%%%%%%%%
\subsubsection*{Fisher exponent \texorpdfstring{$\tau$}{τ}}
%%%%%%%%%%%%%%%%%%%%%%%%%%%%%%%%%%%%%%%%%%%%%%%%%%%%%%%%%%%%%%%%%%%%%%%%%%%%

\begin{figure}[t]
    \centering
    \includegraphics[width=.95\columnwidth]{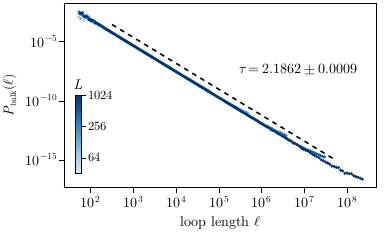}
    \caption{
    \textbf{Fisher exponent for Cardy's L-lattice.}
    We show the bulk length distribution of closed loops $P_\textrm{bulk}(\ell)$ at the critical point of the non-orientable loop model on Cardy's L-lattice along the line $p=q$ in parameter space.
    The critical point is fixed by the result of the FSS of the spanning number in Fig.~\ref{fig:L_lattice_FSS}.
    Data are shown for linear dimensions $L = 32, 64, 128, 256, 512, 1024$.
    The black dashed line corresponds to the power-law $P_\textrm{bulk}(\ell)\sim \ell^{-\tau}$ fit on the data in the interval $10^{5/2} \leq \ell \leq L^3/20$.
    }
    \label{fig:L_lattice_fractal_dim}
\end{figure}

As an additional check, we determine the Fisher exponent $\tau$ at the critical point from the power-law scaling of the bulk loop length distribution $P_\textrm{bulk}(\ell) \sim \ell^{-\tau}$.
Fixing the critical point from the FSS of the spanning number, we plot the distribution $P_\textrm{bulk}(\ell)$ in Fig.~\ref{fig:L_lattice_fractal_dim}.
The results here may be compared with those for the honeycomb lattice shown in Fig.~\ref{fig:HC_fractal_dim}(b), showing comparable values of the exponent $\tau$.
Taken together, the values of $\nu$ and $\tau$ determined here allow for the remaining critical exponents to be determined by hyperscaling relations.
In particular, we report the values of critical exponents $\eta = 5-\tfrac{6}{\tau-1}$, $d_f = 3/(\tau-1)$, and $\beta=3\nu(\tau-2)/(\tau-1)$ in Tab.~\ref{tab:crit_exp_lattice}.
All such exponents are in close agreement with the values found for the honeycomb lattice in the main text.

%\clearpage
%%%%%%%%%%%%%%%%%%%%%%%%%%%%%%%%%%%%%%%%%%%%%%%%%%%%%%%%%%%%%%%%%%%%%%%%%%%%
\section{Supplemental numerical data}\label{app:supplemental_data}
%%%%%%%%%%%%%%%%%%%%%%%%%%%%%%%%%%%%%%%%%%%%%%%%%%%%%%%%%%%%%%%%%%%%%%%%%%%%

\subsection{Entanglement scaling in the metallic phase}

\begin{figure}[h!]
    \centering
    \includegraphics[width=.95\columnwidth]{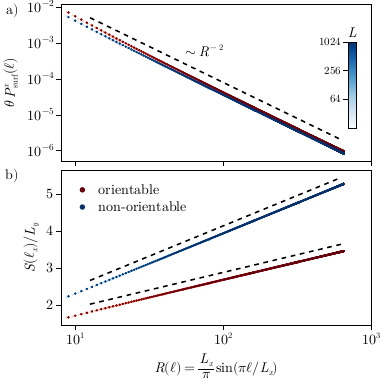}
    \caption{
        {\bf Steady-state entanglement entropy and loop length distribution in the liquid phase.}
        We show data for both symmetry classes at the isotropic point for linear system sizes $L=32, 64, 128, 256, 512, 1024$.
        (a) The distribution $P_\textrm{surf}^x(\ell)$ of surface loop lengths along the $\hat{x}$ direction, plotted against the arc-length $R(\ell) = \tfrac{L_x}{\pi} \sin\left(\tfrac{\pi \ell}{L_x}\right)$ appropriate for PBC.
        The distribution exhibits the expected critical $\ell^{-2}$ scaling (black dashed line) in both the orientable (red) and non-orientable (blue) symmetry classes.
        Normalizing $P_\textrm{surf}(\ell)$ by a factor of the universal quantity $\theta$ brings the magnitude in the two classes into agreement.
        (b) Entanglement entropy density $S(\ell_x) / L_y$ for a cylindrical subsystem of length $\ell_x$ and circumference $L_y$.
        Plotting against the arc length $R(\ell)$ yields a data collapse which highlights the logarithmic entanglement scaling $S(\ell_x) / L_y \sim \tilde{c}_x\log\left[R(\ell)\right]$ observed throughout the metallic phase.
        The ratio of $\tilde{c}^2$ in the two symmetry classes is comparable to the ratio of mean-squared displacement of a random walker on the two lattices after two steps.
    }
    \label{fig:isotropic_Psurf}
\end{figure}

Throughout the metallic phase of both the orientable and non-orientable models, we observe logarithmic entanglement scaling $S(L,L_y) ~ L_y \log(L)$ for a cylindrical subsystem of length $L$ along the $\hat{x}$ direction and circumference $L_y$.
This is accompanied by a power-law scaling of the surface loop length distribution $P_\textrm{surf}^\alpha(\ell) \sim \ell^{-2}$.
In Fig.~\ref{fig:isotropic_Psurf}, we verify the expected scaling of $P_\textrm{surf}^\alpha(\ell)$ and $S(L, L_y)$ at the isotropic point.
Notably, the prefactor of the logarithmic entanglement scaling is larger in the non-orientable symmetry class as alleviating the orientability constraint yields greater mobility for the random walk undergone by loop endpoints.
Moreover, the relative magnitudes of the logarithmic term in the symmetry classes is comparable to that predicted by the ratio of the mean-squared displacement $\langle\abs{\vec r}^2\rangle$ of the random walk.

%%%%%%%%%%%%%%%%%%%%%%%%%%%%%%%%%%%%%%%%%%%%%%%%%%%%%%%%%%%%%%%%%%%%%%%%%%%%
%%%%%%%%%%%%%%%%%%%%%%%%%%%%%%%%%%%%%%%%%%%%%%%%%%%%%%%%%%%%%%%%%%%%%%%%%%%%

\subsection{Bulk length of open arcs}

\begin{figure}[t]
    \centering
    \includegraphics[width=.95\columnwidth]{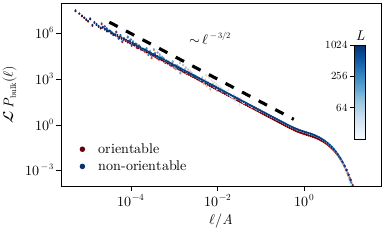}
    \caption{
        {\bf Distribution of bulk loop lengths for open boundaries.}
        $P_\textrm{bulk}(\ell)$ for \emph{open} loops in the critical phase when taking a pure initial state but not projecting out the final state produced by the circuit.
        Data are shown for $L=32,64,128,256,512,1024$ at the isotropic point.
        In both the orientable (red) and non-orientable (blue) symmetry classes, we observe a power-law scaling regime $P_\textrm{bulk}(\ell)\sim\ell^{-3/2}$ followed by an exponential cutoff.
    }
    \label{fig:open_loops_FBC}
\end{figure}

As noted in Sec.~\ref{ss:crit_phase}, an open boundary is \emph{absorbing} for the random walk undergone by loop endpoints.
Thus in the metallic phase, the \emph{bulk} length of open arcs is dictated by the first passage time distribution $P_\textrm{FP}(\ell) \propto \ell^{-3/2}$ of the random walk along the temporal direction.
This is shown clearly in Fig.~\ref{fig:open_loops_FBC} for both symmetry classes.
For finite system size $L$, the first passage time distribution acquires an exponential cutoff which is reflected in Fig.~\ref{fig:open_loops_FBC} for loop length $\ell$ comparable to the cross-sectional area $A$.

%%%%%%%%%%%%%%%%%%%%%%%%%%%%%%%%%%%%%%%%%%%%%%%%%%%%%%%%%%%%%%%%%%%%%%%%%%%%

\clearpage 

\end{appendix}

\end{document}